\documentclass[12pt,technote,onecolumn]{IEEEtran}
\usepackage{cite}

\ifCLASSINFOpdf
   \usepackage[pdftex]{graphicx}
\else
   \usepackage[dvips]{graphicx}
\fi
\usepackage{amsmath}
\usepackage{amssymb}

\newtheorem{theorem}{Theorem}
\newtheorem{corollary}{Corollary}
\newtheorem{lemma}{Lemma}
\newtheorem{assumption}{Assumption}
\newtheorem{proposition}{Proposition}
\newtheorem{definition}{Definition}
\newtheorem{example}{Example}
\newtheorem{fact}{Fact}

\usepackage{algorithm}
\usepackage{algorithmic}

\usepackage{array}
\begin{document}
%
\title{Approximate Value Iteration for Risk-aware Markov Decision Processes}
%
%
%

\author{Pengqian~Yu,
        William~B.~Haskell,
        and~Huan~Xu
\thanks{The work of W.~B.~Haskell was supported by the Ministry of Education of Singapore through grant R-266-000-083-133. The work of H.~Xu was supported by the Ministry of Education of Singapore through Tier-2 grant R-266-000-098-112.}        
\thanks{P.~Yu, W.~B.~Haskell and H.~Xu are with the Department of Industrial Systems Engineering and Management, National University of Singapore, 1 Engineering Drive 2, Singapore 117576, Singapore (e-mail: yupengqian@u.nus.edu; isehwb@nus.edu.sg; isexuh@nus.edu.sg).}}
\maketitle

\begin{abstract}
We consider large-scale Markov decision processes (MDPs) with a risk measure of variability in cost, under the risk-aware MDPs paradigm. Previous studies showed that risk-aware MDPs, based on a minimax approach to handling risk, can be solved using dynamic programming for small to medium sized problems. However, due to the ``curse of dimensionality'', MDPs that model real-life problems are typically prohibitively large for such approach. In this paper, we employ an approximate dynamic programming  approach, and develop a family of simulation-based algorithms to approximately solve large-scale risk-aware MDPs. In parallel, we develop a unified convergence analysis technique to derive sample complexity bounds for this new family of algorithms.
	
		
\end{abstract}


%
\IEEEpeerreviewmaketitle

\section{Introduction}
%
%
%
%
Markov decision processes (MDPs) (e.g., \cite{puterman2014markov,bertsekas1996neuro})
are a well established framework for modeling sequential decision-making
problems. They have been studied and applied extensively. The classical MDPs search for a policy with minimum expected cost. Nonetheless, it turns out that solely considering the expectation is insufficient in various applications (see the motivated example in \cite{ruszczynski2010risk}). In particular, the expected value can fail to be useful when there is significant stochasticity in the MDP transitions, which may lead to significant variability in the cost distribution~\cite{howard1972risk}.

The natural method for dealing with stochasticity, motivated by classical studies in the financial literature, is through the notion of risk, such as its exponential-utility \cite{howard1972risk}, variance \cite{markowitz1968portfolio}, or conditional value-at-risk (CVaR) \cite{rockafellar2000optimization}. Such measures capture the variability of the cost, or quantify the effect of rare but potentially disastrous outcomes. The risk measure is extended to the setting of sequential optimization problems (e.g., \cite{howard1972risk,bauerle2013more}), in which the objective is to minimize a risk measure defined over the whole time horizon. In this setting, the total cost is considered as a standard random variable, without any regard to the temporal nature of the process generating it. In particular, expected utility minimizing MDPs are considered earlier in \cite{kreps1977decisionb}. MDPs with variance-related criteria are studied in \cite{Filar1989}, while CVaR minimizing MDPs are explored in \cite{borkar2014risk}. It was shown that problems of this type can be difficult \cite{le2007robust}, and even for the mean-variance model the Bellman's principle of optimality does not hold and the associated MDPs are NP-hard \cite{mannor2013algorithmic}. Moreover, these problems may lead to ``time inconsistent'' phenomenon, i.e., the analysis of risk in a multi-period setting
can be a treacherous exercise as identical risk preferences can imply vastly different decisions at different time periods \cite{iancu2015tight}. To resolve the issues of these models, time-consistent Markov risk measures were proposed in \cite{ruszczynski2010risk}. The concept of time consistency (e.g., \cite{Shapiro:2009fk}) is usually defined as follows: if a certain outcome is considered less risky in all states of the world at stage $t+1$, then it should also be considered less risky at stage $t$. Markov risk measures capture the multi-period nature of the decision-making process in the definition of the risk, and can be written as compositions of one-step conditional risk measures (these are simply risk measures defined in a conditional setting, analogous to the conditional expectation for the traditional case). In addition, Markov risk measures are notable because they readily yield minimax formulation and the corresponding optimal solution can be obtained using dynamic programming (DP)~\cite{ruszczynski2010risk}, at least for small to medium sized MDPs. Broadly speaking, the risk-aware dynamic programming is useful in settings with either heavy-tailed distributions or rare high-impact events. For example, heavy-tailed distributions arise frequently in finance (e.g., \cite{bystrom2005extreme,kim2011hour}) as well as energy and sustainability \cite{jiang2015approximate}; rare high-impact events may appear in inventory problems \cite{glasserman1996rare} as well as management of high-value assets \cite{enders2010dynamic}. 

This paper considers planning in large risk-aware MDPs with Markov risk measures. It is widely known that, due to the ``curse of dimensionality,'' practical problems modeled as MDPs often have prohibitively large state spaces, under which the previous work \cite{ruszczynski2010risk} with exact DP approach becomes intractable. Many approximation schemes have been proposed to alleviate the curse of dimensionality of large-scale risk-neutral MDPs, among which approximate dynamic programming (ADP) is a popular approach and has been used successfully in large-scale problems with hundreds of state dimensions  \cite{powell2007approximate}. Simulation-based algorithms, algorithms which randomly sample the MDP state space and simulate MDP trajectories, comprise a large part of the work on ADP. They have been shown to give good solutions with high probability for classical MDPs (e.g., \cite{de2003linear,munos2008finite,rust1997using,Haskell_EDP_2015}). 

There is considerable development of simulation-based algorithms for risk-aware MDPs with Markov risk measures in the literature, but the computational and theoretical challenges have not been explored as thoroughly. Specifically, the recent work \cite{jiang2015approximate} proposes a simulation-based ADP algorithm for risk-aware MDPs. However, it has limited use since it only considers a specific choice of Markov risk measures called dynamic quantile-based risk measures. A cutting plane algorithm for time-consistent multistage linear stochastic programming problems is given in \cite{asamov2014time}, but restricted to finite decision horizons. In \cite{tamar2016sequential}, an actor-critic style sampling-based algorithm for Markov risk is developed. Although the sensitivity of approximation error is analyzed, the algorithm can only search for a locally optimal policy. Risk-averse dual dynamic programming is introduced in \cite{Petrik:2012} for MDPs with hybrid continuous-discrete state space. Even though the method yields an output that converges to the optimal solution, the significant weaknesses are that it requires the linearity of state and action spaces, and the convergence criterion is not well defined. Our goal in this paper is to consider the whole class of Markov risk measures, propose a new simulation-based ADP approach, and develop improved convergence results and error bounds.

Our first contribution is a new family of computationally tractable and simulation-based algorithms for
risk-aware MDPs with infinite state space.  We show how to develop risk-aware analogs of several
major simulation-based algorithms for classical MDPs (e.g., \cite{jain2010simulation,munos2008finite}), which cannot optimize Markov risk measures.  In particular, the main novelty of our proposed algorithms is twofold. First, not all existing ADP techniques for classical MDPs are proper for the risk-aware setting. A typical example is the approximate linear programming approach \cite{de2003linear} which yields a non-convex formulation in our setting. Second, the empirical estimation of risk is more complex than the empirical estimation of expectation in  classical ADP algorithms (e.g., \cite{munos2008finite,Haskell_EDP_2015}). We use extensive numerical experiments to verify the validity and effectiveness of our proposed algorithms for risk-aware MDPs. To the best of our knowledge, it is the first time approximate value iteration has been proposed for Markov risk measures in the risk-aware MDPs literature. 

The second contribution of the paper is a unified convergence and sample complexity analysis technique that applies to a broad family of algorithms, including all of the algorithms considered
in this paper. The technique is inspired by the existing convergence analysis for classical MDPs such as weighted $p-$norm performance bounds\cite{munos2008finite}, supremum norm analysis\cite{jain2010simulation} and stochastic dominance framework\cite{Haskell_EDP_2015}. Yet, we must extend the existing convergence analysis to the minimax setting, which covers risk-aware MDPs. The critical difference in our approach is that in the risk-aware setting, we have the added difficulty in bounding approximation errors in both each and final iterations due to the minimax DP formulation.

This paper is organized as follows. In Section \ref{sec2} we review necessary
preliminaries for classical and risk-aware MDPs. Next, in Section
\ref{sec3} we propose and discuss a general family of simulation-based algorithms
for risk-aware MDPs and report their convergence results. Section
\ref{sec4} then focuses on the key issue of empirical estimation of risk functions,
which plays a major role in all of our algorithms. Section \ref{sec5} offers
an alternative convergence analysis based on the technique in \cite{Haskell_EDP_2015}.
In the following Section \ref{sec6} we present the proofs of all of our main
results. Section \ref{sec7} reports numerical experiments that serve to illustrate
the methods in this paper, and we conclude in Section \ref{sec8}. Proofs of all technical results can be found in the Appendix. 


\section{Preliminaries}\label{sec2}

This section reviews important preliminary concepts for both classical
and risk-aware MDPs.

\subsection{Classical MDP}\label{sec2.1}
A discounted MDP is defined as a 5-tuple
$
\left(\mathbb{S},\mathbb{A},P,c,\gamma\right),
$
where $\mathbb{S}$ and $\mathbb{A}$ are the state and action space, $P(\cdot|s,a)$ is the transition probability distribution, $c(s,a)$ is a bounded, deterministic, and state-action dependent cost, and $0<\gamma<1$ is a discount factor. In this paper, we consider continuous state space, finite action MDPs (i.e., the cardinality  $|\mathbb{A}|<+\infty$). For the sake of simplicity, we assume that $\mathbb{S}$ is a bounded, closed subset of a Euclidean space $\mathbb{R}^d$. Let $\mathbb{K}\triangleq\mathbb{S}\times\mathbb{A}$ denote the
set of all state-action pairs. We make the following assumption on the cost function $c$ throughout
this paper.
\begin{assumption}
	\label{assu:Bounded_costs}  $0\leq c(s,a)\leq c_{\max}<+\infty$
	for all $\left(s,a\right)\in\mathbb{K}$.
\end{assumption}

Let $J_{\max}\triangleq c_{\max}/\left(1-\gamma\right)$. We denote the space of bounded measurable functions with domain $\mathbb{S}$ as $B(\mathbb{S})$ and the space of measurable functions $f\mbox{ : }\mathbb{S}\rightarrow\mathbb{R}$
bounded by $J_{\max}$ as $B\left(\mathbb{S};J_{\max}\right)$. Let $\mathcal{B}(\mathbb{S})$ be a Borel $\sigma-$algebra and $\mathcal{P}\left(\mathbb{S}\right)$
be the space of probability measures over $\mathbb{S}$ w.r.t. $\mathcal{B}\left(\mathbb{S}\right)$. For a probability measure $\mu\mathcal{\in P\left(\mathbb{S}\right)}$
and $1\leq p<+\infty$, we let $\mathcal{L}_{p}(\mathbb{S},\mathcal{B}\left(\mathbb{S}\right),\mu)$
be the space of measurable mappings $f\mbox{ : }\mathbb{S}\rightarrow\mathbb{R}$
such that $\|f\|_{p,\mu}\triangleq(\int|f\left(s\right)|^{p}\mu\left(ds\right))^{1/p}<+\infty$. Furthermore, we denote by $\Pi$ the class of \textit{stationary deterministic Markov policies}:
mappings $\pi\mbox{ : }\mathbb{S}\rightarrow\mathbb{A}$ which only
depend on history through the current state. We only consider such
policies since it is well known that there is an optimal policy within
this class for classical MDPs~\cite{puterman2014markov}\footnote{For coherent Markov risk measures studied in this paper, the optimal policies belong to $\Pi$~\cite{ruszczynski2010risk}, while the optimal policies for the risk measures of the total cost may be history-dependent.}. For a given state $s\in\mathbb{S}$,
$\pi\left(s\right)\in\mathbb{A}$ is the action chosen in state $s$
under the policy $\pi$. The deterministic stationary policy $\pi$
defines the transition probability kernel $P^{\pi}$ according to
$P^{\pi}\left(dy\vert s\right)=P\left(dy\vert s,\pi\left(s\right)\right)$.
We define two operators related to $P^{\pi}$. The right-linear operator
$P^{\pi}\left(\cdot\right)\mbox{ : }B(\mathbb{S})\rightarrow B(\mathbb{S})$
is defined as
$
\left(P^{\pi}J\right)\left(s\right)=\int J\left(y\right)P^{\pi}\left(dy|s\right),
$
where $J\in B\left(\mathbb{S}\right),$ and the left-linear
operator $\left(\cdot\right)P^{\pi}:\mathcal{P}\left(\mathbb{S}\right)\rightarrow\mathcal{P}\left(\mathbb{S}\right)$
is defined as
$
\left(\mu P^{\pi}\right)\left(dy\right)=\int P^{\pi}\left(dy|s\right)\mu\left(ds\right),
$
where $\mu\in\mathcal{P}\left(\mathbb{S}\right)$. The product of
two transition kernels is defined in the natural way
$
P^{\pi_{1}}P^{\pi_{2}}\left(dz\vert s\right)=\int P^{\pi_{1}}\left(dy\vert s\right)P^{\pi_{2}}\left(dz\vert y\right).
$

The state and action at time $t\geq0$ are denoted by $s_{t}$ and
$a_{t}$, respectively. Any policy $\pi\in\Pi$ and initial state
$s_{0}\in\mathbb{S}$ determine a probability measure $P_{s_{0}}^{\pi}$
and an associated stochastic process $\left\{ \left(s_{t},a_{t}\right),t\geq0\right\} $
defined on the canonical measurable space of trajectories of state-action
pairs. The expectation operator w.r.t. $P_{s_{0}}^{\pi}$
is denoted $\mathbb{E}_{s_{0}}^{\pi}[\cdot]$. The classical
risk-neutral MDP is
\begin{equation}
\inf_{\pi\in\Pi}\mathbb{E}_{s_{0}}^{\pi}\left[\sum_{t=0}^{\infty}\gamma^{t}c(s_t,a_t)\right].\label{NEUTRAL}
\end{equation}
There are many algorithms available to solve Problem (\ref{NEUTRAL}),
such as value iteration, policy iteration, and linear programming.

\subsection{Risk-aware MDP}\label{sec2.2}

Problem (\ref{NEUTRAL}) does not account for the risk incurred due
to the underlying stochasticity in state transitions. The family of
Markov risk measures was first proposed in \cite{ruszczynski2010risk}
as a way to model and mitigate this risk. As mentioned earlier, this
class of risk measures has a special form based on risk transition
mappings which readily leads to a minimax DP solution approach.

To formalize Markov risk measures~\cite{ruszczynski2010risk}, we define a family of admissible
random variables on the state space $\left(\mathbb{S},\mathcal{B}\left(\mathbb{S}\right)\right)$.
For a fixed probability measure $P_{0}$ on $\left(\mathbb{S},\mathcal{B}\left(\mathbb{S}\right)\right)$,
we can define the space $\mathcal{L}=\mathcal{L}_{\infty}\left(\mathbb{S},\mathcal{B}\left(\mathbb{S}\right),P_{0}\right)$
of essentially bounded measurable mappings on $\mathbb{S}$. A risk measure $\rho\mbox{ : }\mathcal{L}\rightarrow\mathbb{R}$ is
called ``coherent'' if it satisfies convexity, monotonicity, translation equivariance and positive homogeneity properties (see \cite{artzner1999coherent} for details). Mean-deviation, mean-semideviation and CVaR are examples of coherent risk functions. Given the initial state $s_{0}\in\mathbb{S}$ and discount factor $\gamma$, the infinite-horizon risk-aware MDP is
\begin{equation}
\inf_{\pi\in\Pi}J^{\pi}\left(s_{0}\right).\label{RISK}
\end{equation}
Here, the risk-to-go function $J^{\pi}$ for any given $\pi$
is defined as
\begin{equation}\label{iterated risk}
\begin{aligned}
J^{\pi}(s_{0})\triangleq\,c(s_0,a_0)+\rho(\gamma c(s_1,a_1)+\rho(\gamma^{2}c(s_2,a_2)+\cdots)),
\end{aligned}
\end{equation}
where each $\rho$ is a coherent one-step conditional risk measure (see \cite{ruszczynski2006conditional,ruszczynski2010risk}), and the evaluation of $\rho$ is Markov, in the sense that it is not allowed to depend on the whole past, and $s_0,a_0,s_1,a_1,\dots$ is a trajectory drawn of the MDP under policy $\pi$. Note that $J^{\pi}$ is defined through nested and multi-stage compositions of $\rho$ (rather than through a single $\rho(\sum_{t=0}^{\infty}\gamma^t c(s_t,a_t))$) and each stage is a risk-measure of the remaining future risk-to-go (see \cite{ruszczynski2010risk} for details). Given a sequence of discounted costs $c(s_0,a_0),\gamma c(s_1,a_1),\dots$, the intuitive meaning of $J^{\pi}(s_{0})$ is a certainty equivalent cost (i.e., at time $0$, one is indifferent between incurring $J^{\pi}(s_{0})$ and the alternative of being subjected to the stream of stochastic future discounted costs; see \cite{rudloff2014time} for an in-depth discussion regarding the certainty equivalent interpretation in the context of multistage stochastic models).

In the next lemma, we confirm that the risk-to-go functions are uniformly bounded and belong to $B(\mathbb{S};J_{\max})$. 
\begin{lemma}
	\label{lem:Bounded} Let Assumption \ref{assu:Bounded_costs}
	hold. For all $\pi\in\Pi$, we have $\|J^{\pi}\|_{\infty}\leq J_{\max}$.
\end{lemma}

A risk-aware Bellman operator is developed for Problem (\ref{RISK})
in \cite[Theorem 4]{ruszczynski2010risk}. We emphasize that the one-step conditional risk measure $\rho$ depends on the underlying transition kernel, and we define the risk-aware Bellman operator $T\mbox{ : }B(\mathbb{S};J_{\max})\rightarrow B(\mathbb{S};J_{\max})$ as
\begin{equation}
\left[TJ\right]\left(s\right)=\min_{a\in\mathbb{A}}\left\{ c(s,a)+\gamma\rho\left(J\left(Y^{s,a}\right)\right)\right\},\,\forall s\in\mathbb{S}.\label{eq:RISK_Bellman}
\end{equation}
When $\rho(Y^{s,a})=\mathbb{E}_{Y^{s,a}\sim P(\cdot\vert s,a)}[Y^{s,a}]$,
$T$ is just the classical Bellman operator for Problem (\ref{NEUTRAL}).
Coherent risk measures have a special representation via Fenchel duality
\cite{Ruszczynski:2006uq} which lead to minimax DP
equations~\cite{ruszczynski2010risk}. Since $\rho$
is coherent, for all $s\in\mathbb{S}$, by \cite[Theorem 2.2]{Ruszczynski:2006uq}, the risk-aware
Bellman operator $T$ has a minimax structure
\begin{equation}
\left[TJ\right]\left(s\right)=\min_{a\in\mathbb{A}}\left\{ c(s,a)+\gamma\max_{\mu\in\mathcal{Q}\left(s,a\right)}\mathbb{E}_{Y^{s,a}\sim\mu}\left[J\left(Y^{s,a}\right)\right]\right\},\label{eq:RISK_Bellman-1}
\end{equation}
where $\{\mathcal{Q}(s,a)\} _{\left(s,a\right)\in\mathbb{K}}$
is a collection of distributional sets on $(\mathbb{S},\mathcal{B}(\mathbb{S}))$.
The two representations (\ref{eq:RISK_Bellman}) and (\ref{eq:RISK_Bellman-1})
of $T$ are equivalent, but we often find advantage in using one form
over the other.

We define the following notation to capture the dependence on our
sets of distributions $\{\mathcal{Q}(s,a)\}_{(s,a)\in\mathbb{K}}$.
For fixed $\pi\in\Pi,$ we define a stochastic kernel $Q^{\pi}\mbox{ : }B(\mathbb{S};J_{\max})\rightarrow B(\mathbb{S};J_{\max})$ such that $Q^{\pi}(\cdot\vert s)\in\mathcal{Q}(s,\pi(s))$
is an element of the distributional set $\mathcal{Q}(s,a)$
when $a=\pi(s)$, for all $s\in\mathbb{S}$. Note that
$Q^{\pi}(\cdot\vert s)$ is a probability distribution
on $\mathbb{S}$ for all $s\in\mathbb{S}$. The right-linear operator
$Q^{\pi}(\cdot)$ and left-linear operator $(\cdot)Q^{\pi}$
can be defined similarly as those for $P^{\pi}.$ We
say that a policy $\pi$ is greedy w.r.t. the risk-to-go function
$J\in B(\mathbb{S};J_{\max})$ if
\[
\pi\left(s\right)\in\arg\min_{a\in\mathbb{A}}\left\{ c(s,a)+\gamma\rho\left(J\left(Y^{s,a}\right)\right)\right\},\,\forall s\in\mathbb{S}.
\]
We let $J^{*}\in B\left(\mathbb{S};J_{\max}\right)$ be the optimal
risk-to-go function for the risk-aware Bellman operator 
$$T:TJ^{*}=J^{*},$$ 
and $\pi^{*}\mbox{ : }\mathbb{S}\rightarrow\mathbb{A}$ be any optimal
policy satisfying
\[
\pi^{*}\left(s\right)\in\arg\min_{a\in\mathbb{A}}\left\{ c(s,a)+\gamma\rho\left(J^*\left(Y^{s,a}\right)\right)\right\},\,\forall s\in\mathbb{S}.
\]

\subsection{Notation}

For ease of reference, we summarize the notation used in this paper in Table \ref{tab:notations}.

\begin{table}[!tph]
	\centering
	\caption{A summary of notations \label{tab:notations}}
	\begin{tabular}{|c|l|}
		\hline
		Symbol                                                                 & \multicolumn{1}{c|}{Meaning}                                                                         \\ \hline
		$\mathbb{S}$                                                           & State space                                                                                          \\ \hline
		$\mathcal{S}$                                                          & An $\epsilon-$net on $\mathbb{S}$                                                                    \\ \hline
		$B\left(\mathbb{S};J_{\max}\right)$                                   & Space of measurable functions on $\mathbb{S}$ bounded by  $J_{\max}$                                              \\ \hline
		$\mathcal{P\left(\mathbb{S}\right)}$                                   & Space of probability measures over $\mathbb{S}$ with respect to $\mathcal{B}\left(\mathbb{S}\right)$ \\ \hline
		$\mathbb{A}$                                                           & Action space; assumed to be finite                                                                   \\ \hline
		$P$                                                                    & Transition probability kernel                                                                        \\ \hline
		$c$                                                                    & Cost function; assumed to be measurable and bounded                                                  \\ \hline
		$\gamma$                                                               & Discount factor; $0<\gamma<1$                                                                        \\ \hline
		$\pi$                                                                  & Policy; $\pi\in\Pi$                                                                                  \\ \hline
		$\Pi$                                                                  & Class of stationary deterministic Markov policies                                                    \\ \hline
		$J$                                                                    & Risk-to-go function                                                                                  \\ \hline
		$J^{\pi}$                                                              & Risk-to-go function for a given policy $\pi$                                                         \\ \hline
		$J^{*}$                                                                & Optimal risk-to-go function; $J^{*}=\min_{\pi\in\Pi}J^{\pi}$                                         \\ \hline
		$\widehat{J}_{k}$                                                      & Approximate risk-to-go function at iteration $k$                                                     \\ \hline
		$\widehat{\pi}_{k}$                                                    & Greedy policy with respect to $\widehat{J}_{k}$ at iteration $k$                                     \\ \hline
		$T$                                                                    & Risk-aware Bellman operator                                                                          \\ \hline
		$T^{\pi}$                                                              & Risk-aware Bellman operatorfor fixed policy $\pi$                                                    \\ \hline
		$\widehat{T}$                                                          & Random risk-aware Bellman operator                                                                   \\ \hline
		$\varepsilon_{k}$                                                      & Approximation error of the Bellman operator in iteration $k$                                         \\ \hline
		$\epsilon_{g}$                                                         & Granularity for stochastic dominance convergence analysis                                            \\ \hline
	\end{tabular}
\end{table}

\section{The algorithms and main results}\label{sec3}

In this section, we review the framework of our simulation-based
algorithms for risk-aware MDPs. It broadly
consists of three steps:
\begin{enumerate}
	\item[1] A random sampling scheme for $\mathbb{S}$. Using random sampling from a fixed distribution $\mu\in\mathcal{P}\left(\mathbb{S}\right)$ on $\mathbb{S}$, we construct a subset $\{s_1,\dots, s_n\}\subset\mathbb{S}$	at which to approximate the Bellman update\footnote{An alternative choice of the subset $\{s_1,\dots,s_n\}$ appears in our supremum analysis, and it is constructed deterministically as an $\epsilon-$net.}. 
	\item[2] An estimation scheme to approximate the Bellman update at each of the sampled states in $\{s_1,\dots, s_n\}$. This step  depends
	on simulation to generate samples of the next state visited. Here we must use novel technique to estimate the risk-to-go.
	\item[3] A function fitting scheme to extend the estimates on $\{s_1,\dots, s_n\}$
to a function on the entire $\mathbb{S}$.
\end{enumerate}

Simulation-based algorithms for classical MDPs also consist of these
three steps. As we will see, the major difference between simulation-based
algorithms for classical MDPs and those for risk-aware MDPs shows
in the above step 2. We next summarize the general framework for our proposed algorithms in Algorithm \ref{algo}, which closely
resembles the steps of the main algorithm in \cite{munos2008finite}.

\begin{algorithm}[!tph] 	 	 	 	 		 		
	\caption{Simulation-based approximate value iteration} \label{algo}	 	 	 	 		 		
	\begin{algorithmic}[1] 	
	\renewcommand{\algorithmicrequire}{\textbf{Input:}}
	\renewcommand{\algorithmicensure}{\textbf{Output:}}	 		 		 		 			 		
	\REQUIRE Functional family $\mathcal{F}\subset B\left(\mathbb{S};J_{\max}\right)$, initial risk-to-go function $\widehat{J}_0\in \mathcal{F}$ and sample distribution $\mu\in\mathcal{P}\left(\mathbb{S}\right)$.  	
	\FOR {$k=0,1,2,\dots$}	
	\STATE Construct a subset $\{s_1,\dots, s_n\}\subset\mathbb{S}$ where $s_i$'s  are sampled  from distribution $\mu$ independent of each other.		 		 		 			 			
	\FOR {$i=1,2,\dots,n$} 	 			
	\STATE Compute
	\begin{equation*}
	\widetilde{J}(s_i)=\min_{a\in\mathbb{A}}\left\{c(s_i,a)+\gamma\hat{\rho}_{m}\left(\left\{\widehat{J}_k\left(Y_j^{s_i,a}\right)\right\}^m_{j=1}\right)\right\}, 
	\end{equation*}		 		 		
	where $\hat{\rho}_{m}(\{\widehat{J}_k(Y_j^{s_i,a})\}^m_{j=1})$ is the empirical estimation of $\rho(\widehat{J}_k(Y^{s_i,a}))$, and $\{Y_j^{s_i,a}\}_{j=1}^m$ are $m$ i.i.d. samples of  transitions from $P(\cdot|s_i,a)$. 		 			 			
	\ENDFOR 		 		 		 			 			
	\STATE Compute the best fit $\widehat{J}_{k+1}\in\mathcal{F}$ to the data  $\{(s_i,\widetilde{J}(s_i))\}_{i=1}^n$.	
	\ENDFOR 		 		 		 			 			
	\ENSURE A sequence of risk-to-go functions $\{\widehat{J}_k\}_{k\geq0}$.  	 	 	 	 		 		
	\end{algorithmic}     	 	
\end{algorithm}	    	 	

Before we present our main results, a discussion of the
estimated risk value $\hat{\rho}_{m}$ is needed. For the remainder
of this technical note, we let $m\geq1$ be the number of transitions sampled
at each state and we let $\hat{\rho}_{m}$ be the empirical estimation
of $\rho$ using $m\geq1$ samples. We make a key assumption about
risk-to-go estimation. 
\begin{assumption}
	\label{assu:Risk_estimation} For any $s\in\mathbb{{S}}$,
	$a\in\mathbb{A},$ $J\in B\left(\mathbb{S};J_{\max}\right)$, and
	$\varepsilon>0$,	
	\[
	\mathbb{P}\left(\left|\rho\left(J\left(Y^{s,a}\right)\right)-\hat{\rho}_{m}\left(\left\{J\left(Y_j^{s,a}\right)\right\}^m_{j=1}\right)\right|>\varepsilon\right)\leq\theta\left(\varepsilon,m\right),
	\]
	where $Y^{s,a}\sim P\left(\cdot\vert s,a\right)$, $\theta\left(\varepsilon,m\right)\in\left(0,1\right)$,
	and $\theta\left(\varepsilon,m\right)\rightarrow0$ as $m\rightarrow+\infty.$
\end{assumption}

Assumption \ref{assu:Risk_estimation} essentially means that the
empirical risk measure $\hat{\rho}_{m}$ becomes exact as number of
samples $m$ approaches infinity. The specific form of $\theta\left(\varepsilon,m\right)$
depends on the details of the risk measures, which will be discussed
in Section \ref{sec4}.

We are interested in the rate that our risk-to-go estimates approach the optimal risk-to-go in the $p-$norm and the supremum norm, respectively. The key difference comes in the function
fitting step 3. First, a general function fitting scheme in
the $p-$norm is used. This analysis is more difficult than the supremum norm analysis because we
cannot use a contracting property of $T$ w.r.t. this norm.
In addition, the supremum norm is quite conservative and we get much more
optimistic error guarantees w.r.t. $p-$norm, thus justifying
the extra effort required. Second, we analyze convergence in the supremum norm.
This analysis follows readily because $T$ is a contraction operator
in the supremum norm. In both cases, we want to show that our risk-to-go estimates 
get close to the optimal risk-to-go with high
probability as the number of iterations and the number of samples
becomes large. For later use, we make the error in the sequence $\{ \widehat{J}_{k}\} _{k\geq0}$
explicit by writing
\begin{equation}
\widehat{J}_{k+1}=T\widehat{J}_{k}-\varepsilon_{k},\quad\forall k\geq0,\label{eq:Error}
\end{equation}
where $\varepsilon_{k}\in B(\mathbb{S})$ is the
error incurred by one iteration of our algorithm due to sampling and
function fitting.

\subsection{$p-$norm}\label{sec3.2}

In this subsection, we conduct a convergence analysis in the $p-$norm for $1\leq p<+\infty$. We remark that the Bellman operator $T$ is not a contraction operator w.r.t. this family
of norms. Instead, we develop analogs of the point-wise inequalities
developed in \cite{munos2008finite} for the risk-neutral case.

First, we discuss some details of lines $3-6$ in Algorithm \ref{algo}. In the $k^{th}$ iteration, given $\widehat{J}_{k},$ for $i=1,2,\dots,n,$
the function $\widehat{J}_{k+1}$ is computed as follows 
\begin{IEEEeqnarray}{rCl}
	\widetilde{J}\left(s_{i}\right) & = & \min_{a\in\mathbb{A}}\left\{ c(s_{i},a)+\gamma\hat{\rho}_{m}\left(\left\{\widehat{J}_k\left(Y_j^{s_i,a}\right)\right\}^m_{j=1}\right)\right\},\label{eq:Estimation}\\
	\widehat{J}_{k+1} & = & \arg\min_{f\in\mathcal{F}}\sum_{i=1}^{n}|f\left(s_{i}\right)-\widetilde{J}\left(s_{i}\right)|^{p}.\label{eq:Function fitting}
\end{IEEEeqnarray}

Let $\widehat{\pi}_{k}$ be a greedy policy w.r.t. $\widehat{J}_{k}$,
i.e., $T^{\widehat{\pi}_{k}}\widehat{J}_{k}=T\widehat{J}_{k}$.
We are interested in bounding the $\mathcal{L}_{p}$-error of the
optimality gap $\Vert J^{\widehat{\pi}_{k}}-J^{*}\Vert _{p,\varrho}$.
Here $\varrho$ is a distribution whose role is to put more weight
on those parts of state space where performance matters more. When
$p=1$ and $p\rightarrow\infty$, we recover the expected and supremum-norm
loss, respectively. The functional family $\mathcal{F}$ is generally selected to be a finitely parameterized class of functions
$$\mathcal{F}=\left\{f_\theta\in B(\mathbb{S};J_{\max}):\, \theta\in\Theta,\,\text{dim}(\Theta)<+\infty\right\}.$$

Our $p-$norm results apply to both linear ($f_\theta(x)=\theta^\top\phi(x)$) and non-linear ($f_\theta(x)=f(x;\theta)$) parameterizations, such as wavelet based approximations, multi-layer neural networks or kernel-based regression techniques. Given a (positive definite) kernel function
$\mathcal{K}$, another choice of $\mathcal{F}$ is a closed convex subset of the reproducing-kernel Hilbert-space (RKHS) associated to $\mathcal{K}$. 

To continue, we define the metric projection of $f$ onto $\mathcal{F}$
w.r.t. the norm on $\mathcal{L}_{p}\left(\mathbb{S},\mathcal{B}\left(\mathbb{S}\right),\mu\right)$
by 
$$
\Pi_{\mathcal{F}}\left(f\right)\triangleq\arg\min_{g\in\mathcal{F}}\|f-g\|_{p,\mu}.
$$
Similar to \cite{munos2008finite}, the approximation error is defined
by
\[
d_{p,\mu}\left(TJ,\mathcal{F}\right)=\|\Pi_{\mathcal{F}}\left(TJ\right)-TJ\|_{p,\mu}=\inf_{f\in\mathcal{F}}\|f-TJ\|_{p,\mu}.
\]
The inherent Bellman error defined by
\[
d_{p,\mu}\left(T\mathcal{F},\mathcal{F}\right)\triangleq\sup_{f\in\mathcal{F}}d_{p,\mu}\left(Tf,\mathcal{F}\right)
\]
is a key measure of the approximation power of $\mathcal{F}$
w.r.t. the norm on $\mathcal{L}_{p}(\mathbb{S},\mathcal{B}(\mathbb{S}),\mu)$,
this constant will appear throughout our analysis. When $\mathcal{F}$
is infinite, the ``capacity'' of $\mathcal{F}$
can be measured by the (empirical) covering number of $\mathcal{F}.$
Let $\varepsilon>0$, $q\geq1$, $s^{1:n}\triangleq(s_{1},\dots,s_{n})\in\mathbb{R}^{n}$
be fixed. The $\left(\varepsilon,q\right)$-covering number of the
set $\mathcal{F}(s^{1:n})=\{ (f(s_{1}),\dots,f(s_{n}))\vert f\in\mathcal{F}\} $
is the smallest integer $v$ such that $\mathcal{F}(s^{1:n})$
can be covered by $v$ balls of the normed space $(\mathbb{R}^{n},\Vert \cdot\Vert _{q})$
with centers in $\mathcal{F}(s^{1:n})$ and radius $n^{1/q}\varepsilon$.
The $\left(\varepsilon,q\right)$-covering number of the set $\mathcal{F}(s^{1:n})$
is denoted by $\mathcal{N}_{q}(\varepsilon,\mathcal{F}(s^{1:n}))$.
When $q=1$, we use $\mathcal{N}$ instead of $\mathcal{N}_{1}$.
When $s^{1:n}$ are i.i.d. with common underlying distribution $\mu$
then $\mathbb{E}[\mathcal{N}_{q}(\varepsilon,\mathcal{F}(s^{1:n}))]$
shall be denoted by $\mathcal{N}_{q}(\varepsilon,\mathcal{F},n,\mu).$ For specific choices of $\mathcal{F},$ it
is possible to bound covering number as a function of pseudo-dimension
of the function class. 


Let us discuss the condition that allows us to derive $\mathcal{L}_{p}$
error bounds. If the error in any given iteration can be bounded,
it remains to show that the error does not blow up as it is propagated
though the algorithm. Similar to\cite[Assumption A2]{munos2008finite},
we make an assumption about the operator norms of weighted sums of
the product of arbitrary stochastic kernels $Q^{\pi}$ defined in
Section \ref{sec2.2}.
\begin{assumption}
	\label{assu:Absolute_continuity} Given $\varrho,\mu\in\mathcal{P}\left(\mathbb{S}\right)$,
	$M\geq1$, and an arbitrary sequence of policies $\{ \pi_{M}\} _{M\geq1}$. Assume the future-state distribution $\varrho Q^{\pi_{1}}Q^{\pi_{2}}\dots Q^{\pi_{M}}$
	for any such selection $Q^{\pi_{1}},\ldots,Q^{\pi_{M}}$ is absolutely
	continuous w.r.t. $\mu$. Assume	
	\[
	c\left(M\right)\triangleq\sup_{\pi_{1},\dots,\pi_{M}}\left\Vert \frac{d\left(\varrho Q^{\pi_{1}}Q^{\pi_{2}}\dots Q^{\pi_{M}}\right)}{d\mu}\right\Vert _{\infty}
	\]
	satisfies	
	$
	C_{\varrho,\mu}\triangleq\left(1-\gamma\right)^{2}\sum_{M\geq1}M\gamma^{M-1}c\left(M\right)<+\infty.
	$
\end{assumption}

We remind the reader that the selection of $Q^{\pi}$ is not unique.
Rather, for fixed $\pi\in\Pi,$ $Q^{\pi}\mbox{ : } B\left(\mathbb{S};J_{\max}\right)\rightarrow  B\left(\mathbb{S};J_{\max}\right)$
is a linear operator such that $Q^{\pi}\left(\cdot\vert s\right)\in\mathcal{Q}\left(s,\pi\left(s\right)\right)$
is an element of the distributional set $\mathcal{Q}\left(s,a\right)$
when $a=\pi\left(s\right)$, for all $s\in\mathbb{S}$. A remark about
this assumption is in order. For each state $s\in\mathbb{S}$, the
distributional sets $\left\{ \mathcal{Q}\left(s,a\right)\right\} _{a\in\mathbb{A}}$
include transition kernels which may assign positive probability to
finitely many elements of the state space. If the union of all distributional
sets $\{ \mathcal{Q}(s,a)\} _{a\in\mathbb{A}}$
for all $s\in\mathbb{S}$ remains finite, then we may simply choose
$\mu$ to have positive probability on these finitely many points.
However, if this set of distinguished points differs among states
$s\in\mathbb{S}$, then constructing such a $\mu$ that satisfies
our absolute continuity assumption will be challenging.

For $s\in\mathbb{S}$ and $a\in\mathbb{A},$ if any element $Q\left(\cdot\vert s,a\right)\in\mathcal{Q}\left(s,a\right)$
is absolutely continuous w.r.t. $\mu,$ we define a coefficient
$C_{\mu}$ that helps us to verify Assumption \ref{assu:Absolute_continuity}
\[
C_{\mu}\triangleq\sup_{s,a,Q}\left\Vert \frac{dQ\left(\cdot|s,a\right)}{d\mu}\right\Vert _{\infty}.
\]

We claim that if $C_{\mu}<+\infty$ then Assumption \ref{assu:Absolute_continuity}
holds. It suffices to show $c\left(M\right)\leq C_{\mu}$ for any
$M,$ as stated in the lemma below. The proof is given in the Appendix.
\begin{lemma}
	\label{lem:C} $c\left(M\right)\leq C_{\mu}$ for $M\geq1.$
\end{lemma}

To illustrate the idea behind Assumption \ref{assu:Absolute_continuity}
and coefficient $C_{\mu},$ we discuss CVaR and mean-deviation below.
Once the distribution $\mu$ is properly chosen, given state $s\in\mathbb{S}$ and action $a\in\mathbb{A},$
the distributional set $\mathcal{Q}\left(s,a\right)$ in (\ref{eq:RISK_Bellman-1})
for Markovian CVaR at level $\alpha\in[0,1)$ has the form (see
\cite[Example 4.3]{Ruszczynski:2006uq})
\begin{equation}\label{cvar}
\mathcal{Q}(s,a)=\left\{h:\begin{aligned}&0\leq h\left(s'\right)\leq\left(1-\alpha\right)^{-1},\,\text{a.e.}\,s'\in\mathbb{S},\\&\int_{\mathbb{S}}h\left(s'\right)P\left(ds'|s,a\right)=1\end{aligned}\right\}.
\end{equation}
Since the Radon-Nikodym derivatives $h$ of distributions $Q\left(\cdot\vert s,a\right)\in\mathcal{Q}\left(s,a\right)$
w.r.t. $\mu$ are bounded by $\left(1-\alpha\right)^{-1},$
Assumption \ref{assu:Absolute_continuity} automatically holds by
Lemma \ref{lem:C} since $C_{\mu}=\left(1-\alpha\right)^{-1}$ is bounded. Similarly,
under a proper choice of $\mu,$ fix $s\in\mathbb{S},$ $a\in\mathbb{A},$
$p\in(1,+\infty),$ and constant $b\geq0,$ the distributional set
$\mathcal{Q}\left(s,a\right)$ in (\ref{eq:RISK_Bellman-1}) for
mean-deviation risk function becomes (see \cite[Example 4.1]{Ruszczynski:2006uq})
\[
\mathcal{Q}\left(s,a\right)=\left\{ h:h=1+g-\int_{\mathbb{S}}g\left(s'\right)P\left(ds'\vert s,a\right),\left\Vert g\right\Vert _{q,\bar{\mu}}\leq b\right\},
\]
where $q=p/(p-1)$ and $\bar{\mu}=P(\cdot|s,a).$ Since the Radon-Nikodym derivatives $h$ of distributions $Q\left(\cdot\vert s,a\right)\in\mathcal{Q}\left(s,a\right)$
w.r.t. $\mu$ are bounded by $\left\Vert h\right\Vert _{\infty}\leq1+2\left\Vert g\right\Vert _{\infty}\leq1+2Bb$
where $B$ is a positive real number, Assumption \ref{assu:Absolute_continuity}
holds by Lemma \ref{lem:C} if $b<+\infty.$ We will design a suitable
sample distribution $\mu$ in our numerical experiments.

The following theorem states that with high probability the final
performance of the policy found by the algorithm can be made as close
as to a constant times the inherent Bellman error of the function
space $\mathcal{F}$ as desired by selecting
a sufficiently high number of samples. Hence, the sampling-based algorithm
can be used to find near-optimal policies if $\mathcal{F}$
is sufficiently rich.
\begin{theorem}
	\label{thm:PAC bound} Consider an MDP satisfying Assumption \ref{assu:Bounded_costs},
	\ref{assu:Risk_estimation} and \ref{assu:Absolute_continuity}. Fix
	$1\leq p<\infty$, $\mu\in\mathcal{P}\left(\mathbb{S}\right)$ and let $\widehat{J}_{0}\in\mathcal{F}\subset B\left(\mathbb{S};J_{\max}\right)$.
	Then for any $\varepsilon,\delta>0$, there exists integers $K,$
	$m$ and $n$ such that $K$ is linear in $\log\left(1/\varepsilon\right)$,
	$\log J_{\max}$ and $\log\left(1/\left(1-\gamma\right)\right)$,
	$n$ is polynomial in $\log(\mathcal{N}(8^{-1}[\varepsilon(1-\gamma)^{2}/(16\gamma C_{\varrho,\mu}^{1/p})]^{p},\mathcal{F},n,\mu)),$
	$1/\varepsilon$, $\log\left(1/\delta\right),$ $J_{\max}$ and $m$
	is chosen according to $$\theta\left(\frac{\varepsilon\left(1-\gamma\right)^{2}}{16\gamma C_{\varrho,\mu}^{1/p}},m\right)\leq\frac{\delta}{4n\left|\mathbb{A}\right|K},$$
	such that if the sampling-based algorithm is run with parameters $\left(n,m,\mu,\mathcal{F}\right)$
	and $\widehat{\pi}_{K}$ is a policy greedy w.r.t. the $K^{th}$ iterate
	then w.p. at least $1-\delta$,	
	\[
	\left\Vert J^{\widehat{\pi}_{K}}-J^{*}\right\Vert _{p,\varrho}\leq\frac{2\gamma}{\left(1-\gamma\right)^{2}}C_{\varrho,\mu}^{1/p}d_{p,\mu}\left(T\mathcal{F},\mathcal{F}\right)+\varepsilon.
	\]
\end{theorem}

We can control the error term $\varepsilon$ in the preceding theorem
through the number of samples, but we can only control the constant
term $d_{p,\mu}(T\mathcal{F},\mathcal{F})$ through
the choice of the approximating family $\mathcal{F}$.

\subsection{Supremum norm}\label{sec3.1}

Our supremum norm analysis is inspired by \cite{jain2010simulation}.
In \cite{jain2010simulation}, an $\epsilon-$net
over the space of policies is constructed, each policy in the $\epsilon-$net
is evaluated by simulation, and then the optimal policy from the $\epsilon-$net
is chosen. It is shown that the resulting policy is close to the true
optimal policy with high probability. We now use the idea of an $\epsilon-$net
to perform approximate value iteration for MDPs with continuous state
spaces. For this setting, the subset $\{s_1,\dots,s_n\}$ in Algorithm \ref{algo} is constructed deterministically as an $\epsilon-$net $\mathcal{S}\subset\mathbb{S}$ and $|\mathcal{S}|=n.$

Similar to \cite{rust1997using}, we assume the following regularity conditions.
\begin{assumption}
	\label{assu:Epsilon_net_regularity} 
	\begin{enumerate}
		\item There exists $\kappa_{c}<\infty$ such that $|c(s,a)-c(s',a)|\leq\kappa_{c}||s-s'||_\infty$
		for all $s,s'\in\mathbb{S}$ and $a\in\mathbb{A}$.
		\item There exists $\kappa_{\mu}<\infty$ such that $\int|\mu(dy\vert s,a)-\mu^{'}(dy\vert s',a)|\leq\kappa_{\mu}||s-s'||_\infty$
		for all $\mu(\cdot\vert s,a)\in\mathcal{Q}(s,a),$
		$\mu^{'}(\cdot\vert s',a)\in\mathcal{Q}(s',a),$
		$s,s'\in\mathbb{S}$, and $a\in\mathbb{A}$.
	\end{enumerate}	
\end{assumption}

Assumption \ref{assu:Epsilon_net_regularity} part 1) states that the
cost function $s\mapsto c(s,a)$ is Lipschitz continuous
for all fixed $a\in\mathbb{A}$. Assumption \ref{assu:Epsilon_net_regularity} part
2) ensures regularity of the distributions in the distributional
sets $\mathcal{Q}\left(s,a\right)$ w.r.t. the total variation norm.

The main idea in this subsection is to use a finite partition of the
state space $\mathbb{S}$. Let $\mathcal{S}$ be a finite subset of
$\mathbb{S}$, and let $\{ B_{s}\} _{s\in\mathcal{S}}\subset\mathcal{B}\left(\mathbb{S}\right)$
be a corresponding partition of $\mathbb{S}$ such that $s\in B_{s}$
for all $s\in\mathcal{S}$ ($s$ is a representative element of the
set $B_{s}$ for all $s\in\mathcal{S}$). The diameter of a set $B\subset\mathbb{S}$
is
$$
\mbox{diam}\left(B\right)\triangleq\sup_{s,s'\in B}\|s-s'\|_{\infty}.
$$
We make the following assumption on the fineness of the partition
$\{ B_{s}\} _{s\in\mathcal{S}}$.
\begin{assumption}
	\label{assu:Epsilon_net_partition} For an accuracy $\epsilon>0$,
	there is a set $\mathcal{S}\subset\mathbb{S}$ and a partition $\{ B_{s}\} _{s\in\mathcal{S}}\subset\mathcal{B}\left(\mathbb{S}\right)$
	such that $\mbox{diam}(B_{s})\leq\epsilon$ for all
	$s\in\mathcal{S}$.
\end{assumption}

For the rest of this subsection, when we refer to $\{ B_{s}\} _{s\in\mathcal{S}}$
we mean the specific partition in Assumption \ref{assu:Epsilon_net_partition}
with accuracy $\epsilon$. This partition is closely related to
the idea of an $\epsilon-$net. Since the state space $\mathbb{S}$ is a compact subset of a Euclidean space, we can construct an $\epsilon-$net $\mathcal{S}$ for $\mathbb{S}$
such that for every $s\in\mathbb{S}$ there is an $s'\in\mathcal{S}$ with
$\|s-s'\|_{\infty}\leq\epsilon$. By construction of $\{ B_{s}\} _{s\in\mathcal{S}}$,
$\mathcal{S}$ is an $\epsilon-$net for $\mathbb{S}$ because all
$s'\in\mathbb{S}$ belong to $B_{s}$ for some $s\in\mathcal{S}$
and $\|s-s'\|_{\infty}\leq\epsilon$ since $\mbox{diam}\left(B_{s}\right)\leq\epsilon$.

Assumption \ref{assu:Epsilon_net_partition} suggests a finite state
space MDP that approximates the continuous state space MDP, where
the states are the elements of $\mathcal{S}$. To be specific, the functional family $\mathcal{F}$ in Algorithm \ref{algo} is chosen as\footnote{In this paper, we only consider linear approximation by piecewise constants. We remark that other types of approximation by piecewise constants are possible (e.g., nonlinear or adaptive approximation, see \cite[Section 3]{devore1998nonlinear}).}
\[
\mathcal{F}=\left\{ f\in B(\mathbb{S};J_{\max}):\,f\mbox{ is piecewise constant on }\{ B_{s}\} _{s\in\mathcal{S}}\right\},
\]
which only appears in our supremum norm analysis. The function fitting scheme, i.e., line $6$ in Algorithm \ref{algo}, is to let the approximate risk-to-go function $\widehat{J}_{k+1}$ be piecewise constant on the partition $\{B_s\}_{s\in\mathcal{S}}$ of $\mathbb{S}$. In other words, for $s\in\mathbb{S}$ and $s'\in\mathcal{S}$,
$$\widehat{J}_{k+1}(s)=\widetilde{J}(s')\,\,\text{if}\,\,\|s-s'\|_{\infty}\leq\epsilon.$$

The theorem below provides a finite-sample
error bound for approximate value iteration on the finite state space
MDP.
\begin{theorem}
	\label{thm:infinite_convergence} Let $\varepsilon>0.$ Under Assumption
	\ref{assu:Bounded_costs}, \ref{assu:Risk_estimation}, \ref{assu:Epsilon_net_regularity}
	and \ref{assu:Epsilon_net_partition}, if the $\epsilon-$net $\mathcal{S}$ is chosen such that
	$$
	\epsilon\leq\frac{\varepsilon}{2\left(\kappa_{c}+\gamma\kappa_{\mu}J_{\max}\right)},
	$$
	we have	
	\[
	\mathbb{P}\left(\|\widehat{J}_{K}-J^{*}\|_{\infty}\leq\gamma^{K}J_{\max}+\frac{\varepsilon}{1-\gamma}\right)\geq1-Kp_{m,n}\left(\varepsilon\right)
	\]
	where $p_{m,n}\left(\varepsilon\right)=n\left|\mathbb{A}\right|\theta(\varepsilon/(2\gamma),m)$ is an upper bound on the probability that the approximation errors exceed $\varepsilon$ in any iteration.
\end{theorem}


Finally, we remark that the sample analysis in this section assumes
that the approximation errors $\varepsilon_{k}$ defined in \eqref{eq:Error} are bounded above
by some $\varepsilon>0$ in \emph{every} iteration $k=0,\ldots,K-1$
for some fixed $K$.

\section{Risk-to-go estimation}\label{sec4}

In classical MDPs, estimation of cost-to-go function can
be a standard sample average approximation, which has well known convergence
guarantees (e.g., \cite{Haskell_EDP_2015,cooper2012performance}).
Our current setting is more subtle because we must consider empirical
estimates of the risk-to-go. In this section we discuss several examples
of one-step risk measure for which such empirical estimation is possible, and give specific
form of $\theta(\varepsilon,m)$ in Assumption \ref{assu:Risk_estimation}.
In particular, we consider: mean-deviation, mean-semideviation, optimized
certainty equivalent, and conditional value-at-risk. For
the next example, let $\mu$ be a probability distribution on the
state space $\mathbb{S}$. We then let $\|f\|_{p,\hat{\mu}}^{p}\triangleq\sum_{j=1}^{m}\left|f(Y_{j})\right|^{p}/m$
denote an empirical estimation of $\|f\|_{p,\mu}^{p}$ where the
samples $\{ Y_{j}\} _{j=1}^{m}$ are drawn according to
$\mu$. Similarly, $\|f\|_{\hat{\mu}}\triangleq\sum_{j=1}^{m}f(Y_{j})/m$
is the usual sample average approximation. In addition, for any real number $z$, we denote $(z)_+\triangleq\max\{0,z\}.$ We emphasize that the symbol $\rho$ in this section denotes an one-step conditional risk measure in the iterated compositions \eqref{iterated risk}, not a risk measure of the total cost.

\begin{example}[Mean-deviation and mean-semideviation risk functions] 
	\begin{enumerate}
	\item The mean-deviation risk function \cite[Example 4.1]{Ruszczynski:2006uq} of a random variable $Y\sim\mu$ is 	
	\[
	\rho\left(Y\right)\triangleq\mathbb{E}\left[Y\right]+b(\|Y-\mathbb{E}\left[Y\right]\|_{p,\mu}^{p})^{1/p},
	\]
	where $p\in[1,+\infty)$ and $b\geq0$ are given constants. The corresponding empirical estimation of $\rho(Y)$ is given by 	
	\begin{equation*}
	\begin{aligned}
	\hat{\rho}_{m}\left(\{Y_j\}^m_{j=1}\right)
	=\|Y\|_{\hat{\mu}}	+b(\|Y-\|Y\|_{\hat{\mu}}\|_{p,\hat{\mu}}^{p})^{1/p}.		
	\end{aligned}
	\end{equation*}
	\item The mean-semideviation risk function \cite[Example 4.2]{Ruszczynski:2006uq} of a random variable $Y\sim\mu$ is 		
	\[
	\rho\left(Y\right)\triangleq\mathbb{E}\left[Y\right]+b(\|\left(Y-\mathbb{E}\left[Y\right]\right)_{+}\|_{p,\mu}^{p})^{1/p},
	\]
	where $p\in[1,+\infty)$ and $b\geq0$ are given constants. The corresponding empirical estimation of $\rho(Y)$ is given by 	
	\begin{equation*}
	\begin{aligned}	
	\hat{\rho}_{m}\left(\{Y_j\}^m_{j=1}\right)
	=\|Y\|_{\hat{\mu}}
	+b(\|(Y
	-\|Y\|_{\hat{\mu}})_{+}\|_{p,\hat{\mu}}^{p})^{1/p}.
	\end{aligned}
	\end{equation*}	
	\end{enumerate}
\end{example}	

The mean-deviation and mean-semideviation risk functions are analyzed in \cite{ogryczak1999stochastic,ogryczak2001consistency,Ruszczynski:2006uq,ruszczynski2006conditional}. Both risk functions are known to belong to the class of mean-risk models \cite{shapiro2014lectures}. The main idea of the models is to characterize the uncertain outcome $Y$ by two scalar characteristics: the mean $\mathbb{E}[Y]$, describing the expected outcome, and the risk (dispersion measure) $\mathbb{D}[Y]$, which measures the uncertainty of the outcome. Specifically, the models can be written in a form of composite objective functional 
$\rho(Y)\triangleq\mathbb{E}[Y]+b\mathbb{D}[Y],$ where coefficient $b\geq0$ plays the role of the price of risk. This mean-risk approach has many advantages: it allows one to formulate a corresponding parametric optimization problem and it facilitates the trade-off analysis between mean and risk.	When the dispersion measure has the form $\mathbb{D}[Y]=(||(Y-\mathbb{E}[Y]||^p_{p,\mu})^{1/p},$ we obtain the mean-deviation risk function \cite[Example 4.1]{Ruszczynski:2006uq}.	Note that for $p=2$, the function $\rho(\cdot)$ corresponds to the Markowitz mean-variance model \cite{markowitz1968portfolio}, which has drawn continuing and resurgent attention for several decades \cite{markowitz2000mean,bielecki2005continuous,costa2010sampled,yin2004markowitz}. When the dispersion measure is chosen to be the semideviation of order $p$, $\mathbb{D}[Y]=(||(Y-\mathbb{E}[Y])_+||^p_{p,\mu})^{1/p},$ we obtain the mean-semideviation risk function \cite[Example 4.2]{Ruszczynski:2006uq}, which is appropriate for minimization problems where $Y$ represents a cost. It is aimed at penalization of an excess of $Y$ over its mean.
	
	\begin{example}[Optimized certainty equivalent (OCE)]
		The coherent optimized certainty equivalent \cite{ben2007old} of a random variable $Y\sim\mu$ is 		
		\[
		\rho\left(Y\right)\triangleq\inf_{\eta\in\mathbb{R}}\left\{ \eta+\mathbb{E}\left[u\left(Y-\eta\right)\right]\right\} ,
		\]
		where $u$ is a piecewise linear function given by $u(x)=\beta_1(x)_+-\beta_2(-x)_+$	for some $0\leq\beta_{1}<1<\beta_{2}.$ The corresponding empirical estimation of $\rho(Y)$ is given by 
		$$\hat{\rho}_{m}\left(\{Y_j\}^m_{j=1}\right)=
		\inf_{\eta\in\mathbb{R}}\left\{ \eta+\|u\left(Y-\eta\right)\|_{\hat{\mu}}\right\} .
		$$
	\end{example}

The optimized certainty equivalent is first introduced in \cite{ben1986expected} and further studied in \cite{ben2007old}. In the definition of OCE, the term $\mathbb{E}[u(Y)]$ is interpreted as the sure present value of a future uncertain income $Y$. The rational behind the definition of the OCE is as follows: suppose a decision maker expects a future uncertain income of $Y$ dollars, and can consume part of $Y$ at present. If he chooses to consume $\eta$ dollars, the resulting present value of $Y$ is then $\eta + \mathbb{E}[u(Y-\eta)]$. Thus, the sure (present) value of $Y$, (i.e., its certainty equivalent $\rho(Y)$) is the result of an {\em optimal} allocation of $Y$ between present and future consumption. The latter also motivates the name OCE. The OCE has wide applications, such as portfolio theory \cite{ben1991portfolio},  production, investment, inventory and insurance problems \cite{ben1997duality,ben1991recourse}. 
	
	\begin{example}[Conditional value-at-risk (CVaR)]
		Conditional value-at-risk  \cite{rockafellar2000optimization}
		is a special case of OCE by
		choosing the utility function $u\left(x\right)=(x-\eta)_{+}/(1-\alpha)$
		where $\alpha\in[0,1)$. The CVaR at level $\alpha$
		of a random variable $Y\sim\mu$ is 	
		\[
		\rho\left(Y\right)=\text{CVaR}_{\alpha}\left(Y\right)\triangleq\inf_{\eta\in\mathbb{R}}\left\{ \eta+\frac{1}{1-\alpha}\mathbb{E}\left[\left(Y-\eta\right)_{+}\right]\right\}.
		\]
		The corresponding empirical estimation of $\rho(Y)$ is given by
		$$\hat{\rho}_{m}\left(\{Y_j\}^m_{j=1}\right)=
		\inf_{\eta\in\mathbb{R}}\left\{ \eta+\frac{1}{1-\alpha}\|\left(Y-\eta\right)_{+}\|_{\hat{\mu}}\right\}.
		$$
	\end{example}

\begin{figure}[!tph]
	\begin{centering}
		\includegraphics[scale=0.5]{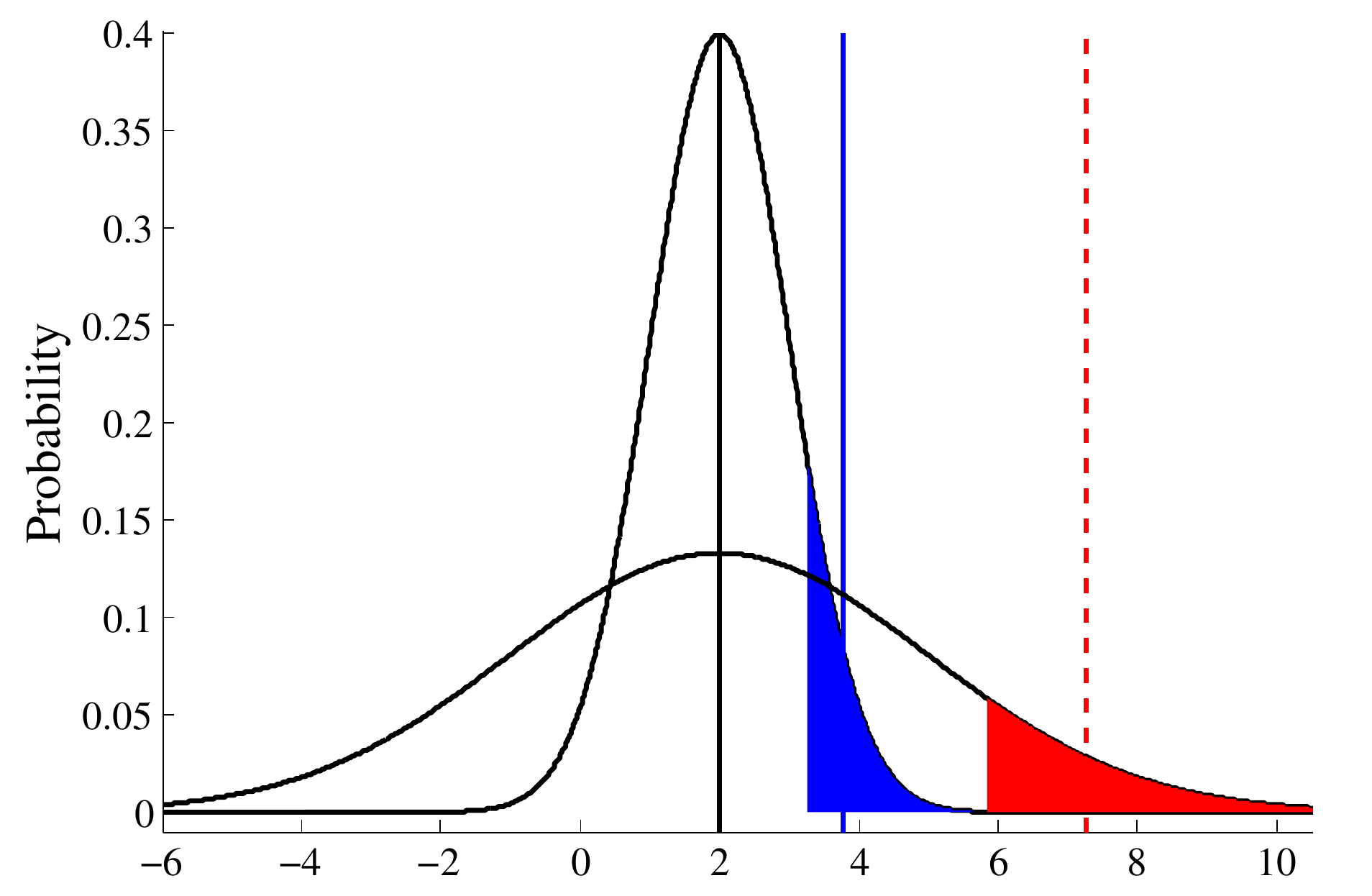}
		\par\end{centering}	
	\caption{Comparison of two distributions with identical expectations and different $\text{CVaR}_{0.9}$ values. The
		filled-in regions illustrate the quantiles, while the vertical lines indicate the expectations and conditional values-at-risk.
		\label{fig:CVaR}}
\end{figure}

As a special case of coherent OCE, the conditional value-at-risk is a prominent risk measure that has found extensive use in stochastic optimization (see \cite{rockafellar2000optimization}
for example). Mathematically, for a random variable $Y$, we define
$F_{Y}$ to be the cumulative distribution function of $Y$, $\text{VaR}_{\alpha}\left(Y\right)\triangleq\inf\left\{ t\mbox{ : }F_{Y}\left(t\right)\geq\alpha\right\} $
to be the value-at-risk of $Y$ at level $\alpha\in[0,1)$. The CVaR of $Y$ at level $\alpha\in[0,1)$ can be equivalently defined as	$\text{CVaR}_{\alpha}(Y)\triangleq(1-\alpha)^{-1}\int_{\alpha}^{1}\text{VaR}_{\tau}(Y)d\tau.$
It is easy to see that $\text{CVaR}_{0}=\mathbb{E}(Y)$ and $\text{CVaR}_{\alpha}$ is the worst-case (or robust) realization as $\alpha\rightarrow1.$ Put simply, the CVaR is the expected $1-\alpha$ worst-cases of the return, and it assigns a higher overall cost to a scenario with heavier tails even if the expected value stays the same. Thus, by
appropriately tuning $\alpha$, the CVaR may be tuned to be sensitive to rare, but
very low returns, which makes it particularly attractive as a risk measure. Fig. \ref{fig:CVaR} illustrates how a $\text{CVaR}_{\alpha}$ is computed in comparison with a plain expectation. 
The CVaR has been studied extensively \cite{rockafellar2000optimization,shapiro2014lectures}, and is known to have favorable mathematical properties such as coherence \cite{MAFI:MAFI068}. It has also been used in many practical applications, in finance and other domains \cite{uryasev2010var}.

The next lemma gives sample complexity results for the preceding four
risk measures.
\begin{lemma}
	\label{lem:Risk_estimation} Given  $s\in\mathbb{{S}}$,
	$a\in\mathbb{A},$ $J\in B\left(\mathbb{S};J_{\max}\right)$, $\varepsilon>0$
	and $m\geq1$. Denote $d(\rho,\hat{\rho}_m)=|\rho(J(Y^{s,a}))-\hat{\rho}_{m}(\{J(Y_j^{s,a})\}^m_{j=1})|$.
	\begin{enumerate}
		\item For mean-deviation or mean-semideviation, we have	
		\begin{equation*}
		\begin{aligned}
		\mathbb{P}(d(\rho,\hat{\rho}_m)>\varepsilon)
		\leq2(e^{-x}+e^{-y}+e^{-z}),
		\end{aligned}
		\end{equation*}	
		where $x=m\varepsilon^{2}/(\sqrt{2}J_{\max})^{2},$ $y=m\varepsilon^{2}/(\sqrt{2}bp(1+C)J_{\max}^{p})^{2}$
		and $z=m\varepsilon^{2}/(\sqrt{2}bp(1+C)J_{\max}^{2p-1})^{2}$
		with constant $C>0.$ 
		\item For coherent optimized certainty equivalent, we have	
		\begin{equation*}
		\begin{aligned}
		\mathbb{P}(d(\rho,\hat{\rho}_m)>\varepsilon)
		\leq2\left(1+\frac{4\beta_{2}}{\varepsilon}\right)\exp\left[\frac{-m\varepsilon^{2}}{(\sqrt{2}u(J_{\max}))^{2}}\right]
		.
		\end{aligned}
		\end{equation*}
		\item For conditional value-at-risk, we have	
		\begin{equation*}
		\begin{aligned}
		\mathbb{P}(d(\rho,\hat{\rho}_m)>\varepsilon)
		\leq2\left(1+\frac{4}{\varepsilon(1-\alpha)}\right)\exp\left[\frac{-m(\varepsilon(1-\alpha))^{2}}{(\sqrt{2}(2-\alpha)J_{\max})^{2}}\right].
		\end{aligned}
		\end{equation*}
	\end{enumerate}	
\end{lemma}

\section{Convergence analysis via stochastic dominance}\label{sec5}

In this section, we expand upon our convergence analysis to explore
the tradeoff between sample complexity and convergence rate. In Section
\ref{sec3} we computed the required number of iterations to reach a desired
accuracy given a certain error tolerance, and then computed the number
of samples required to stay within this error tolerance in every iteration.
We relax this idea in this section and instead we allow the approximation
error in iterations to exceed this error tolerance. In \cite{Haskell_EDP_2015},
a stochastic dominance technique is developed to study this situation.
The original work in \cite{Haskell_EDP_2015} was specific to finite
state and action space MDPs. Now extend this argument to show that
this method is applicable to our present setting.

Theorems \ref{thm:PAC bound} and \ref{thm:infinite_convergence} give
an estimate for the error $\|J^{\widehat{\pi}_{k}}-J^{*}\|_{p,\varrho}$ and $\|\widehat{J}_{K}-J^{*}\|_{\infty}$ based on fixing $\varepsilon>0$
and assuming $\|\varepsilon_{k}\|\leq\varepsilon$ for \textit{all}
iterations $k=0,\ldots,K-1$. The next sample complexity result
allows for a smaller number of samples in each iteration, but requires
a larger overall number of iterations.
\begin{theorem}
	\label{thm:Sample complexity} Let Assumption \ref{assu:Bounded_costs}
	and \ref{assu:Risk_estimation} hold. Given $\epsilon_{g}\in\left(0,1\right),$
	$\delta\in\left(0,1\right)$ and let $\delta_{1}+2\delta_{2}\leq\delta.$
	Choose $K$ such that 
	$$
	K\geq\log\left(\frac{1}{\delta_{2}\mu_{\min}}\right),
	$$
	where $\mu_{\min}\triangleq\min_{\eta}\mu\left(\eta\right)$ with
	$\mu\left(\eta\right)$ given in Lemma \ref{lem:Stationary distribution}. 
	\begin{enumerate}
		\item Under Assumption \ref{assu:Epsilon_net_regularity} and \ref{assu:Epsilon_net_partition},
		select $\varepsilon<\epsilon_{g},$ $\epsilon$ and $m$ such
		that	
		\[
		\epsilon\leq\frac{\varepsilon}{2\left(\kappa_{c}+\gamma\kappa_{\mu}J_{\max}\right)}\quad\text{and}\quad\theta\left(\frac{\varepsilon}{2\gamma},m\right)\leq\frac{\delta_{1}}{\left|\mathbb{A}\right|\left|\mathcal{S}\right|}.
		\]
		Then we have $\mathbb{P}(\|\widehat{J}_{K}-J^{*}\|_{\infty}>\epsilon_{g})\leq\delta$.
		\item Under Assumption \ref{assu:Absolute_continuity}, select $\varepsilon<\epsilon_{g}-d_{p,\mu}(T\mathcal{F},\mathcal{F}),$
		$n$ and $m$ such that 		
		\[
		n>128\left(\frac{8J_{\max}}{\varepsilon}\right)^{2p}\left(\log\left(1/\delta_{1}\right)+\log\left(32\mathcal{N}_{0}\left(n\right)\right)\right)
		\]
		and
		$$
		\theta\left(\varepsilon/4,m\right)\leq\frac{\delta_{1}}{4n\left|\mathbb{A}\right|}
		$$
		where $\mathcal{N}_{0}(n)=\mathcal{N}(8^{-1}({\varepsilon}/{4})^{p},\mathcal{F},n,\mu).$
		Then we have $\mathbb{P}(\|J^{\widehat{\pi}_{K}}-J^{*}\|_{p,\varrho}>\epsilon_{g})\leq\delta$.
	\end{enumerate}
\end{theorem}

Theorem \ref{thm:infinite_convergence} and Theorem \ref{thm:Sample complexity} part 1)
offer two different convergence analysis. We now confirm our claim
that the stochastic dominance analysis requires a smaller number of
samples in each iteration. First, we take $\theta\left(\varepsilon,m\right)=Ce^{-m\varepsilon^{2}}$
where $C>0$ is some constant, and compute the minimal number of samples
$m_{1}$ required by Theorem \ref{thm:infinite_convergence} and samples
$m_{2}$ required for the stochastic dominance analysis:
\begin{equation*}
\begin{aligned}
	&   m_{1}=\frac{4\gamma^{2}}{\left[\varepsilon\left(1-\gamma\right)-2\gamma^{K}J_{\max}\left(1-\gamma\right)\right]^{2}}\log\frac{\left|\mathbb{A}\right|\left|\mathcal{S}\right|C}{1-\left(1-\delta\right)^{1/K}},\\
	&   m_{2}=\frac{4\gamma^{2}}{\varepsilon^{2}}\log\frac{\left|\mathbb{A}\right|\left|\mathcal{S}\right|C}{\delta-2/e^{K}\mu_{\min}},
\end{aligned}
\end{equation*}
where
$
\mu_{\min}=\min\left\{\mu_1,\mu_2\right\},
$
with $\mu_1=(1-\delta)^{(\lceil J_{\max}/\varepsilon\rceil -1)/K}$ and $\mu_2=[1-(1-\delta)^{1/K}](1-\delta)^{(\lceil J_{\max}/\varepsilon\rceil -2)/K}$.

To verify our claim, we next allow $K$ to be arbitrarily large and
show $m_{1}\geq m_{2}$ in the following. First, we need to show $e^{K}\mu_{\min}\rightarrow+\infty$
as $K\rightarrow+\infty.$ Note that for constant $d\in\left(0,1\right)$
and $K>0,$ we have
$$
(1-d^{1/K})(1+d^{1/K}+d^{2/K}+\dots+d^{\left(K-1\right)/K})=1-d,
$$
and thus $1-d^{1/K}\geq{(1-d)}/{K}.$ We then obtain that $e^{K}(1-d^{1/K})\geq{[e^{K}(1-d)]}/{K}\rightarrow+\infty$
as $K\rightarrow+\infty.$ Since $\mu_{\min}\in\left(0,1\right),$
we conclude that $e^{K}\mu_{\min}\rightarrow+\infty$ as $K\rightarrow+\infty.$
Finally, by letting $K$ be arbitrarily large, we have $m_{1}/m_{2}\approx{1}/{(1-\gamma)^{2}},$
which implies the stochastic dominance analysis requires a smaller
number of samples in each iteration given sufficiently large amount
of iterations $K.$

The sample comparison for analysis in Theorem \ref{thm:PAC bound}
and Theorem \ref{thm:Sample complexity} 2) is nontrivial because
it does not follow from a contraction argument. This discussion is
left for future work.

\section{Proofs of main results}\label{sec6}

This section is organized as follows. In Section \ref{sec6.1} and \ref{sec6.2}, we provide
details for two types of analysis, i.e., $p-$norm and supremum analysis,
followed by their alternative stochastic dominance convergence analysis
in Section \ref{sec6.3}. Proofs of all technical results can be found in the Appendix.

\subsection{Analysis in $p-$norm}\label{sec6.2}

The idea of analysis in $p-$norm is to show that (i) the approximation
errors stay small with high probability in each iteration provided
that $m,n$ are sufficiently large, and (ii) if the errors in each
iteration are small then the final error will be small when $K$,
the number of iterations, is big enough. To show (i), we provide a
lemma which gives us a probabilistic guarantee on the approximation
error introduced in a single iteration of our algorithm. 
\begin{lemma}
	\label{lem:Error each iteration} Let Assumption \ref{assu:Bounded_costs}
	and \ref{assu:Risk_estimation} hold. Fix real number $p\in[1,+\infty),$
	integers $n,m\geq1,$ $\mu\in\mathcal{P}\left(\mathbb{S}\right)$
	and $\mathcal{F}\subset B\left(\mathbb{S};J_{\max}\right).$
	Pick any $J\in B\left(\mathbb{S};J_{\max}\right)$ and let $\widehat{J}_{k+1}=\widehat{J}_{k+1}\left(J,n,m,\mu,\mathcal{F}\right)$
	be defined by Equation (\ref{eq:Function fitting}). Let $\mathcal{N}_{0}\left(n\right)=\mathcal{N}\left(8^{-1}\left({\varepsilon}/{4}\right)^{p},\mathcal{F},n,\mu\right).$
	Then for any $\varepsilon,\delta>0,$	
	\[
	\|\widehat{J}_{k+1}-T\widehat{J}_{k}\|_{p,\mu}\leq d_{p,\mu}(T\widehat{J}_{k},\mathcal{F})+\varepsilon
	\]
	holds w.p. at least $1-\delta$ provided that	
	\[
	n>128\left({8J_{\max}}/{\varepsilon}\right)^{2p}\left(\log\left(1/\delta\right)+\log\left(32\mathcal{N}_{0}\left(n\right)\right)\right)
	\]
	and $m$ satisfying	
	$$
	\theta\left(\frac{\varepsilon}{4},m\right)\leq\frac{\delta}{4n\left|\mathbb{A}\right|}.
	$$	
\end{lemma}

Lemma \ref{lem:Error each iteration} shows that with high probability,
$\widehat{J}_{k+1}$ is a good approximation to $T\widehat{J}_{k}$
provided that some element of $\mathcal{F}$
is close to $T\widehat{J}_{k}$ and if the number of samples is sufficiently
large. In other words, the lemma states the finite-sample bound for
a single iterate. 

For risk measures defined in Section \ref{sec4}, we have the
following error bound for each iteration.
\begin{corollary}
	\label{cor:Error each iteration} Let Assumptions \ref{assu:Bounded_costs}
	and \ref{assu:Risk_estimation} hold. Given a real number $p\in[1,+\infty),$
	$\mu\in\mathcal{P}(\mathbb{S}),$ $\mathcal{F}\subset B(\mathbb{S};J_{\max}).$
	Pick any $J\in B\left(\mathbb{S};J_{\max}\right)$ and let $\widehat{J}_{k+1}=\widehat{J}_{k+1}\left(J,n,m,\mu,\mathcal{F}\right)$
	be defined by Equation (\ref{eq:Function fitting}). Let $\mathcal{N}_{0}(n)=\mathcal{N}(8^{-1}({\varepsilon}/{4})^{p},\mathcal{F},n,\mu).$
	Then for any $\varepsilon,\delta>0,$	
	\[
	\|\widehat{J}_{k+1}-T\widehat{J}_{k}\|_{p,\mu}\leq d_{p,\mu}(T\widehat{J}_{k},\mathcal{F})+\varepsilon
	\]
	holds w.p. at least $1-\delta$ provided that	
	\[
	n>128\left(\frac{8J_{\max}}{\varepsilon}\right)^{2p}\left(\log\left(1/\delta\right)+\log\left(32\mathcal{N}_{0}\left(n\right)\right)\right)
	\]
	and $m$ satisfying
	\begin{enumerate}
		\item $$
		m>\frac{32m'}{\varepsilon^{2}}\left(\log\left(1/\delta\right)+\log\left(8n\left|\mathbb{A}\right|\right)\right),
		$$
		where 
		$m'=\min\{ ((1+C)J_{\max}^{p})^{2},(bp(1+C)J_{\max}^{2p-1})^{2},J_{\max}^{2}\} $
		with constant $C>0$ for mean-deviation or mean-semideviation risk
		function $\rho.$
		\item 
		\begin{equation*}
		\begin{aligned}
		m>32(u(J_{\max})/\varepsilon)^{2}(\log\left(1/\delta\right)+\log\left(8n\left|\mathbb{A}\right|\right)
		+\log(1+16\beta_{2}/\varepsilon))
		\end{aligned}
		\end{equation*}
		for coherent optimized certainty equivalent $\rho.$
		\item 
		\begin{equation*}
		\begin{aligned}
		m>32\left(\frac{\left(2-\alpha\right)J_{\max}}{\left(1-\alpha\right)\varepsilon}\right)^{2}(\log\left(1/\delta\right)+\log\left(8n\left|\mathbb{A}\right|\right)
		+\log\left(1+16/{\varepsilon\left(1-\alpha\right)}\right))
		\end{aligned}
		\end{equation*}
		given CVaR with parameter $\alpha\in[0,1).$
	\end{enumerate}	
\end{corollary}

The proof below puts (i) and (ii)
together and gives the main result.
\begin{IEEEproof}[Proof of Theorem \ref{thm:PAC bound}]
	The proof essentially follows the proof of \cite[Theorem 2]{munos2008finite}
	and states PAC-bounds on the sample size of sampling-based approximate
	value iteration. First, we state the key piece of the derivation of
	the error bounds. Recall that a stochastic kernel is $Q\mbox{ : }B\left(\mathbb{S};J_{\max}\right)\rightarrow B\left(\mathbb{S};J_{\max}\right)$
	such that $[QJ](s)$ is an expectation of
	$J\left(Y\right)$ w.r.t. some probability distribution,
	for all states $s\in\mathbb{S}$.
	\begin{lemma}
		\label{lem:p-norm} 
		\begin{enumerate}
			\item For any $\widehat{J}_{k}\in B\left(\mathbb{S};J_{\max}\right)$,
			there is a stochastic kernel $Q^{\pi_{k}}$ such that $T^{\pi^{*}}\widehat{J}_{k}-T^{\pi^{*}}J^{*}\leq\gamma Q^{\pi_{k}}(\widehat{J}_{k}-J^{*})$.
			\item For any $\widehat{J}_{k}\in B\left(\mathbb{S};J_{\max}\right)$,
			there is a stochastic kernel $Q^{\pi_{k}^{*}}$ such that $T^{\hat{\pi}_{k}}\widehat{J}_{k}-T^{\hat{\pi}_{k}}J^{*}\geq\gamma Q^{\pi_{k}^{*}}(\widehat{J}_{k}-J^{*})$.
		\end{enumerate}
	\end{lemma}
	
	Next, we apply Lemma \ref{lem:p-norm} and adapt \cite[Lemma 3]{munos2008finite} to obtain point-wise
	error bounds (i.e., bounds hold for any state $s\in\mathbb{S}$)
	for $\{ \widehat{J}_{k}\} _{k\geq0}$ relative to $J^{*}$
	with the approximation errors $\varepsilon_{k}$ defined in (\ref{eq:Error}). 
	\begin{lemma}
		\label{lem:Point error bounds} Choose $K\geq1$. We have
		\begin{equation*}		
		\begin{aligned}
		J^{\widehat{\pi}_{K}}-J^{*}\leq2Q_0
		\left\{ \sum_{k=0}^{K-1}\gamma^{K-k}Q_{1}|\varepsilon_{k}|
		+\gamma^{K+1}Q_{2}|\widehat{J}_{0}-J^{*}|\right\} 
		\end{aligned}
		\end{equation*}
		where  
		\begin{equation*}		
		\begin{aligned}
		Q_0=&(I-\gamma Q^{\hat{\pi}_{K}})^{-1},\\
		Q_{1}=&(Q^{\pi_{K}}Q^{\pi_{K-1}}\dots Q^{\pi_{k+1}}
		+Q^{\hat{\pi}_{K}}Q^{\pi_{K-1}^{*}}Q^{\pi_{K-2}^{*}}\dots Q^{\pi_{k+1}^{*}})/2,\\
		Q_{2}=&(Q^{\pi_{K}}Q^{\pi_{K-1}}\dots Q^{\pi_{0}}
		+Q^{\hat{\pi}_{K}}Q^{\pi_{K-1}^{*}}Q^{\pi_{K-2}^{*}}\dots Q^{\pi_{0}^{*}})/2.
		\end{aligned}
		\end{equation*}
	\end{lemma}
	
	We need to adapt \cite[Lemma 3]{munos2008finite} to get the previous
	lemma because the Bellman operator $T$ is not a contraction operator
	w.r.t. the $\mathcal{L}_{p}$ norm for $1\leq p<\infty$.
	The preceding point-wise error bounds suggest that if the sequence
	of errors $\{ \varepsilon_{k}\} _{k\geq0}$ is small then
	$\widehat{J}_{K}$ should be close to $J^{*}$ and the greedy policy
	$\widehat{\pi}_{K}$ w.r.t. $\widehat{J}_{K}$ should be
	close to optimal. The next lemma gives $\mathcal{L}_{p}$ bounds by
	using the point-wise error bounds in Lemma \ref{lem:Point error bounds}.
	\begin{lemma}
		\label{lem:Lp bounds} Let Assumption \ref{assu:Absolute_continuity}
		hold. For any $\eta>0$, there exists $K$ that is linear in $\log\left(1/\eta\right)$
		and $\log J_{\max}$ such that, if the $\mathcal{L}_{p}\left(\mu\right)-$norm
		of the approximation errors is bounded by some $\varepsilon$ ($\Vert \varepsilon_{k}\Vert _{p,\mu}\leq\varepsilon$
		for all $0\leq k<K$) then		
		$$
		\left\Vert J^{\widehat{\pi}_{K}}-J^{*}\right\Vert _{p,\varrho}\leq\frac{2\gamma}{\left(1-\gamma\right)^{2}}C_{\varrho,\mu}^{1/p}\varepsilon+\eta.
		$$
	\end{lemma}
	
	Next, we state a technical lemma without its proof. 
		\begin{lemma}[{\cite[Lemma 5]{munos2008finite}}]
			\label{lem:Conditional expectation} Assume that $X,Y$ are independent
			random variables taking values in the respective measurable spaces,
			$\mathcal{X}$ and $\mathcal{Y}$. Let $g:\mathcal{X}\times\mathcal{Y}\rightarrow\mathbb{R}$
			be a Borel-measurable function such that $\mathbb{E}[g\left(X,Y\right)]$
			exists. Assume that for all $y\in\mathcal{Y},$ $\mathbb{E}[g\left(X,y\right)]\geq0$.
			Then $\mathbb{E}[g(X,Y)|Y]\geq0$ holds, too,
			w.p. $1.$
		\end{lemma}
		
		Fix $\varepsilon,\delta>0.$ The aim is to show that by selecting
		the number of iterates $K$, and the number of samples $m,n$ large
		enough, the bound	
		\begin{equation}
		\left\Vert J^{\widehat{\pi}_{K}}-J^{*}\right\Vert _{p,\varrho}\leq\frac{2\gamma}{\left(1-\gamma\right)^{2}}C_{\varrho,\mu}^{1/p}d_{p,\mu}\left(T\mathcal{F},\:\mathcal{F}\right)+\varepsilon\label{eq:Final bound}
		\end{equation}
		holds w.p. at least $1-\delta.$ Note that by construction the iterates
		$\widehat{J}_{k}$ remain bounded by $J_{\max}.$ By Lemma \ref{lem:Lp bounds},
		under Assumption \ref{assu:Absolute_continuity}, for all those events,
		where the error $\varepsilon_{k}=T\widehat{J}_{k}-\widehat{J}_{k+1}$
		of the $k^{th}$ iterate is below (in $\mathcal{L}_{p}\left(\mu\right)-$norm)
		some level $\varepsilon_{0},$ we have		
		\begin{equation}
		\left\Vert J^{\widehat{\pi}_{K}}-J^{*}\right\Vert _{p,\varrho}\leq\frac{2\gamma}{\left(1-\gamma\right)^{2}}C_{\varrho,\mu}^{1/p}\varepsilon_{0}+\eta,\label{eq:K_th iterate}
		\end{equation}
		provided that $K=\Omega\left(\log\left(1/\eta\right)\right).$ Now
		choose 	
		\[
		\varepsilon'=\frac{\varepsilon\left(1-\gamma\right)^{2}}{4\gamma C_{\varrho,\mu}^{1/p}}\quad\mbox{and }\quad\eta=\frac{\varepsilon}{2}.
		\]
		Let $f\left(\varepsilon,\delta\right)$ denote the function that
		gives lower bounds on $m,n$ in Lemma \ref{lem:Error each iteration}
		based on the value of the desired estimation error $\varepsilon$ and
		confidence $\delta.$ Let $\left(n,m\right)\geq f\left(\varepsilon',\delta/K\right).$
		Let us denote the collection of random variables used in $k^{th}$
		step by $s_{k}.$ Hence, $s_{k}$ consists of the $n$ sampled states,
		as well as $\left|\mathbb{A}\right|\times n\times m$ next states.
		Further, introduce the notation $\widehat{J}\left(J,s_{k}\right)$
		to denote the result of solving the optimization problems (\ref{eq:Estimation})
		and (\ref{eq:Function fitting}) based on the samples $s_{k}$ and
		starting from the risk-to-go $J\in B\left(\mathbb{S};J_{\max}\right).$
		By Lemma \ref{lem:Error each iteration},	
		\[
		\mathbb{P}(\|\widehat{J}(J,s_{k})-TJ\|_{p,\mu}\leq d_{p,\mu}(TJ,\mathcal{F})+\varepsilon')\geq1-\delta/K.
		\]
		Apply Lemma \ref{lem:Conditional expectation} with $X=s_{k},$ $Y=\widehat{J}_{k}$
		and $g\left(s,J\right)=\mathbb{I}_{\left\{ \|\widehat{J}\left(J,s\right)-TJ\|_{p,\mu}\leq d_{p,\mu}\left(TJ,\mathcal{F}\right)+\varepsilon'\right\} }-\left(1-\delta/K\right).$
		Since $s_{k}$ is independent of $\widehat{J}_{k},$ the lemma can
		be applied. Therefore,	
		\[
		\mathbb{P}(\|\widehat{J}(\widehat{J}_{k},s_{k})-T\widehat{J}_{k}\|_{p,\mu}\leq d_{p,\mu}(T\widehat{J}_{k},\mathcal{F})+\varepsilon'\big|\widehat{J}_{K})\geq1-\delta/K.
		\]
		Taking expectation of both sides gives	
		\[
		\mathbb{P}(\|\widehat{J}(\widehat{J}_{k},s_{k})-T\widehat{J}_{k}\|_{p,\mu}\leq d_{p,\mu}(T\widehat{J}_{k},\mathcal{F})+\varepsilon')\geq1-\delta/K.
		\]
		Since $\widehat{J}(\widehat{J}_{k},s_{k})=\widehat{J}_{k+1}$
		and $\varepsilon_{k}=T\widehat{J}_{k}-\widehat{J}_{k+1},$ we thus
		have	
		$$
		\left\Vert \varepsilon_{k}\right\Vert _{p,\mu}\leq d_{p,\mu}\left(TJ,\mathcal{F}\right)+\varepsilon'
		$$
		holds except for a set of bad events $B_{k}$ of measure at most $\delta/K.$
		Hence, above inequality holds simultaneously for $k=1,\dots,K$ except
		for the events in $B=\cup_{k}B_{k}.$ Note that 	
		$$
		\mathbb{P}\left(B\right)\leq\sum_{k=1}^{K}\mathbb{P}\left(B_{k}\right)\leq\delta.
		$$	
		Now pick any event in the complement of \textbf{$B.$ }Thus, for such
		an event (\ref{eq:K_th iterate}) holds when $\varepsilon_{0}=d_{p,\mu}\left(TJ,\mathcal{F}\right)+\varepsilon'.$
		Plugging in the definition of $\varepsilon'$ and $\eta$ we obtain
		(\ref{eq:Final bound}).
\end{IEEEproof}

\subsection{Analysis in supremum norm}\label{sec6.1}

The convergence analysis in the supremum norm follows from the fact
that $T$ is a contracting operator as shown below.
\begin{lemma}
	\label{lem:Contraction} $|[TJ_{1}]\left(s\right)-[TJ_{2}]\left(s\right)|\leq\gamma\|J_{1}-J_{2}\|_{\infty}$
	for all $s\in\mathbb{S}$ and $J_{1},$ $J_{2}\in B\left(\mathbb{S};J_{\max}\right)$.
\end{lemma}

It follows that $\|TJ-J^{*}\|_{\infty}\leq\gamma\|J-J^{*}\|_{\infty}$
for all $J\in B\left(\mathbb{S};J_{\max}\right)$. Next, given the
true risk value $\rho(J(Y^{s,a}))$, we define
$\overline{T}\mbox{ : }B\left(\mathbb{S};J_{\max}\right)\rightarrow \mathcal{F}$
as the Bellman operator corresponding to the finite state space MDP
\[
\left[\overline{T}J\right]\left(s\right)=\min_{a\in\mathbb{A}}\left\{ c(s,a)+\gamma\rho\left(J\left(Y^{s,a}\right)\right)\right\},\forall s\in\mathcal{S}.
\]
The operator $\widetilde{T}\mbox{ : }B\left(\mathbb{S};J_{\max}\right)\rightarrow B\left(\mathbb{S};J_{\max}\right)$
is defined as an extension of $\overline{T}:$ for $s\in\mathbb{S},$
we can find $s'\in\mathcal{S}$ and $\Vert s-s'\Vert _{\infty}\leq\epsilon,$
such that
$$
[\widetilde{T}J](s)=[\overline{T}J](s')
$$
by Assumption \ref{assu:Epsilon_net_partition}. Moreover, we use
a random operator $\widehat{T}: B\left(\mathbb{S};J_{\max}\right)\rightarrow B\left(\mathbb{S};J_{\max}\right)$ to represent steps 1, 2, and 3 of
Algorithm \ref{algo}, i.e., the state space sampling over an $\epsilon-$net $\mathcal{S}$,
risk-to-go estimation $\{ \widetilde{J}_{k+1}\left(s\right)\} _{s\in\mathcal{S}}$
from $\hat{J}_{k}$, and function extension to produce $\widehat{J}_{k+1}\in B\left(\mathbb{S};J_{\max}\right)$
(we leave the dependence on the sample size $m\geq1$ in $\widehat{T}$
implicit for cleaner notation). The iterates $\{ \widehat{J}_{k}\} _{k\geq0}$
of our approximate value iteration algorithm then satisfy $\widehat{J}_{k+1}=\widehat{T}\widehat{J}_{k}$
for all $k\geq0$. 

Under Assumption \ref{assu:Bounded_costs}, the risk-to-go functions
are uniformly bounded by $J_{\max},$ and thus the worst error satisfies
$
\|\widehat{J}_{K}-J^{*}\|_{\infty}\leq J_{\max}.
$
When we use the random operator $\widehat{T}$, the error $\|\widehat{T}J-TJ\|_{\infty}$
is incurred and we have 
\begin{equation}
\begin{aligned}
\|\widehat{T}J-J^{*}\|_{\infty}&\leq\|TJ-J^{*}\|_{\infty}+\|\widehat{T}J-TJ\|_{\infty}\\
&\leq\gamma\|J-J^{*}\|_{\infty}+\|\widehat{T}J-TJ\|_{\infty}.\label{eq:infinity_norm}
\end{aligned}
\end{equation}
If the stochastic error term $\|\widehat{T}J-TJ\|_{\infty}$ is
small then $\widehat{T}$ is \textit{nearly} a contraction operator.
Based on this observation, inequality (\ref{eq:infinity_norm}) yields
our $\infty-$norm convergence analysis. The following lemma bounds
$\|\widehat{T}J-TJ\|_{\infty}.$ Its proof relies on the fact
$\|\widehat{T}J-TJ\|_{\infty}\leq\|\widehat{T}J-\widetilde{T}J\|_{\infty}+\|\widetilde{T}J-TJ\|_{\infty}.$

\begin{lemma}
	\label{lem:Contraction operators error} Let $\varepsilon>0.$
	Under Assumption \ref{assu:Bounded_costs}, \ref{assu:Risk_estimation},
	\ref{assu:Epsilon_net_regularity} and \ref{assu:Epsilon_net_partition},
	if the $\epsilon-$net is chosen such that
	$
	\epsilon\leq{\varepsilon}/{\left(2\kappa_{c}+2\gamma\kappa_{\mu}J_{\max}\right)},
	$
	we have	
	$$
	\mathbb{P}(\|\widehat{T}J-TJ\|_{\infty}\leq\varepsilon)\geq1-n|\mathbb{A}|\theta({\varepsilon}/({2\gamma}),m).
	$$	
\end{lemma}

The convergence result for the supremum norm algorithm then follows
immediately, as shown below.
\begin{IEEEproof}[Proof of Theorem \ref{thm:infinite_convergence}]
	Let the approximation errors $\varepsilon_{k}$ defined in \eqref{eq:Error} satisfy $\left\Vert \varepsilon_{k}\right\Vert _{\infty}\leq\varepsilon$
for all $k=0,\dots,K-1,$ and denote $p_{m,n}\left(\varepsilon\right)=n\left|\mathbb{A}\right|\theta\left({\varepsilon}/(2\gamma),m\right).$
Starting with $K=1$, we have
\begin{equation*}
\begin{aligned}
\|\widehat{J}_{1}-J^{*}\|_{\infty}&=\|\widehat{T}\widehat{J}_{0}-J^{*}\|_{\infty}\\
&\leq\|T\widehat{J}_{0}-J^{*}\|_{\infty}+\|\widehat{T}\widehat{J}_{0}-T\widehat{J}_{0}\|_{\infty}\\
&\leq\gamma\|\widehat{J}_{0}-J^{*}\|_{\infty}+\varepsilon,
\end{aligned}
\end{equation*}
with probability at least $1-p_{m,n}\left(\varepsilon\right)$ by Lemma
\ref{lem:Contraction} and \ref{lem:Contraction operators error}.
For $K=2$, by the union bound of probability, we have
\begin{equation*}
\begin{aligned}
\|\widehat{J}_{2}-J^{*}\|_{\infty}&\leq\gamma\|\widehat{J}_{1}-J^{*}\|_{\infty}+\varepsilon\\
&\leq\gamma^{2}\|\widehat{J}_{0}-J^{*}\|_{\infty}+\gamma\varepsilon+\varepsilon,
\end{aligned}
\end{equation*}
with probability at least $1-2p_{m,n}\left(\varepsilon\right)$. By
induction, for $K\geq1$, 
\begin{equation*}
\begin{aligned}
\|\widehat{J}_{K}-J^{*}\|_{\infty}&\leq\gamma\|\widehat{J}_{K-1}-J^{*}\|_{\infty}+\varepsilon\\
&\leq\gamma^{K}\|\widehat{J}_{0}-J^{*}\|_{\infty}+\sum_{k=0}^{K-1}\gamma^{K-k-1}\varepsilon
\end{aligned}
\end{equation*}
with probability at least $1-Kp_{m,n}\left(\varepsilon\right)$.	

Note that $\sum_{k=0}^{K-1}\gamma^{K-k-1}\leq1/\left(1-\gamma\right)$
for all $K\geq1$ and $\|\widehat{J}_{0}-J^{*}\|_{\infty}\leq J_{\max}.$
We obtain	
$$
\mathbb{P}\left(\|\widehat{J}_{K}-J^{*}\|_{\infty}\leq\gamma^{K}J_{\max}+{\varepsilon}/{(1-\gamma)}\right)\geq1-Kp_{m,n}\left(\varepsilon\right).
$$
\end{IEEEproof}

\subsection{Analysis via stochastic dominance}\label{sec6.3}

We first recall the approximation error $\varepsilon_{k}$ defined
in Equation (\ref{eq:Error}) that appears in approximate value iteration:
\[
\widehat{J}_{k+1}=T\widehat{J}_{k}-\varepsilon_{k},\quad\forall k\geq0.
\]
The following inequalities form the foundation of our stochastic dominance
analysis, they give bounds on the approximation error for both the
supremum and $p-$norms.
\begin{lemma}
	\label{lem:Stoc_approximation error} 
		\begin{enumerate}
			\item Let Assumption \ref{assu:Bounded_costs}
			hold. If $\left\Vert \varepsilon_{k}\right\Vert _{\infty}\leq\varepsilon$
			for all $0\leq k<K,$ we have
			\begin{equation}
			\Vert \widehat{J}_{K}-J^{*}\Vert _{\infty}\leq\gamma^{K}J_{\max}+\frac{\varepsilon}{1-\gamma}.\label{eq:Error bound-1}
			\end{equation}
			\item Let Assumption \ref{assu:Bounded_costs} and \ref{assu:Absolute_continuity}
			hold. If $\left\Vert \varepsilon_{k}\right\Vert _{p,\mu}\leq\varepsilon$
			for all $0\leq k<K,$ we have			
			\begin{equation}
			\begin{aligned}
			\Vert J^{\widehat{\pi}_{K}}-J^{*}\Vert _{p,\varrho}\leq&\frac{2\gamma}{\left(1-\gamma\right)^{2}}[C_{\varrho,\mu}^{1/p}\varepsilon
			+\gamma^{K/p}(1-\gamma)^{1/p}
			(1-\gamma^{K+1})^{1-1/p}J_{\max}].\label{eq:Error bound-2}
			\end{aligned}
			\end{equation}
		\end{enumerate}	
\end{lemma}

We remark that the above results hold by assuming that the approximation
error in\textbf{ }\emph{all}\textbf{ }iterations $k=0,\dots,K-1$
falls below the tolerance $\varepsilon.$ The iteration count $K$
is chosen to control the error $J_{\max}.$ 

Note that the RHS of the inequalities (\ref{eq:Error bound-1}) and
(\ref{eq:Error bound-2}) do not depend on the initial error between
$\widehat{J}_{0}$ and $J^{*}$, it depends on the \textit{worst-case}
error $J_{\max}$. For the rest of this section, let us fix an error
tolerance $\varepsilon>0$. Once $\varepsilon>0$ is fixed, we consider
an iteration ``good'' if the error falls below $\varepsilon$, and
we consider an iteration to be ``bad'' if the error exceeds $\varepsilon$.
Once $\varepsilon>0$ is fixed, inequalities (\ref{eq:Error bound-1})
and (\ref{eq:Error bound-2}) give us guidance on how many ``good''
iterations $K$ are required to reach a desired approximation error.
Again, this number $K$ can be chosen directly from the inequalities
(\ref{eq:Error bound-1}) and (\ref{eq:Error bound-2}), even though
the latter inequality does not follow from a contraction argument
as it is originally done in $\infty-$norm in \cite{Haskell_EDP_2015}.

Now we are in a position to use our stochastic dominance convergence
analysis. Consider a probability space $(\Omega,\mathcal{B}(\Omega),P)$
where $\Omega$ is a sample space with elements denoted $\omega\in\Omega$,
$\mathcal{B}\left(\Omega\right)$ is the Borel $\sigma-$algebra on
$\Omega$, and $P$ is a probability distribution on $\left(\Omega,\mathcal{B}(\Omega)\right)$.
In our upcoming algorithms, $\left(\Omega,\mathcal{B}(\Omega),P\right)$
corresponds to the randomness used to drive one round of simulation.
We are interested in repeated samples from $\left(\Omega,\mathcal{B}(\Omega),P\right)$,
so we define the space of sequences $(\Omega^{\infty},\mathcal{B}(\Omega^{\infty}),\mathcal{P})$
where $\Omega^{\infty}=\times_{k=0}^{\infty}\Omega$ with elements
denoted $\boldsymbol{\omega}=(\omega_{k})_{k\geq0}$, $\mathcal{B}(\Omega^{\infty})=\times_{k=0}^{\infty}\mathcal{B}(\Omega)$,
and $\mathcal{P}$ is the probability measure on $(\Omega^{\infty},\mathcal{B}(\Omega^{\infty}))$
guaranteed by the Kolmogorov extension theorem applied to $P$. Let
$\{ X_{k}\} _{k\geq0}$ be a stochastic process on $(\Omega^{\infty},\mathcal{B}(\Omega^{\infty}),\mathcal{P})$
with the integer-valued state space $\{ 0,1,\ldots,K^{*}\} $
where $K^{*}$ is an upper bound on $\{ X_{k}\} _{k\geq0}$.

Let $\left\lceil x\right\rceil $ denote the smallest integer greater
than or equal to $x\in\mathbb{R}$ and $\epsilon_{g}>0$ be a granularity.
The stochastic process $\{ X_{k}\} _{k\geq0}$ on a discrete
and finite state space is defined by 
\[
X_{k}=\left\lceil \|J^{k}-J^{*}\|/\epsilon_{g}\right\rceil ,
\]
where $J^{k}=\widehat{J}_{k}$ when $\|\cdot\|$ is the $\infty-$norm
and $J^{k}=J^{\widehat{\pi}_{k}}$ when $\|\cdot\|$ is the $p-$norm.
Since $\|J^{k}-J^{*}\|\leq J_{\max},$ we define a constant 
$$
K^{*}\triangleq\left\lceil {J_{\max}}/{\epsilon_{g}}\right\rceil .
$$
Notice that $K^{*}$ is the smallest number of intervals of length
$\epsilon_{g}$ needed to cover the interval $[0,J_{\max}].$
By construction, the stochastic process $\{ X_{k}\} _{k\geq0}$
is restricted to the finite state space $\{ \eta\in\mathbb{N}:0\leq\eta\leq K^{*}\} .$
If we could understand the behavior of the stochastic process $\{ X_{k}\} _{k\geq0},$
then we could analysis the convergence of $\{\|J^{k}-J^{*}\|\} _{k\geq0}.$
Throughout this paper, $\{ X_{k}\} _{k\geq0}$ will represent
the error between a risk-to-go function estimate and the optimal risk-to-go
function in the simulation-based approximate value iteration algorithms.

Consider the state space
$
\{ 1,2,\ldots,K^{*}\} ,
$
where state $K^{*}$ corresponds to the worst case starting error
$J_{\max}$ and state $1$ corresponds to the desired approximation
error. In other words, if we have a string of $K^{*}$ ``good'' iterations,
we are able to reach our desired performance. We are thus interested
in studying the convergence of $\{ X_{k}\} _{k\geq0}$
to zero. We next make an assumption about the behavior of $\{ X_{k}\} _{k\geq0}$.
\begin{assumption}
	\label{assu:Probability} For $\varepsilon>0$ and all $k\geq0$,
	$\mathbb{P}(\Vert J^{k}-J^{*}\Vert \leq\varepsilon)\geq p$
	with probability $p\in(0,1)$. Here $J^{k}=\widehat{J}_{k}$
	when $\|\cdot\|$ is the $\infty-$norm and $J^{k}=J^{\widehat{\pi}_{k}}$
	when $\|\cdot\|$ is the $p-$norm. 
\end{assumption}

The choice of $p$ in Assumption \ref{assu:Probability} depends on
the specifics of $\left\{ X_{k}\right\} _{k\geq0}$, and we now discuss:
in the supremum-norm analysis, we choose $p\triangleq1-n|\mathbb{A}|\theta({\varepsilon}/{(2\gamma)},m)$
since 
$$
\mathbb{P}(\|\widehat{T}\widehat{J}_{k}-T\widehat{J}_{k}\|_{\infty}\leq\varepsilon)\geq1-n\left|\mathbb{A}\right|\theta({\varepsilon}/{(2\gamma)},m)
$$
from Lemma \ref{lem:Contraction operators error}; in the $p-$norm
analysis, we set $p\triangleq1-\delta$ since 
$$
\mathbb{P}(\|\widehat{J}_{k+1}-T\widehat{J}_{k}\|_{p,\mu}\leq d_{p,\mu}(T\widehat{J}_{k},\mathcal{F})+\varepsilon)\geq1-\delta
$$
from Lemma \ref{lem:Error each iteration}. As shown before, we are
able to control $p$ in Assumption \ref{assu:Probability} by improving
the quality of our simulation-based approximate value iteration algorithms
with more samples and also by choosing a richer functional family.

Based on Assumption \ref{assu:Probability}, we can construct a ``dominating''
Markov chain $\left\{ Y_{k}\right\} _{k\geq0}$ to help us analyze
the behavior of $\left\{ X_{k}\right\} _{k\geq0}$. We construct $\left\{ Y_{k}\right\} _{k\geq0}$
on $\left(\mathbb{N}^{\infty},\mathcal{N}\right)$, the canonical
measurable space of trajectories on $\mathbb{N}$, so $Y_{k}\mbox{ : }\mathbb{N}^{\infty}\rightarrow\mathbb{N}$.
We will use $\mathcal{Q}$ to denote the probability measure of $\{ Y_{k}\} _{k\geq0}$
on $\left(\mathbb{N}^{\infty},\mathcal{N}\right)$. Since $\left\{ Y_{k}\right\} _{k\geq0}$
will be a Markov chain by construction, the probability measure $\mathcal{Q}$
is completely determined by an initial distribution on $\mathbb{N}$
and a transition kernel for $\left\{ Y_{k}\right\} _{k\geq0}$ denoted
$\mathfrak{Q}$. We restrict $\left\{ Y_{k}\right\} _{k\geq0}$ to
the finite state space $\left\{ 1,2,\ldots,K^{*}-1,K^{*}\right\} $.
We then define:
\[
Y_{k+1}=\begin{cases}
\max\left\{ Y_{k}-1,1\right\} , & \mbox{w.p. }p,\\
K^{*}, & \mbox{w.p. }1-p,
\end{cases}
\]
where $p$ is the same one from Assumption \ref{assu:Probability}
and is the probability of a ``good'' iteration (where the error falls
bellow $\epsilon_{g}$). In words, $\left\{ Y_{k}\right\} _{k\geq0}$
moves one unit closer to state 1 with probability $p$ (corresponding
to a ``good'' iteration) or moves back to the starting worst-case
error $J_{\max}$ with probability $1-p$, corresponding to a ``bad''
iteration. Notice that this bound is extremely conservative, because
we always assume that a ``bad'' iteration is so bad that it resets
the entire process. Moreover, any time we are in state $1$, we know
that we have reached the desired performance level: if we are in state
$1$, and we have a ``good'' iteration, then we remain in state $1$
(since the RHS of both inequalities (\ref{eq:Error bound-1}) and
(\ref{eq:Error bound-2}) is decreasing in $K$, so more ``good''
iterations than we need does not increase the approximation error).

We now show that $\{ X_{k}\} _{k\geq0}$ and $\{ Y_{k}\} _{k\geq0}$
have a stochastic dominance relationship. The following definition
gives the notion of (first-order) stochastic dominance (see \cite{Shaked2007}).
\begin{definition}
	Let $X$ and $Y$ be two real-valued random variables. Y stochastically
	dominates $X,$ written as $X\leq_{st}Y,$ when $\mathbb{P}\left(X\geq\theta\right)\leq\mathbb{P}\left(Y\geq\theta\right)$
	for all $\theta$ in the support of $Y.$
\end{definition}

The theorem below compares the marginal distributions of $\{ X_{k}\} _{k\geq0}$
and $\{ Y_{k}\} _{k\geq0}$ at all times $k\geq0$ when
the two stochastic processes $\{ X_{k}\} _{k\geq0}$ and
$\{ Y_{k}\} _{k\geq0}$ start from the same state.
\begin{lemma}
	\label{lem:Stochastic dominance} Under Assumption \ref{assu:Probability}.
	If $X_{0}=Y_{0},$ then $X_{k}\leq_{st}Y_{k}$ for all $k\geq0.$
\end{lemma}

Next we compute the steady state distribution of the Markov chain
$\{ Y_{k}\} _{k\geq0}.$ Let $\mu$ denote the steady state
distribution of $Y=_{d}\lim_{k\rightarrow\infty}Y_{k},$ whose existence
is guaranteed since $\{ Y_{k}\} _{k\geq0}$ is an irreducible
Markov chain on a finite state space. Denote $\mu\left(i\right)=\mathcal{Q}\left(Y=i\right)$
for $i\in\left\{ 1,2,\ldots,K^{*}\right\} .$ The next lemma gives
$\{\mu(i)\} _{i=1}^{K^{*}}.$
\begin{lemma}
	\label{lem:Stationary distribution} Under Assumption \ref{assu:Probability}.
	The values of $\{\mu\left(i\right)\} _{i=1}^{K^{*}}$
	are 	
	$
	\mu\left(1\right)=p^{K^{*}-1},\quad\mu\left(K^{*}\right)=1-p,$ and   $\mu\left(i\right)=\left(1-p\right)p^{K^{*}-i},$ $i=2,\dots,K^{*}-1.$
\end{lemma}

We are now ready to prove Theorem \ref{thm:Sample complexity}.
\begin{IEEEproof}[Proof of Theorem \ref{thm:Sample complexity}]

	First, we use Lemma \ref{lem:Stochastic dominance} and \ref{lem:Stationary distribution}
	to derive an asymptotic result.
	\begin{proposition}
		\label{prop:Asymptotic bound} For any $\delta_{1}\in\left(0,1\right).$
		\begin{enumerate}
			\item Select $\varepsilon<\epsilon_{g},$ $\epsilon$ and $m$ such
			that			
			\[
			\epsilon\leq\frac{\varepsilon}{2\left(\kappa_{c}+\gamma\kappa_{\mu}J_{\max}\right)}
			\]
			and 			
			\[
			\theta\left(\frac{\epsilon_{g}}{2\gamma},m\right)\leq\frac{\delta_{1}}{\left|\mathbb{A}\right|\left|\mathcal{S}\right|},
			\]
			then $\limsup_{k\rightarrow\infty}\mathbb{P}(\|\widehat{J}_{k}-J^{*}\|_{\infty}>\epsilon_{g})\leq\delta_{1}.$
			\item Select $\varepsilon<\epsilon_{g}-d_{p,\mu}\left(T\mathcal{F},\mathcal{F}\right),$
			$n$ and $m$ such that 			
			\[
			n>128\left(\frac{8J_{\max}}{\varepsilon}\right)^{2p}\left(\log\left(1/\delta_{1}\right)+\log\left(32\mathcal{N}_{0}\left(n\right)\right)\right)
			\]
			and			
			\[
			\theta\left(\varepsilon/4,m\right)\leq\frac{\delta_{1}}{4n\left|\mathbb{A}\right|},
			\]
			then $\limsup_{k\rightarrow\infty}\mathbb{P}(\|J^{\widehat{\pi}_{k}}-J^{*}\|_{p,\varrho}\geq\epsilon_{g})\leq\delta_{1}.$
		\end{enumerate}
	\end{proposition}
	
	Our earlier Lemma \ref{lem:Stationary distribution} gives the stationary
	distribution of $\{ Y_{k}\} _{k\geq0}$. To continue, we
	will use a mixing time argument to find out how ``close'' $\{ Y_{k}\} _{k\geq0}$
	is to its stationary distribution as a function of time. The total
	variation distance between two probability measures $\mu$ and $\nu$
	on $\mathbb{S}$ as 
	\[
	\bigl\Vert\mu-\nu\bigr\Vert_{TV}=\max_{S\subset\mathbb{S}}\bigl|\mu\left(S\right)-\nu\left(S\right)\bigr|=\frac{1}{2}\int_{\mathbb{S}}\bigl|\mu\left(ds\right)-\nu\left(ds\right)\bigr|.
	\]
	Let $\mathcal{Q}_{k}$ be the marginal distribution of $Y_{k}$ on
	$\mathbb{N}$ at stage $k$ and $d\left(k\right)=\Vert\mathcal{Q}_{k}-\mu\Vert_{TV}$
	be the total variation distance between $\mathcal{Q}_{k}$ and the
	steady state distribution $\mu.$ For $\delta_{2}>0,$ we define 	
	\[
	t_{mix}\left(\delta_{2}\right)=\min\left\{ k:d\left(k\right)\leq\delta_{2}\right\} 
	\]
	to be the minimum length of time needed for the marginal distribution
	of $Y_{k}$ to be within $\delta_{2}$ of the steady state distribution
	in total variation norm. By \cite[Theorem 12.3]{levin2009markov},
	$t_{mix}\left(\delta_{2}\right)$ can be bounded as below.
	\begin{lemma}
		\label{lem:Mixing time} For any $\delta_{2}>0,$	we have	
		$$
		t_{mix}\left(\delta_{2}\right)\leq\log\left(\frac{1}{\delta_{2}\mu_{\min}}\right)
		$$
		where $\mu_{\min}\triangleq\min_{\eta}\mu\left(\eta\right).$
	\end{lemma}
	
	Next, we use the above bound on mixing time to get a non-asymptotic
	bound.
	\begin{proposition}
		\label{prop:Non-asymptotic bound} For $k\geq\log\left({1}{(\delta_{2}\mu_{\min})}\right),$
		we have
		\begin{enumerate}
			\item $\mathbb{P}(\|\widehat{J}_{k}-J^{*}\|_{\infty}>\epsilon_{g})\leq1+2\delta_{2}-\mu\left(1\right).$
			\item $\mathbb{P}(\|J^{\widehat{\pi}_{k}}-J^{*}\|_{p,\varrho}>\epsilon_{g})\leq1+2\delta_{2}-\mu\left(1\right).$
		\end{enumerate}
	\end{proposition}
	
	Finally, combing Proposition \ref{prop:Asymptotic bound} and \ref{prop:Non-asymptotic bound},
	we prove Theorem \ref{thm:Sample complexity} 1) and 2).
	\begin{enumerate}
		\item Let $\delta_{1},\delta_{2}>0$ and $\delta_{1}+2\delta_{2}\leq\delta.$
		By the choice of $\varepsilon,\epsilon$ and $n$ in Proposition
		\ref{prop:Asymptotic bound} 1), we have
		\[
		\limsup_{k\rightarrow\infty}\mathbb{P}(\|\widehat{J}_{k}-J^{*}\|_{\infty}\geq\epsilon_{g})\leq1-\mu(1)\leq\delta_{1}.
		\]
		For $k\geq\log({1}/{(\delta_{2}\mu_{\min})}),$ by Proposition
		\ref{prop:Non-asymptotic bound} 1), we have
		\[
		\mathbb{P}(\|\widehat{J}_{k}-J^{*}\|_{\infty}>\epsilon_{g})\leq1+2\delta_{2}-\mu\left(1\right)\leq\delta_{1}+2\delta_{2}.
		\]
		Combining both inequalities, we obtain $\mathbb{P}(\|\widehat{J}_{k}-J^{*}\|_{\infty}>\epsilon_{g})\leq\delta$.
		\item Let $\delta_{1},\delta_{2}>0$ and $\delta_{1}+2\delta_{2}\leq\delta.$
		By the choice of $\varepsilon,m$ and $n$ in Proposition \ref{prop:Asymptotic bound}
		2), we have
		\[
		\limsup_{k\rightarrow\infty}\mathbb{P}(\|J^{\widehat{\pi}_{k}}-J^{*}\|_{p,\varrho}\geq\epsilon_{g})\leq1-\mu\left(1\right)\leq\delta_{1}.
		\]
		For $k\geq\log\left({1}{(\delta_{2}\mu_{\min})}\right),$ by Proposition
		\ref{prop:Non-asymptotic bound} 2), we have
		\[
		\mathbb{P}(\|J^{\widehat{\pi}_{k}}-J^{*}\|_{p,\varrho}>\epsilon_{g})\leq1+2\delta_{2}-\mu\left(1\right)\leq\delta_{1}+2\delta_{2}.
		\]
		Combining both inequalities, we obtain $\mathbb{P}\left(\|J^{\widehat{\pi}_{k}}-J^{*}\|_{p,\varrho}>\epsilon_{g}\right)\leq\delta$.
	\end{enumerate}	
\end{IEEEproof}

\section{Numerical experiments}\label{sec7}

In this section, we report some simulation results that illustrate
the performance of the methods developed in this paper.

\subsection{An optimal maintaining problem}

We consider a continuous one-dimensional optimal maintaining problem
which is similar in spirit to the one in \cite{munos2008finite}.
The state variable $s_{t}\in\mathbb{R}_{+}$ measures the accumulated
utilization of a piece of equipment. The larger the value of the state,
the worse the condition of the product; $s_{t}=0$ represents a brand
new equipment. In addition, there is an absorbing ``bad'' state
$s^{\text{bad}}$ that corresponds to broken equipment. 

At each time $t\geq0$, one can either keep ($a_{t}=\text{K}$) or repair
($a_{t}=\text{R}$) the existing equipment. The bad state models the situation
where the equipment is broken and cannot be operated or repaired,
and so $P(s_{t+1}^{\text{bad}}|s_{t}^{\text{bad}},a)=1$. When action
$\text{K}$ is chosen at time step $t$, the transition to a new state has
a mixture distribution: with probability $q$ the new state is $s_{t+1}^{\text{bad}},$
and with probability $1-q$ next state follows the exponential density:
\[
P\left(s_{t+1}|s_{t},\text{K}\right)=\begin{cases}
\beta e^{-\beta\left(s_{t+1}-s_{t}\right)}, \quad& \text{if}\,\,\mbox{\ensuremath{s_{t+1}\geq s_{t}};}\\
0, \quad& \text{otherwise.}
\end{cases}
\]
When action $\text{R}$ is taken at time step $t\geq0$, the next state follows
\[
P\left(s_{t+1}|s_{t},\text{R}\right)=\begin{cases}
\beta e^{-\beta s_{t+1}}, \quad& \text{if}\,\,\mbox{\ensuremath{s_{t+1}\geq0};}\\
0, \quad& \text{otherwise.}
\end{cases}
\]
The cost function is $c(s,\text{K})=f(s)$ where
the monotonically increasing function $f\left(s\right)$ is the cost
of operating the equipment when its condition is $s$. The cost associated
with the repair of the equipment is independent of the state and is
given by $c(s,\text{R})=C_{1}+f(0).$ Finally, the
penalty of breaking the equipment is $c(s^{\text{bad}},a)=C_{2}.$ 

We consider both risk-neutral and risk-aware decision makers, where
the risk-aware decision maker seeks to minimize the Markovian conditional
value-at-risk (CVaR) of his discounted cost. In the risk-neutral case,
the optimal policy $\pi_{\text{neutral}}^{*}$ solves Problem (\ref{NEUTRAL})
and satisfies
\begin{equation*}
\begin{aligned}
\pi_{\text{neutral}}^{*}\left(s\right)\in\arg\min_{a\in\mathbb{A}}\bigg\{c(s,a)
+\gamma\int_{0}^{\infty}P\left(ds'|s,a\right)J_{\text{neutral}}^{*}\left(s'\right)\bigg\} ,
\end{aligned}
\end{equation*}
where $J_{\text{neutral}}^{*}\left(s\right)$ is the classical cost-to-go
function representing the optimal expected total discounted cost when
the process is started from state $s.$ Given a confidence level $\alpha\in[0,1),$
a Markovian CVaR minimizing risk-aware decision maker chooses
\begin{equation*}
\begin{aligned}
\pi_{\text{CVaR}}^{*}\left(s\right)\in\arg\min_{a\in\mathbb{A}}\bigg\{c(s,a)
+\gamma\min_{\eta\in\left[0,J_{\max}\right]}[\eta
+\frac{1}{1-\alpha}\int_{0}^{\infty}P\left(ds'|s,a\right)\left(J^{*}\left(s'\right)-\eta\right)_{+}]\bigg\} .
\end{aligned}
\end{equation*}
We next compare the performance these two decision makers.

\subsection{Result}

We choose values $\gamma=0.6,$ $\beta=0.5,$ $q=0.2$, $C_{1}=30,$
$C_{2}=120$ and $f\left(s\right)=4s.$ Similar to \cite{munos2008finite},
we use state space truncation. In order to make the state space bounded,
we fix an upper bound $s_{\max}=30$ for the state. We then modify
the problem definition so that if the next state is outside the interval
$\left[0,s_{\max}\right]$, then the equipment is immediately repaired,
and then a new state is drawn as if the action $\text{R}$ were chosen in
the previous step. By the choice of $s_{\max},$ the probability $\int_{s_{\max}}^{\infty}P\left(ds'|s,a\right)$
is negligible and hence $J_{\text{\text{neutral}}}^{*}$ and $J^{*}$ of the modified
problem closely match that of the original problem. We let $s^{\text{bad}}=30$
denote the bad state where the equipment is broken.

For both the risk-neutral and risk-aware cases, we consider approximations
of risk-to-go-functions using polynomials of degree $l=4$ and we
choose the distribution $\mu$ to be uniform over the state space
$\left[0,s_{\max}\right].$ The number of iterations is set to $K=30$
and the number of samples is fixed at $m=n=100.$ We compute the best
fit in functional family $\mathcal{F}$ (for
$l=4$) by minimizing the least square error to the data, i.e., $p=2.$

We take the sampling distribution $\mu$ to be a mixture of a uniform
distribution on the state space with a point mass on $s^{\text{bad}}$. For
our experiments, we choose the uniform distribution with probability
$0.95$ and choose $s^{\text{bad}}$ with probability $0.05$. As discussed
in Section \ref{sec3.2}, fix state $s\in\mathbb{S},$ action $a\in\mathbb{A}$
and $\alpha\in[0,1),$ the distributional set $\mathcal{Q}\left(s,a\right)$
for Markovian CVaR is given by
$$\mathcal{Q}(s,a)=\left\{h:\begin{aligned}&0\leq h\left(s'\right)\leq\left(1-\alpha\right)^{-1},\,\text{a.e.}\, s'\in\mathbb{S},\\&\int_{0}^{\infty}h\left(s'\right)P\left(ds'|s,a\right)=1\end{aligned}\right\}.$$ 
Since the Radon-Nikodym derivatives $h$ of distributions $Q\left(\cdot\,|s,a\right)\in\mathcal{Q}\left(s,a\right)$
with respect to $\mu$ are bounded by $\left(1-\alpha\right)^{-1},$
Assumption \ref{assu:Absolute_continuity} holds with $C_{\mu}=\left(1-\alpha\right)^{-1}$
by Lemma \ref{lem:C}.

Let the initial state be $s_{0}=0.$ Table \ref{tab:decision boundaries}
shows the decision boundaries of the stationary policies $\pi_{\text{neutral}}^{*}$
and $\pi_{\text{CVaR}}^{*}.$ It can be seen that the decision boundaries
of the risk-neutral and Markovian CVaR policies begin to match as
$\alpha$ approaches zero.

\begin{table}[!tph]
	\centering
	\caption{Decision boundaries of policies
		$\pi_{\text{\textnormal {neutral}}}^{*}$ and $\pi_{\text{CV\textnormal{a}R}}^{*}$. 
		\label{tab:decision boundaries}}
	\begin{tabular}{|c|c|}
		\hline
		Policies            & Decision boundaries                                                     \\ \hline
		Risk-Neutral        & $\pi^*_{\text{neutral}}\left(s\right)=\text{K}$ if $s\leq5.3$                    \\ \hline
		$\text{CVaR}_{0.1}$ & $\pi^*_{\text{CVaR}}\left(s\right)=\text{K}$ if $s\leq5.1$                       \\ \hline
		$\text{CVaR}_{0.2}$ & $\pi^*_{\text{CVaR}}\left(s\right)=\text{K}$ if $s\leq5.0$                       \\ \hline
		$\text{CVaR}_{0.3}$ & $\pi^*_{\text{CVaR}}\left(s\right)=\text{K}$ if $s\leq3.5$                       \\ \hline
		$\text{CVaR}_{0.4}$ & $\pi^*_{\text{CVaR}}\left(s\right)=\text{K}$ if $s\leq3.2$                       \\ \hline
		$\text{CVaR}_{0.5}$ & $\pi^*_{\text{CVaR}}\left(s\right)=\text{K}$ if $s\leq3.0$                       \\ \hline
		$\text{CVaR}_{0.6}$ & $\pi^*_{\text{CVaR}}\left(s\right)=\text{K}$ if $s\leq2.6$                       \\ \hline
		$\text{CVaR}_{0.7}$ & $\pi^*_{\text{CVaR}}\left(s\right)=\text{K}$ if $s\leq0.8$                       \\ \hline
		$\text{CVaR}_{0.8}$ & $\pi^*_{\text{CVaR}}\left(s\right)=\text{R}$ for $s\in\left[0,s_{\max}\right]$ \\ \hline
		$\text{CVaR}_{0.9}$ & $\pi^*_{\text{CVaR}}\left(s\right)=\text{R}$ for $s\in\left[0,s_{\max}\right]$ \\ \hline
	\end{tabular}
\end{table}

\begin{figure}[!tph]
	\begin{centering}
		\includegraphics[scale=0.5]{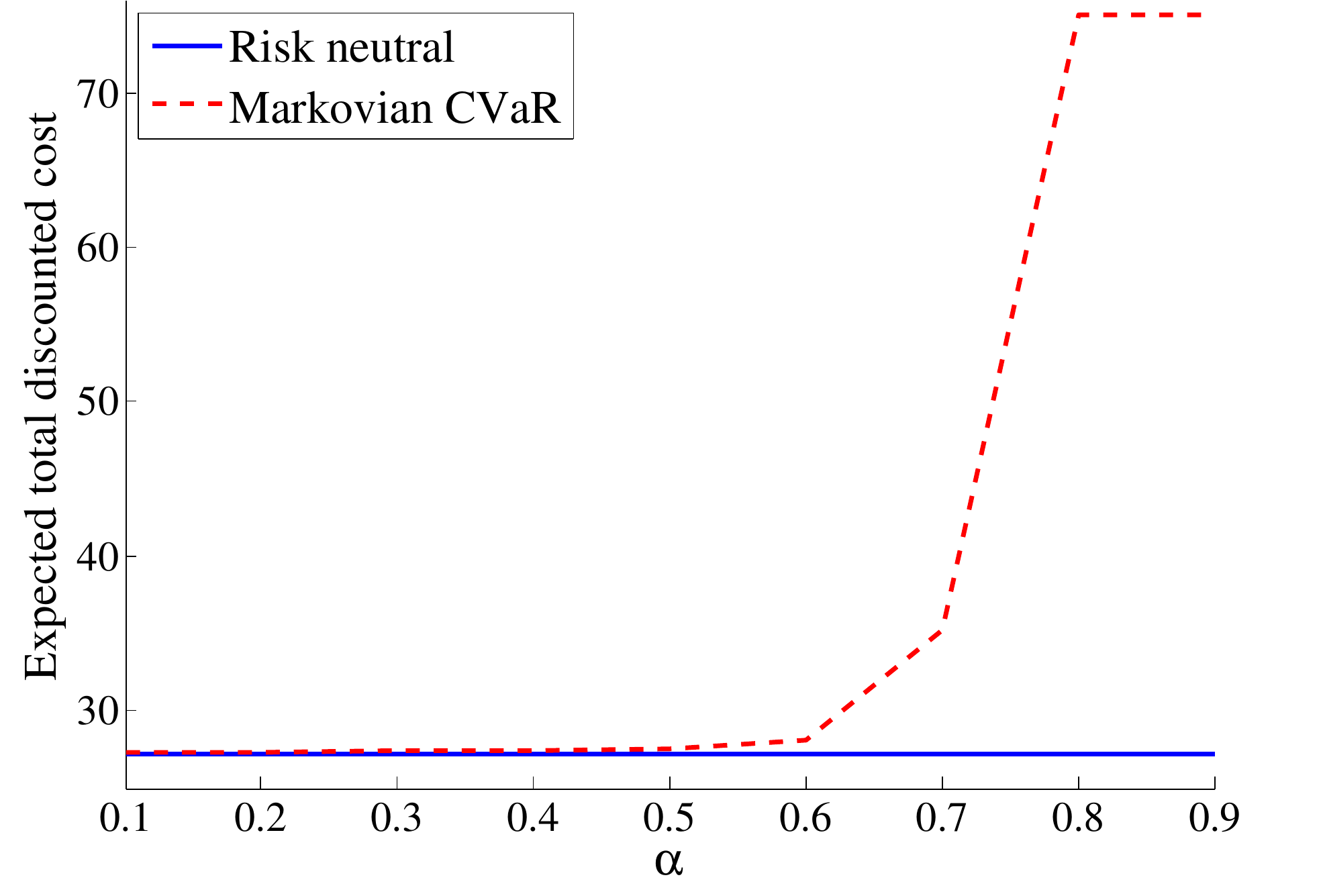}
		\par\end{centering}
		\caption{Expected total discounted cost by following risk-neutral and Markovian
		CVaR policies.\label{fig:expected cost}}
\end{figure}

Fig. \ref{fig:expected cost} illustrates the expected total discounted
cost (averaged over $5,000$ runs) incurred by following policies
$\pi_{\text{neutral}}^{*}$ and $\pi_{\text{CVaR}}^{*}.$ Since both policies are
similar when $\alpha$ is small, the performances of the two is close
as expected. From Table \ref{tab:decision boundaries}, when $\alpha$
is large (say $\alpha=0.9$) the Markovian CVaR policy becomes conservative
and chooses to repair in every state. This choice leads to a huge
expected total cost as observed in Fig. \ref{fig:expected cost}.

\begin{figure}[!tph]
	\begin{centering}
		\includegraphics[scale=0.5]{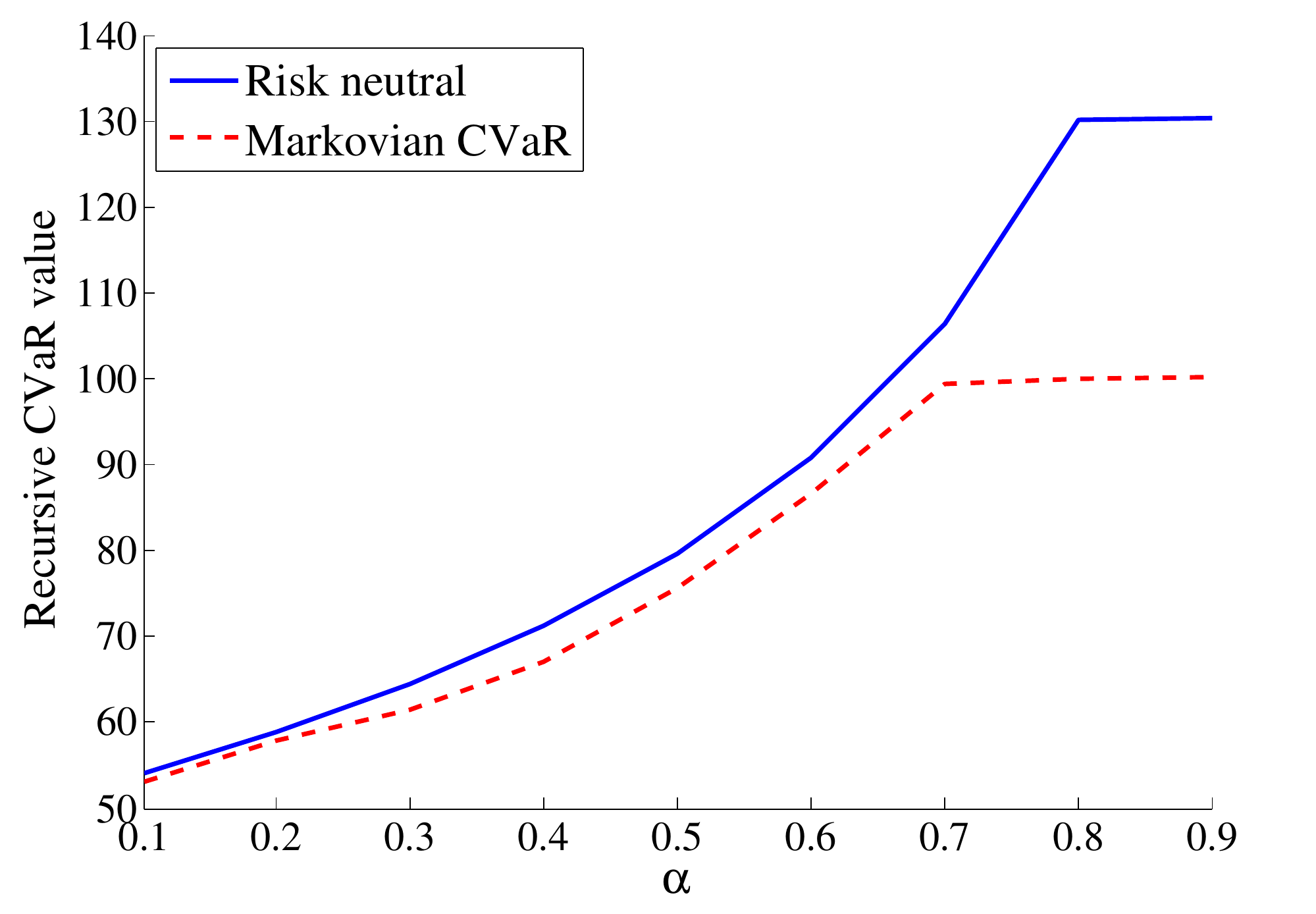}
		\par\end{centering}	
		\caption{Recursive CVaR value of risk-neutral and Markovian CVaR policies.
		\label{fig:CVaR value}}
\end{figure}

Fig. \ref{fig:CVaR value} shows the recursive CVaR value for stationary
policies $\pi_{\text{neutral}}^{*}$ and $\pi_{\text{CVaR}}^{*}.$ From Table \ref{tab:decision boundaries},
when $\alpha$ is large, $\pi_{\text{CVaR}}^{*}$ prevents the decision maker
from keeping the equipment (i.e., $a=\text{K}$), thus reducing the chance
of reaching the bad state $s^{\text{bad}}$ and incurring a large cost.

\section{Conclusion}\label{sec8}
In this paper, we have extended simulation-based approximate value
iteration algorithms for classical risk-neutral MDPs to the risk-aware
setting. This work is significant because it shows that, under mild
technical assumptions, risk-aware sequential decision-making can be
done efficiently on large scales. Our algorithms apply to the whole class of Markov risk measures, and generalize several recent studies that focused on specific risk measures. Most importantly, we are able to
give finite time bounds (instead of asymptotic bounds) in both supremum and $p-$norms on the solution
quality of our algorithms so that decision makers may know the quality
of the resulting policies as a function of computational effort. 

We have two main directions for future work. First, Markov
risk measures developed in \cite{ruszczynski2010risk}  naturally lead to dynamic programming formulations. Yet, there
are still many risk-aware MDPs (e.g., \cite{bauerle2013more})
that do not satisfy the time consistency axiom. We wish to develop simulation-based algorithms for those models as well. Second,
we are interested in creating \emph{online} algorithms for risk-aware
MDPs, such as variants of Q-learning (e.g., \cite{bertsekas1996neuro}).
This work would have high impact because it would allow controllers
of complex systems to manage risk in real time.

\bibliographystyle{IEEEtran}
\bibliography{references}

\newpage
\appendix  

%

The following well known fact will be used throughout our analysis
in this paper, we mention it here for ease of reference.
\begin{fact}
	\label{Fact} Let $X$ be a given set, and $f_{1}\mbox{ : }X\rightarrow\mathbb{R}$
	and $f_{2}\mbox{ : }X\rightarrow\mathbb{R}$ be two real-valued functions
	on $X$. Then,
	\begin{enumerate}
		\item $|\inf_{x\in X}f_{1}\left(x\right)-\inf_{x\in X}f_{2}\left(x\right)|\leq\sup_{x\in X}|f_{1}\left(x\right)-f_{2}\left(x\right)|$,
		\item $|\sup_{x\in X}f_{1}\left(x\right)-\sup_{x\in X}f_{2}\left(x\right)|\leq\sup_{x\in X}|f_{1}\left(x\right)-f_{2}\left(x\right)|$.
	\end{enumerate}
\end{fact}

For a fixed probability measure $P_{0}$ on $\left(\mathbb{S},\mathcal{B}\left(\mathbb{S}\right)\right)$,
we define the space $\mathcal{L}=\mathcal{L}_{\infty}\left(\mathbb{S},\mathcal{B}\left(\mathbb{S}\right),P_{0}\right)$
of essentially bounded measurable mappings on $\mathbb{S}$. The following
four properties of coherent risk measures are important throughout our analysis:
\begin{enumerate}
	\item [(A1)] Convexity: $\rho\left(\lambda X+\left(1-\lambda\right)Y\right)\leq\lambda\rho\left(X\right)+\left(1-\lambda\right)\rho\left(Y\right)$
	for all $X,Y\in\mathcal{L}$ and $\lambda\in\left[0,1\right]$.
	\item [(A2)] Monotonicity: If $X,Y\in\mathcal{L}$ and $X\leq Y$, then
	$\rho\left(X\right)\leq\rho\left(Y\right)$.
	\item [(A3)] Translation equivariance: If $\alpha\in\mathbb{R}$ and $X\in\mathcal{L}$,
	then $\rho\left(X+\alpha\right)=\rho\left(X\right)+\alpha$.
	\item [(A4)] Positive homogeneity: If $\alpha>0$ and $X\in\mathcal{L}$,
	then $\rho\left(\alpha X\right)=\alpha\rho\left(X\right)$.
\end{enumerate}

\section*{Proof of Lemma \ref{lem:Bounded}}
\begin{IEEEproof}[Proof of Lemma \ref{lem:Bounded}]
	Fix $\pi\in\Pi$, for any $k\geq0$ we have
	\begin{equation*}
	\begin{aligned}
	0\leq & c(s_{0},a_{0})+\rho(\gamma c(s_{1},a_{1})
	+\rho(\gamma^{2}c(s_{2},a_{2})+\cdots+\gamma^{k}\rho(c(s_{k},a_{k}))))\\
	\leq& c_{\max}+\rho\left(\gamma c_{\max}+\rho\left(\gamma^{2}c_{\max}+\cdots\gamma^{k}\rho\left(c_{\max}\right)\right)\right)\\
	= & c_{\max}\left(1-\gamma^{k}\right)/\left(1-\gamma\right),
	\end{aligned}
	\end{equation*}
	where the inequalities follows by monotonicity of $\rho$ and the
	equality follows by translation equivariance. Taking the limit as
	$k\rightarrow\infty$ gives the desired result.
\end{IEEEproof}

\section*{Proof of Lemma \ref{lem:C}}
\begin{IEEEproof}[Proof of Lemma \ref{lem:C}]
	For any distribution $\nu\in\mathcal{P}\left(\mathbb{S}\right)$
	and a set $B\in\mathcal{B}\left(\mathbb{S}\right)$, we have
	\begin{align*}
	\nu\,Q^{\pi}\left(B\right) & =\int_{\mathbb{S}}Q^{\pi}\left(B|s\right)\nu\left(ds\right)\\
	& =\int_{\mathbb{S}}\left[\int_{B}Q^{\pi}\left(dy|s\right)\right]\nu\left(ds\right)\\
	& =\int_{\mathbb{S}}\left[\int_{B}\frac{dQ^{\pi}}{d\mu}\left(y\right)\mu\left(dy\right)\right]\nu\left(ds\right)\\
	& \leq\int_{\mathbb{S}}\int_{B}C_{\mu}\mu\left(dy\right)\nu\left(ds\right)\\
	& =\int_{B}C_{\mu}\mu\left(dy\right)\\
	& =C_{\mu}\mu\left(B\right),
	\end{align*}
	by the definition of $C_{\mu}$ and the condition $\int_{\mathbb{S}}\nu\left(ds\right)=1.$
	Now let $\nu=\varrho\,Q^{\pi_{1}}Q^{\pi_{2}}\dots Q^{\pi_{M-1}}$
	and $\pi=\pi_{M},$ and we obtain $\varrho\,Q^{\pi_{1}}Q^{\pi_{2}}\dots Q^{\pi_{M}}\leq C_{\mu}\mu$,
	which implies $c\left(M\right)\leq C_{\mu}.$ 
\end{IEEEproof}

\section*{Proof of Lemma \ref{lem:Risk_estimation}}
\begin{IEEEproof}[Proof of Lemma \ref{lem:Risk_estimation}]
	Given $s\in\mathbb{{S}},$ $a\in\mathbb{A},$ $J\in B\left(\mathbb{S};J_{\max}\right)$,  $\varepsilon>0$ and $m\geq1$.
	\begin{enumerate}
		\item For notation convenience, we let $A=A_{1}-A_{2}$ where 
		\begin{IEEEeqnarray*}{rCl}
			&  & A_{1}=\left\{ \mathbb{E}\left[\left(J\left(Y^{s,a}\right)-\mathbb{E}\left[J\left(Y^{s,a}\right)|s\right]\right)^{p}|s\right]\right\} ^{1/p},\\
			&  & A_{2}=\left\{ \frac{1}{m}\sum_{j=1}^{m}\left|J\left(Y_{j}^{s,a}\right)-\frac{1}{m}\sum_{j=1}^{m}J\left(Y_{j}^{s,a}\right)\right|^{p}\right\} ^{1/p},
		\end{IEEEeqnarray*}
		and 
		\[
		B=\mathbb{E}\left[J\left(Y^{s,a}\right)|s\right]-\frac{1}{m}\sum_{j=1}^{m}J\left(Y_{j}^{s,a}\right).
		\]
		First, we need a technical lemma.
		\begin{lemma}
			\label{lem:tech-1} Given $p\in[1,+\infty).$ For $x\geq0$ and $y\in\left(0,1\right),$
			we have
			\[
			\left(x+y\right)^{p}\leq x^{p}+y\left[(1+x)^{p}-x^{p}\right].
			\]
		\end{lemma}
		\begin{IEEEproof}
			We have
			\begin{eqnarray*}
				& \left(x+y\right)^{p} & =x^{p}+C_{n}^{1}x^{p-1}y+C_{n}^{2}x^{p-2}y^{2}+\dots+y^{p}\\
				&  & \leq x^{p}+y\left(C_{n}^{1}x^{p-1}+C_{n}^{1}x^{p-2}+\dots+1\right)\\
				&  & =x^{p}+y\left[(1+x)^{p}-x^{p}\right].
			\end{eqnarray*}
		\end{IEEEproof}
		Using Lemma \ref{lem:tech-1}, we have
		\begin{eqnarray*}
			&  & \frac{1}{m}\sum_{j=1}^{m}\left|J\left(Y_{j}^{s,a}\right)-\frac{1}{m}\sum_{j=1}^{m}J\left(Y_{j}^{s,a}\right)\right|^{p}\\
			& \leq & \frac{1}{m}\sum_{j=1}^{m}\left(\left|A_{21}\left(j\right)\right|+\left|B\right|\right)^{p}\\
			& \leq & \frac{1}{m}\sum_{j=1}^{m}\left|A_{21}\left(j\right)\right|^{p}+C\left|B\right|
		\end{eqnarray*}
		where $A_{21}(j)=J(Y_{j}^{s,a})-\mathbb{E}[J(Y^{s,a})|s]$
		and constant $C\triangleq\left(1+J_{\max}\right)^{p}-J_{\max}^{p}>0.$
		We thus obtain $\left|A_{2}^{p}-\frac{1}{m}\sum_{j=1}^{m}\left|A_{21}\left(j\right)\right|^{p}\right|\leq C\left|B\right|.$
		Since $\mathbb{P}\left(\left|B\right|<\kappa\right)\geq1-2\exp[-2m\kappa^{2}/J_{\max}]$
		by Hoeffding's inequality, we have
		\begin{equation*}
		\begin{aligned}
		\mathbb{P}\left(
		\left|A_{2}^{p}-\frac{1}{m}\sum_{j=1}^{m}|A_{21}(j)|^{p}\right|<C\kappa|s\right)
		\geq1-2\exp\left[\frac{-2m\kappa^{2}}{J_{\max}^{2}}\right].
		\end{aligned}
		\end{equation*}
		By Hoeffding's inequality, we have
		\begin{equation*}
		\begin{aligned}
		\mathbb{P}\left(\left|A_{1}^{p}-\frac{1}{m}\sum_{j=1}^{m}\left|A_{21}\left(j\right)\right|^{p}\right|<\kappa|s\right)
		\geq1-2\exp\left[\frac{-2m\kappa^{2}}{\left(J_{\max}\right)^{2p}}\right].
		\end{aligned}
		\end{equation*}
		By a union bounding argument, we have 
		\begin{equation*}
		\begin{aligned}
		\mathbb{P}\left(\left|A_{1}^{p}-A_{2}^{p}\right|<\left(1+C\right)\kappa|s\right)
		\leq1-2\left(\exp\left[\frac{-2m\kappa^{2}}{J_{\max}^{2}}\right]+\exp\left[\frac{-2m\kappa^{2}}{\left(J_{\max}\right)^{2p}}\right]\right).
		\end{aligned}
		\end{equation*}
		To proceed, we need another technical lemma.
		\begin{lemma}
			\label{lem:tech-2} For $x,y\in\mathbb{R},$ $p\in[1,+\infty)$,
			and let $\max\left\{ \left|x\right|,\left|y\right|\right\} =c$
			where $c$ is a nonnegative constant. We have
			\begin{equation*}
			\begin{aligned}
			\left|x^{p}-y^{p}\right|=\left|x-y\right|\left|x^{p-1}+x^{p-2}y+\dots+y^{p-1}\right|
			\leq pc^{p-1}\left|x-y\right|.
			\end{aligned}
			\end{equation*}
		\end{lemma}
		Since 
		\begin{equation*}
		\begin{aligned}
		\max\big\{\big|(\|J(Y^{s,a})-\|J(Y^{s,a})\|_{\mu}\|_{p,\mu}^{p})^{1/p}\big|,
		\big|(\|J(Y^{s,a})-\|J(Y^{s,a})\|_{\hat{\mu}}\|_{p,\hat{\mu}}^{p})^{1/p}\big|\big\}
		=J_{\max},
		\end{aligned}
		\end{equation*}
		we have
		\begin{equation*}
		\begin{aligned}
			&\mathbb{P}\left(\left|A\right|>\frac{\varepsilon}{2b}|s\right)\\
			\leq&\mathbb{P}\left(\left|A_{1}^{p}-A_{2}^{p}\right|>\frac{\varepsilon}{2bpJ_{\max}^{p-1}}|s\right)\\
			\leq&2\big(\exp[{(-m\varepsilon^{2})}/{(\sqrt{2}bp\left(1+C\right)J_{\max}^{p})^{2}}]
			+\exp[{(-m\varepsilon^{2})}/{(\sqrt{2}bp\left(1+C\right)J_{\max}^{2p-1})^{2}}]\big),
		\end{aligned}
		\end{equation*}
		where the second inequality holds due to Lemma \ref{lem:tech-2}.
		Note that 
		\[
		\mathbb{P}\left(\left|B\right|>\frac{\varepsilon}{2}|s\right)\leq2\exp\left[\frac{-m\varepsilon^{2}}{\left(\sqrt{2}J_{\max}\right)^{2}}\right].
		\]
		Denote
		\begin{equation*}
		\begin{aligned}
			&   \delta_{1}^{m}=2\big(\exp[-m\varepsilon^{2}/(\sqrt{2}bp(1+C)J_{\max}^{p})^{2}]
			+\exp[-m\varepsilon^{2}/(\sqrt{2}bp(1+C)J_{\max}^{2p-1})^{2}]\big)\\
			&   \delta_{2}^{m}=2\exp[-m\varepsilon^{2}/(\sqrt{2}J_{\max})^{2}].
		\end{aligned}
	\end{equation*}
		We have
		\begin{equation*}
		\begin{aligned} &\mathbb{P}\left(\left|bA+B\right|>\varepsilon|s\right)\\
		\leq&\mathbb{P}\left(b\left|A\right|+\left|B\right|>\varepsilon|s\right)\\
		\leq&\mathbb{P}\left(b\left|A\right|>\frac{\varepsilon}{2}|s\right)+\mathbb{P}\left(\left|B\right|>\frac{\varepsilon}{2}|s\right)\\
		\leq&\delta_{1}^{m}+\delta_{2}^{m}.
		\end{aligned}
		\end{equation*}
		\item Let 
		\begin{eqnarray*}
			&  & A=\inf_{\eta\in\left[0,J_{\max}\right]}\eta+\mathbb{E}\left[u\left(J\left(Y^{s,a}\right)-\eta\right)|s\right],\\
			&  & B=\inf_{\eta\in\left[0,J_{\max}\right]}\eta+\frac{1}{m}\sum_{j=1}^{m}u\left(J\left(Y_{j}^{s,a}\right)-\eta\right).
		\end{eqnarray*}
		Using Fact \ref{Fact}, we have
		\begin{equation*}
		\begin{aligned}
		\left|A-B\right|
		\leq\sup_{\eta\in\left[0,J_{\max}\right]}|\mathbb{E}\left[u\left(J\left(Y^{s,a}\right)-\eta\right)|s\right]
		-\frac{1}{m}\sum_{j=1}^{m}u\left(J\left(Y_{j}^{s,a}\right)-\eta\right)|.
		\end{aligned}
		\end{equation*}
		By Hoeffding's inequality, for $\eta\in\left[0,J_{\max}\right],$
		we have
		\begin{equation*}
		\begin{aligned}
		&\mathbb{P}\bigg(\bigg|\mathbb{E}[u(J(Y^{s,a})-\eta)|s]-
		\frac{1}{m}\sum_{j=1}^{m}u(J(Y_{j}^{s,a})-\eta)\bigg|>\frac{\varepsilon}{2}|s\bigg)\\
		&\leq2\exp\left[-m\varepsilon^{2}/\left[\sqrt{2}u\left(J_{\max}\right)\right]^{2}\right].
		\end{aligned}
		\end{equation*}
		Since the piecewise linear function $u$ is Lipschitz continuous with
		the Lipschitz constant equal to $\beta_{2},$ we construct an $\varepsilon/\left(2\beta_{2}\right)-$covering
		net $\mathcal{N}([0,J_{\max}],\varepsilon/(2\beta_{2}))$
		on $[0,J_{\max}]\subseteq\mathbb{R}$. For any $\eta\in\left[0,J_{\max}\right],$
		we can find $\eta'\in\mathcal{N}([0,J_{\max}],\varepsilon/(2\beta_{2}))$
		such that
		\[
		\frac{1}{m}\left|\sum_{j=1}^{m}u\left(J\left(Y_{j}^{s,a}\right)-\eta\right)-\sum_{j=1}^{m}u\left(J\left(Y_{j}^{s,a}\right)-\eta'\right)\right|\leq\frac{\varepsilon}{2}.
		\]
		Here the inequality follows by the Lipschitz continuity of $u.$ Therefore,
		we conclude
		\begin{equation*}
		\begin{aligned}
		\mathbb{P}\left(\left|A-B\right|>\varepsilon|s\right)
		\leq2\left(1+\frac{4\beta_{2}}{\varepsilon}\right)\exp\left[\frac{-m\varepsilon^{2}}{\left[\sqrt{2}u\left(J_{\max}\right)\right]^{2}}\right].
		\end{aligned}
		\end{equation*}
		\item Let 
		\begin{equation*}
		\begin{aligned}
			&   A=\inf_{\eta\in\left[0,J_{\max}\right]}\eta+\frac{1}{1-\alpha}\mathbb{E}\left[\left(J\left(Y^{s,a}\right)-\eta\right)_{+}|s\right],\\
			&   B=\inf_{\eta\in\left[0,J_{\max}\right]}\eta+\frac{1}{m\left(1-\alpha\right)}\sum_{j=1}^{m}\left(J\left(Y_{j}^{s,a}\right)-\eta\right)_{+}.
		\end{aligned}
		\end{equation*}
		Using Fact \ref{Fact}, we have
		\begin{equation*}
		\begin{aligned}
		\left|A-B\right|
		\leq\sup_{\eta\in\left[0J_{\max}\right]}\frac{1}{1-\alpha}\bigg|\mathbb{E}\left[\left(J\left(Y^{s,a}\right)-\eta\right)_{+}|s\right]
		-\frac{1}{m}\sum_{j=1}^{m}\left(J\left(Y_{j}^{s,a}\right)-\eta\right)_{+}\bigg|.
		\end{aligned}
		\end{equation*}
		By Hoeffding's inequality, for $\eta\in\left[0,J_{\max}\right],$
		we have	
		\begin{equation*}
		\begin{aligned}
			&\quad \mathbb{P}\bigg(\frac{1}{1-\alpha}\bigg|\mathbb{E}\left[\left(J\left(Y^{s,a}\right)-\eta\right)_{+}|s\right]
			-\frac{1}{m}\sum_{j=1}^{m}\left(J\left(Y_{j}^{s,a}\right)-\eta\right)_{+}\bigg|>\frac{\varepsilon}{2}|s\bigg)\\
			& \leq  2\exp\left[\frac{-m\left(\varepsilon\left(1-\alpha\right)\right)^{2}}{\left(\sqrt{2}\left(2-\alpha\right)J_{\max}\right)^{2}}\right].
		\end{aligned}
		\end{equation*}
		Since $\left(x\right)_{+}$ has Lipschitz constant $1,$ we construct
		an $\varepsilon\left(1-\alpha\right)/2-$covering net $$\mathcal{N}\left(\left[0,J_{\max}\right],\varepsilon\left(1-\alpha\right)/2\right)$$
		on $\left[0,J_{\max}\right]\subseteq\mathbb{R}$. For any $\eta\in\left[0,J_{\max}\right],$
		we can find $\eta'\in\mathcal{N}\left(\left[0,J_{\max}\right],\varepsilon\left(1-\alpha\right)/2\right)$
		such that
		\begin{equation*}
		\begin{aligned}
		\quad\frac{\left|\sum_{j=1}^{m}\left(J\left(Y_{j}^{s,a}\right)-\eta\right)_{+}-\sum_{j=1}^{m}\left(J\left(Y_{j}^{s,a}\right)-\eta'\right)_{+}\right|}{m\left(1-\alpha\right)}
		\leq\frac{\varepsilon}{2}.
		\end{aligned}
		\end{equation*}
		Therefore, we conclude
		\begin{equation*}
		\begin{aligned}
		\mathbb{P}\left(\left|A-B\right|>\varepsilon|s\right)
		\leq2\left(1+\frac{4}{\varepsilon\left(1-\alpha\right)}\right)\exp\left[\frac{-m\left(\varepsilon\left(1-\alpha\right)\right)^{2}}{\left(\sqrt{2}\left(2-\alpha\right)J_{\max}\right)^{2}}\right].
		\end{aligned}
		\end{equation*}
	\end{enumerate}
\end{IEEEproof}

\section*{Proof of Lemma \ref{lem:Error each iteration}}
\begin{IEEEproof}[Proof of Lemma \ref{lem:Error each iteration}]
	The proof follows the proof of \cite[Lemma 1]{munos2008finite}.
	Let $\Omega$ denote the sample space underlying the random variables.
	Let $\varepsilon''>0$ be arbitrary and let $f^{*}$ be such that
	$\Vert f^{*}-T\widehat{J}_{k}\Vert _{p,\mu}\leq\inf_{f\in\mathcal{F}}\Vert f-T\widehat{J}_{k}\Vert _{p,\mu}+\varepsilon''.$
	We prove the lemma by showing the sequence of inequalities hold simultaneously
	on a set of events of measure not smaller than $1-\delta$.
	\begin{align}
	\|\widehat{J}_{k+1}-T\widehat{J}_{k}\|_{p,\mu} & \leq\|\widehat{J}_{k+1}-T\widehat{J}_{k}\|_{p,\hat{\mu}}+\varepsilon'\label{eq:Seq-1}\\
	& \leq\|\widehat{J}_{k+1}-\widetilde{J}\|_{p,\hat{\mu}}+2\varepsilon'\label{eq:Seq-2}\\
	& \leq\|f^{*}-\widetilde{J}\|_{p,\hat{\mu}}+2\varepsilon'\label{eq:Seq-3}\\
	& \leq\|f^{*}-T\widehat{J}_{k}\|_{p,\hat{\mu}}+3\varepsilon'\label{eq:Seq-4}\\
	& \leq\|f^{*}-T\widehat{J}_{k}\|_{p,\mu}+4\varepsilon'\label{eq:Seq-5}\\
	& =d_{p,\mu}(T\widehat{J}_{k},\mathcal{F})+4\varepsilon'+\varepsilon''.
	\end{align}
	It then follows that $\|\widehat{J}_{k+1}-T\widehat{J}_{k}\|_{p,\mu}\leq\inf_{f\in\mathcal{F}}\|f-T\widehat{J}_{k}\|_{p,\mu}+4\varepsilon'+\varepsilon''$
	w.p. at least $1-\delta.$ Since $\varepsilon''>0$ is arbitrary,
	it also true that $\|\widehat{J}_{k+1}-T\widehat{J}_{k}\|_{p,\mu}\leq\inf_{f\in\mathcal{F}}\|f-T\widehat{J}_{k}\|_{p,\mu}+4\varepsilon'$
	w.p. at least $1-\delta.$ The lemma follows by choosing $\varepsilon'=\varepsilon/4.$
	
	First, observe that (\ref{eq:Seq-3}) holds for all functions $f\in\mathcal{F}$
	and thus the same inequality holds for $f^{*}\in\mathcal{F},$
	too. Therefore, (\ref{eq:Seq-1}) - (\ref{eq:Seq-5}) will hold if
	(\ref{eq:Seq-1}), (\ref{eq:Seq-2}), (\ref{eq:Seq-4}) and (\ref{eq:Seq-5})
	hold w.p. at least $1-\delta'$ with $\delta'=\delta/4.$ Let
	\begin{equation*}
	\begin{aligned}
	W=\max\big(\big|\|f^{*}-T\widehat{J}_{k}\|_{p,\mu}-\|f^{*}-T\widehat{J}_{k}\|_{p,\hat{\mu}}\big|,
	\big|\|\widehat{J}_{k+1}-T\widehat{J}_{k}\|_{p,\mu}-\|\widehat{J}_{k+1}-T\widehat{J}_{k}\|_{p,\hat{\mu}}\big|\big).
	\end{aligned}
	\end{equation*}
	
	Next we show $\mathbb{P}\left(W>\varepsilon'\right)\leq\delta',$
	which implies (\ref{eq:Seq-1}) and (\ref{eq:Seq-5}) hold. Note that
	for all $\omega\in\Omega,$ $\widehat{J}_{k+1}=\widehat{J}_{k+1}\left(\omega\right)\in\mathcal{F}.$
	Hence, 
	\begin{equation*}
	\begin{aligned}
	\sup_{f\in\mathcal{F}}\left|\|f-T\widehat{J}_{k}\|_{p,\mu}-\|f-T\widehat{J}_{k}\|_{p,\hat{\mu}}\right|
	\geq\left|\|\widehat{J}_{k+1}-T\widehat{J}_{k}\|_{p,\mu}-\|\widehat{J}_{k+1}-T\widehat{J}_{k}\|_{p,\hat{\mu}}\right|
	\end{aligned}
	\end{equation*}
	holds point-wise in $\Omega.$ Therefore, the inequality
	\begin{equation*}
	\begin{aligned}
	\sup_{f\in\mathcal{F}}\left|\|f-T\widehat{J}_{k}\|_{p,\mu}-\|f-T\widehat{J}_{k}\|_{p,\hat{\mu}}\right|\geq W
	\end{aligned}
	\end{equation*}
	holds point-wise in $\Omega,$ and hence
	\begin{equation*}
	\begin{aligned}
	\mathbb{P}\left(W>\varepsilon'\right)
	\leq\mathbb{P}\left(\sup_{f\in\mathcal{F}}\left|\|f-T\widehat{J}_{k}\|_{p,\mu}-\|f-T\widehat{J}_{k}\|_{p,\hat{\mu}}\right|>\varepsilon'\right).
	\end{aligned}
	\end{equation*}
	
	We claim that
	\begin{equation*}
	\begin{aligned}
	&\mathbb{P}\left(\sup_{f\in\mathcal{F}}\left|\|f-T\widehat{J}_{k}\|_{p,\mu}-\|f-T\widehat{J}_{k}\|_{p,\hat{\mu}}\right|>\varepsilon'\right)\\
	\leq&\mathbb{P}\left(\sup_{f\in\mathcal{F}}\left|\|f-T\widehat{J}_{k}\|_{p,\mu}^{p}-\|f-T\widehat{J}_{k}\|_{p,\hat{\mu}}^{p}\right|>\left(\varepsilon'\right)^{p}\right).
	\end{aligned}
	\end{equation*}
	For any event $\omega$ such that
	\[
	\sup_{f\in\mathcal{F}}\left|\|f-T\widehat{J}_{k}\|_{p,\mu}-\|f-T\widehat{J}_{k}\|_{p,\hat{\mu}}\right|>\varepsilon'.
	\]
	For such event $\omega,$ there exists a function $f'\in\mathcal{F}$
	such that
	\[
	\left|\|f'-T\widehat{J}_{k}\|_{p,\mu}-\|f'-T\widehat{J}_{k}\|_{p,\hat{\mu}}\right|>\varepsilon'.
	\]
	Pick such function. Assume that $\|f'-T\widehat{J}_{k}\|_{p,\hat{\mu}}\leq\|f'-T\widehat{J}_{k}\|_{p,\mu}$.
	We obtain $\|f'-T\widehat{J}_{k}\|_{p,\hat{\mu}}+\varepsilon'<\|f'-T\widehat{J}_{k}\|_{p,\mu}.$
	Since $p\geq1,$ $x^{p}+y^{p}\leq\left(x+y\right)^{p}$ for $x,y\geq0,$
	we have $\|f'-T\widehat{J}_{k}\|_{p,\hat{\mu}}^{p}+\left(\varepsilon'\right)^{p}\leq(\|f'-T\widehat{J}_{k}\|_{p,\hat{\mu}}+\varepsilon')^{p}<\|f'-T\widehat{J}_{k}\|_{p,\mu}^{p}$
	and
	\[
	\left|\|f'-T\widehat{J}_{k}\|_{p,\mu}^{p}-\|f'-T\widehat{J}_{k}\|_{p,\hat{\mu}}^{p}\right|>\left(\varepsilon'\right)^{p}.
	\]
	A similar argument can be developed when $\|f'-T\widehat{J}_{k}\|_{p,\hat{\mu}}>\|f'-T\widehat{J}_{k}\|_{p,\mu}.$
	The claim follows since
	\begin{equation*}
	\begin{aligned}
	\bigg|\sup_{f\in\mathcal{F}}\|f-T\widehat{J}_{k}\|_{p,\mu}^{p}-\|f-T\widehat{J}_{k}\|_{p,\hat{\mu}}^{p}\bigg|
	\geq\left|\|f'-T\widehat{J}_{k}\|_{p,\mu}^{p}-\|f'-T\widehat{J}_{k}\|_{p,\hat{\mu}}^{p}\right|.
	\end{aligned}
	\end{equation*}
	
	Next, we state a concentration inequality derived due to Pollard.
	\begin{theorem}[Pollard, 1984]
		\label{thm:Pollard inequality} Let $\mathcal{F}$ be a set of measurable
		functions $f:\mathcal{X}\rightarrow\left[0,K\right]$ and let $\varepsilon>0,$
		$m$ be arbitrary. If $X_{i},$ $i=1,\dots,n$ is i.i.d. sequence taking
		values in the space $\mathcal{X}$ then
		\begin{equation*}
		\begin{aligned}
		\mathbb{P}\left(\sup_{f\in\mathcal{F}}\left|\frac{1}{n}\sum_{i=1}^{n}f\left(X_{i}\right)-\mathbb{E}\left[f\left(X_{1}\right)\right]\right|>\varepsilon\right)
		\leq8\mathbb{E}[\mathcal{N}(\varepsilon/8,\mathcal{F}(X^{1:n}))]\exp[-n\varepsilon^{2}/128K^{2}].
		\end{aligned}
		\end{equation*}
	\end{theorem}
	
	Now, observe that $\|f-T\widehat{J}_{k}\|_{p,\mu}^{p}=\mathbb{E}[|(f(s_{1})-(TJ)(s_{1}))|^{p}],$
	and $\|f-T\widehat{J}_{k}\|_{p,\hat{\mu}}^{p}$ is just the sample
	average approximation of $\|f-T\widehat{J}_{k}\|_{p,\mu}^{p}.$
	Hence, by noting that the covering number associated with $\left\{ f-TJ|f\in\mathcal{F}\right\} $
	is the same as the covering number of $\mathcal{F}$,
	we apply Theorem \ref{thm:Pollard inequality} and obtain
	\begin{equation*}
	\begin{aligned}
	&\mathbb{P}\left(\sup_{f\in\mathcal{F}}\left|\|f-T\widehat{J}_{k}\|_{p,\mu}^{p}-\|f-T\widehat{J}_{k}\|_{p,\hat{\mu}}^{p}\right|>\left(\varepsilon'\right)^{p}\right)\\
	\leq&8\mathbb{E}\left[\mathcal{N}\left(\left(\varepsilon'\right)^{p}/8,\mathcal{F}\left(s^{1:n}\right)\right)\right]\exp\left[-\frac{n}{2}\left(\frac{1}{8}\left(\frac{\varepsilon'}{2J_{\max}}\right)^{p}\right)^{2}\right].
	\end{aligned}
	\end{equation*}
	By making the right-hand side upper bounded by $\delta'=\delta/4$
	we get a lower bound on $n$
	\[
	n>128\left(\frac{8J_{\max}}{\varepsilon}\right)^{2p}\left(\log\left(1/\delta\right)+\log\left(32\mathcal{N}_{0}\left(n\right)\right)\right).
	\]
	
	Next, we prove inequalities (\ref{eq:Seq-2}) and (\ref{eq:Seq-4}).
	Let $f$ denote an arbitrary random function such that $f=f\left(s;\omega\right)$
	is measurable for each $s_{i}\in{\mathbb{S}}$ and assume
	that $f$ is uniformly bounded by $J_{\max}.$ By triangle inequality,
	we have
	\begin{equation}
	\left|\|f-T\widehat{J}_{k}\|_{p,\hat{\mu}}-\|f-\widetilde{J}\|_{p,\hat{\mu}}\right|\leq\|T\widehat{J}_{k}-\widetilde{J}\|_{p,\hat{\mu}}.\label{eq:Triangle inequality}
	\end{equation}
	It suffices to show that $\|T\widehat{J}_{k}-\widetilde{J}\|_{p,\hat{\mu}}\leq\varepsilon'$
	holds w.p. $1-\delta'.$ Under Assumption \ref{assu:Risk_estimation},
	we have
	\[
	\mathbb{P}\left(\left|\rho\left(\widehat{J}_k\left(Y^{s_i,a}\right)\right)-\hat{\rho}_{m}\left(\left\{\widehat{J}_k\left(Y_j^{s_i,a}\right)\right\}^m_{j=1}\right)\right|>\varepsilon'|s^{1:n}\right)\leq\theta\left(\varepsilon',m\right),
	\]
	Let $\theta\left(\varepsilon',m\right)$ upper bounded by $\delta'/\left(n\left|\mathbb{A}\right|\right),$
	we get a lower bound on $m.$
	
	Since
	\[
	\left|T\widehat{J}_{k}\left(s_{i}\right)-\widetilde{J}\left(s_{i}\right)\right|\leq\max_{a\in\mathbb{A}}\left|\rho\left(\widehat{J}_k\left(Y^{s_i,a}\right)\right)-\hat{\rho}_{m}\left(\left\{\widehat{J}_k\left(Y_j^{s_i,a}\right)\right\}^m_{j=1}\right)\right|
	\]
	by Fact \ref{Fact} 1), it follows by a union bounding argument that
	\[
	\mathbb{P}\left(\left|T\widehat{J}_{k}\left(s_{i}\right)-\widetilde{J}\left(s_{i}\right)\right|>\varepsilon'\big|s^{1:n}\right)\leq\delta'/n,
	\]
	and hence another union bounding argument yields
	\[
	\mathbb{P}\left(\max_{i=1,\dots,n}\left|T\widehat{J}_{k}\left(s_{i}\right)-\widetilde{J}\left(s_{i}\right)\right|^{p}>\left(\varepsilon'\right)^{p}\big|s^{1:n}\right)\leq\delta'.
	\]
	Taking the expectation of both sides of this inequality gives
	\[
	\mathbb{P}\left(\max_{i=1,\dots,n}\left|T\widehat{J}_{k}\left(s_{i}\right)-\widetilde{J}\left(s_{i}\right)\right|^{p}>\left(\varepsilon'\right)^{p}\right)\leq\delta'.
	\]
	Hence,
	\[
	\mathbb{P}\left(\frac{1}{n}\sum_{i=1}^{n}\left|T\widehat{J}_{k}\left(s_{i}\right)-\widetilde{J}\left(s_{i}\right)\right|^{p}>\left(\varepsilon'\right)^{p}\right)\leq\delta'.
	\]
	Therefore by (\ref{eq:Triangle inequality}), we have
	\[
	\mathbb{P}\left(\left|\|f-T\widehat{J}_{k}\|_{p,\hat{\mu}}-\|f-\widetilde{J}\|_{p,\hat{\mu}}\right|>\varepsilon'\right)\leq\delta'.
	\]
	Using this with $f=\widehat{J}_{k+1}$ and $f=f^{*}$ shows that inequalities
	(\ref{eq:Seq-2}) and (\ref{eq:Seq-4}) each hold w.p. at least $1-\delta'$. 
\end{IEEEproof}

\section*{Proof of Corollary \ref{cor:Error each iteration}}
\begin{IEEEproof}[Proof of Corollary \ref{cor:Error each iteration}]
	The lower bound for $n$ can be found similarly in the proof of
	Lemma \ref{lem:Error each iteration}. Let $\varepsilon'=\varepsilon/4$
	and $\delta'=\delta/4.$ In the following, we derive the lower bound
	for $m$ such that $\|T\widehat{J}_{k}-\widetilde{J}\|_{p,\hat{\mu}}\leq\varepsilon'$
	holds w.p. $1-\delta'.$ 
	\begin{enumerate}
		\item By Lemma \ref{lem:Risk_estimation} 1), we have
		\begin{equation*}
		\begin{aligned}
		&\mathbb{P}\left(\left|bA+B\right|>\varepsilon'|s^{1:n}\right)\\
		\leq&\mathbb{P}\left(b\left|A\right|+\left|B\right|>\varepsilon'|s^{1:n}\right)\\
		\leq&\mathbb{P}\left(b\left|A\right|>\frac{\varepsilon'}{2}|s^{1:n}\right)+\mathbb{P}\left(\left|B\right|>\frac{\varepsilon'}{2}|s^{1:n}\right)\\
		\leq&\delta_{1}^{m}+\delta_{2}^{m},
		\end{aligned}
		\end{equation*}
		where $A=A_{1}-A_{2}$ with 
		\begin{equation*}
		\begin{aligned}
		&A_{1}=\{ \mathbb{E}[\left(J\left(Y^{s_{i},a}\right)-\mathbb{E}\left[J\left(Y^{s_{i},a}\right)|s^{1:n}\right]\right)^{p}|s^{1:n}]\} ^{1/p},\\
		&A_{2}=\left\{ \frac{1}{m}\sum_{j=1}^{m}\left|J\left(Y_{j}^{s_{i},a}\right)-\frac{1}{m}\sum_{j=1}^{m}J\left(Y_{j}^{s_{i},a}\right)\right|^{p}\right\} ^{1/p},\\
		&B=\mathbb{E}\left[J\left(Y^{s_{i},a}\right)|s^{1:n}\right]-\frac{1}{m}\sum_{j=1}^{m}J\left(Y_{j}^{s_{i},a}\right),\\
		&\delta_{1}^{m}=2\big(\exp[-m(\varepsilon')^{2}/(\sqrt{2}bp(1+C)J_{\max}^{p})^{2}]
		+\exp[-m(\varepsilon')^{2}/(\sqrt{2}bp(1+C)J_{\max}^{2p-1})^{2}]\big),\\
		&\delta_{2}^{m}=2\exp[-m(\varepsilon')^{2}/(\sqrt{2}J_{\max})^{2}].
		\end{aligned}
		\end{equation*}
		Making $\delta_{1}^{m}+\delta_{2}^{m}$ upper bounded by $\delta'/\left(n\left|\mathbb{A}\right|\right),$
		we get a lower bound on $m.$ The rest of proof is same as the one
		for Lemma \ref{lem:Error each iteration}, thus omitted. The proof
		for mean-semideviation risk function can be developed in a similar
		way.
		\item By Lemma \ref{lem:Risk_estimation} 2), we have
		\begin{equation*}
		\begin{aligned}
		\mathbb{P}(\left|A-B\right|>&\varepsilon'\big|s^{1:n})
		\leq2\left(1+\frac{4\beta_{2}}{\varepsilon'}\right)\exp\left[\frac{-m\left(\varepsilon'\right)^{2}}{\left[\sqrt{2}u\left(J_{\max}\right)\right]^{2}}\right],
		\end{aligned}
		\end{equation*}
		where 
		\begin{equation*}
		\begin{aligned}
		&   A=\inf_{\eta\in\left[0,J_{\max}\right]}\eta+\mathbb{E}\left[u\left(J\left(Y^{s_{i},a}\right)-\eta\right)|s^{1:n}\right],\\
		&   B=\inf_{\eta\in\left[0,J_{\max}\right]}\eta+\frac{1}{m}\sum_{j=1}^{m}u\left(J\left(Y_{j}^{s_{i},a}\right)-\eta\right).
		\end{aligned}
		\end{equation*}
		Making the right hand side upper bounded by $\delta'/\left(n\left|\mathbb{A}\right|\right),$
		we get a lower bound on $m.$ The rest of proof is same as the one
		for Lemma \ref{lem:Error each iteration}, thus omitted. 
		\item By Lemma \ref{lem:Risk_estimation} 3), we have
		\begin{equation*}
		\begin{aligned}
		\mathbb{P}(\left|A-B\right|>\varepsilon'\big|s^{1:n})
		\leq2\left(1+\frac{4}{\varepsilon'\left(1-\alpha\right)}\right)\exp\left[\frac{-m\left(\varepsilon'\left(1-\alpha\right)\right)^{2}}{\left(\sqrt{2}\left(2-\alpha\right)J_{\max}\right)^{2}}\right],
		\end{aligned}
		\end{equation*}
		where
		\begin{equation*}
		\begin{aligned}
		&   A=\inf_{\eta\in\left[0,J_{\max}\right]}\eta+\frac{1}{1-\alpha}\mathbb{E}\left[\left(J\left(Y^{s_{i},a}\right)-\eta\right)_{+}|s^{1:n}\right],\\
		&   B=\inf_{\eta\in\left[0,J_{\max}\right]}\eta+\frac{1}{m\left(1-\alpha\right)}\sum_{j=1}^{m}\left(J\left(Y_{j}^{s_{i},a}\right)-\eta\right)_{+}.
		\end{aligned}
		\end{equation*}
		Making the right hand side upper bounded by $\delta'/\left(n\left|\mathbb{A}\right|\right),$
		we get a lower bound on $m.$ The rest of proof is same as the one
		for Lemma \ref{lem:Error each iteration}, thus omitted. 
	\end{enumerate}
\end{IEEEproof}

\section*{Proof of Lemma \ref{lem:p-norm}}
\begin{IEEEproof}[Proof of Lemma \ref{lem:p-norm}]
	\begin{enumerate}
	\item For all $s\in\mathbb{S}$, we have
	\begin{equation*}		
	\begin{aligned}
	&[T^{\pi^{*}}\widehat{J}_{k}]\left(s\right)-[T^{\pi^{*}}J^{*}]\left(s\right)\\
	=&\,  \gamma\max_{\mu\in\mathcal{Q}\left(s,\pi^{*}\left(s\right)\right)}\mathbb{E}_{Y\sim\mu}[\widehat{J}_{k}\left(Y\right)]-\gamma\max_{\mu\in\mathcal{Q}\left(s,\pi^{*}\left(s\right)\right)}\mathbb{E}_{Y\sim\mu}[J^{*}\left(Y\right)]\\
	\leq&\,\gamma Q^{\pi_{k}}(\widehat{J}_{k}-J^{*}),
	\end{aligned}
	\end{equation*}
	where $Q^{\pi_{k}}\mbox{ : }B\left(\mathbb{S};J_{\max}\right)\rightarrow B\left(\mathbb{S};J_{\max}\right)$
	is a linear operator such that 
	$$Q^{\pi_{k}}(\cdot\vert s)\in\arg\max_{\mu\in\mathcal{Q}(s,\pi^{*}(s))}\mathbb{E}_{Y\sim\mu}[\widehat{J}_{k}(Y)]$$
	is an element of distributional set $\mathcal{Q}\left(s,\pi^{*}\left(s\right)\right)$
	for all $s\in\mathbb{S}.$
	\item 	For all $s\in\mathbb{S}$, we have
	\begin{equation*}		
	\begin{aligned}
	&[T^{\hat{\pi}_{k}}\widehat{J}_{k}]\left(s\right)-\left[T^{\hat{\pi}_{k}}J^{*}\right]\left(s\right)\\
	=&\, \gamma\max_{\mu\in\mathcal{Q}\left(s,\hat{\pi}_{k}\left(s\right)\right)}\mathbb{E}_{Y\sim\mu}[\widehat{J}_{k}\left(Y\right)]-\gamma\max_{\mu\in\mathcal{Q}\left(s,\hat{\pi}_{k}\left(s\right)\right)}\mathbb{E}_{Y\sim\mu}[J^{*}\left(Y\right)]\\
	\geq&\,\gamma Q^{\pi_{k}^{*}}(\widehat{J}_{k}-J^{*}),
	\end{aligned}
	\end{equation*}
	where $Q^{\pi_{k}^{*}}\mbox{ : }B\left(\mathbb{S};J_{\max}\right)\rightarrow B\left(\mathbb{S};J_{\max}\right)$
	is a linear operator such that 
	$$Q^{\pi_{k}^{*}}(\cdot\vert s)\in\arg\max_{\mu\in\mathcal{Q}(s,\hat{\pi}_{k}(s))}\mathbb{E}_{Y\sim\mu}[J^{*}(Y)]$$
	is an element of distributional set $\mathcal{Q}\left(s,\hat{\pi}_{k}\left(s\right)\right)$
	for all $s\in\mathbb{S}.$
	\end{enumerate}
\end{IEEEproof}

\section*{Proof of Lemma \ref{lem:Point error bounds}}
\begin{IEEEproof}[Proof of Lemma \ref{lem:Point error bounds}]
	Recall that $\pi^{*}$ is the optimal policy. For $k\geq0$, we have
	$T\widehat{J}_{k}\leq T^{\pi^{*}}\widehat{J}_{k}$ and 
	\begin{equation*}		
	\begin{aligned}
	\widehat{J}_{k+1}-J^{*}=&T\widehat{J}_{k}-T^{\pi^{*}}\widehat{J}_{k}+T^{\pi^{*}}\widehat{J}_{k}-T^{\pi^{*}}J^{*}-\varepsilon_{k}\\\leq& T^{\pi^{*}}\widehat{J}_{k}-T^{\pi^{*}}J^{*}-\varepsilon_{k}
	\end{aligned}
	\end{equation*}
	By Lemma \ref{lem:p-norm} 1), there exists a stochastic kernel $Q^{\pi_{k}}$
	such that $T^{\pi^{*}}\widehat{J}_{k}-T^{\pi^{*}}J^{*}\leq\gamma Q^{\pi_{k}}(\widehat{J}_{k}-J^{*})$.
	Therefore, we have		
	\[
	\widehat{J}_{k+1}-J^{*}\leq\gamma Q^{\pi_{k}}(\widehat{J}_{k}-J^{*})-\varepsilon_{k},
	\]
	from which we deduce by induction		
	\begin{equation}\label{lem8-1}		
	\begin{aligned}
	\widehat{J}_{K}-J^{*}\leq\gamma^{K}(Q^{\pi_{K-1}}Q^{\pi_{K-2}}\dots Q^{\pi_{0}})(\widehat{J}_{0}-J^{*})
	-\sum_{k=0}^{K-1}\gamma^{K-k-1}\left(Q^{\pi_{k+1}}Q^{\pi_{k+2}}\dots Q^{\pi_{K-1}}\right)\varepsilon_{k}.
	\end{aligned}
	\end{equation}		
	From definition of $\widehat{\pi}_{k}$, we have $TJ^{*}=T^{\pi^{*}}J^{*}\leq T^{\widehat{\pi}_{k}}J^{*}$
	and
	\begin{equation*}		
	\begin{aligned}
	\widehat{J}_{k+1}-J^{*}&=T^{\widehat{\pi}_{k}}\widehat{J}_{k}-T^{\widehat{\pi}_{k}}J^{*}+T^{\widehat{\pi}_{k}}J^{*}-T^{\pi^{*}}J^{*}-\varepsilon_{k}\\
	&\geq T^{\widehat{\pi}_{k}}\widehat{J}_{k}-T^{\widehat{\pi}_{k}}J^{*}-\varepsilon_{k}.
	\end{aligned}
	\end{equation*}	
	By Lemma 5 2), there is a stochastic kernel $Q^{\pi_{k}^{*}}$
	such that $T^{\widehat{\pi}_{k}}J_{k}-T^{\widehat{\pi}_{k}}J^{*}\geq\gamma Q^{\pi_{k}^{*}}(\widehat{J}_{k}-J^{*})$.
	Therefore,		
	\[
	\widehat{J}_{k+1}-J^{*}\geq\gamma Q^{\pi_{k}^{*}}(\widehat{J}_{k}-J^{*})-\varepsilon_{k}.
	\]
	By induction, we obtain
	\begin{equation}\label{lem8-2}		
	\begin{aligned}
	\widehat{J}_{K}-J^{*}\geq\gamma^{K}(Q^{\pi_{K-1}^{*}}Q^{\pi_{K-2}^{*}}\dots Q^{\pi_{0}^{*}})(\widehat{J}_{0}-J^{*})
	-\sum_{k=0}^{K-1}\gamma^{K-k-1}(Q^{\pi_{K-1}^{*}}Q^{\pi_{K-2}^{*}}\dots Q^{\pi_{k+1}^{*}})\varepsilon_{k}.
	\end{aligned}
	\end{equation}	
	
	We observe that $T^{\widehat{\pi}_{K}}\widehat{J}_{K}=T\widehat{J}_{K}\leq T^{\pi^{*}}\widehat{J}_{K}$
	by definition of $\hat{\pi}_{K}$ and $T$, and note that $J^{\widehat{\pi}_{K}}=T^{\widehat{\pi}_{K}}J^{\pi_{K}}$
	and $TJ^{*}=T^{\pi^{*}}J^{*}=J^{*}$ gives
	\begin{align*}
	&J^{\widehat{\pi}_{K}}-J^{*}\\
	= & T^{\widehat{\pi}_{K}}J^{\widehat{\pi}_{K}}-T^{\widehat{\pi}_{K}}\widehat{J}_{K}+T^{\widehat{\pi}_{K}}\widehat{J}_{K}-T^{\pi^{*}}\widehat{J}_{K}+T^{\pi^{*}}\widehat{J}_{K}-T^{\pi^{*}}J^{*}\\
	\leq & T^{\widehat{\pi}_{K}}J^{\widehat{\pi}_{K}}-T^{\widehat{\pi}_{K}}\widehat{J}_{K}+T^{\pi^{*}}\widehat{J}_{K}-T^{\pi^{*}}J^{*}\\
	\leq & \gamma Q^{\hat{\pi}_{K}}(J^{\widehat{\pi}_{K}}-\widehat{J}_{K})+\gamma Q^{\pi_{K}}(\widehat{J}_{K}-J^{*})\\
	= & \gamma Q^{\hat{\pi}_{K}}(J^{\widehat{\pi}_{K}}-J^{*}+J^{*}-\widehat{J}_{K})+\gamma Q^{\pi_{K}}(\widehat{J}_{K}-J^{*}),
	\end{align*}
	where $Q^{\hat{\pi}_{K}}\mbox{ : }B\left(\mathbb{S};J_{\max}\right) \rightarrow B\left(\mathbb{S};J_{\max}\right)$
	is a stochastic kernel such that 
	$$Q^{\hat{\pi}_{K}}(\cdot| s)\in\arg\max_{\mu\in\mathcal{Q}(s,\hat{\pi}_{K}(s))}\mathbb{E}_{Y\sim\mu}[J^{\widehat{\pi}_{K}}(Y)]$$
	is an element of the distributional set $\mathcal{Q}\left(s,\hat{\pi}_{k}\left(s\right)\right)$
	for all $s\in\mathbb{S},$ and the second inequality is by Lemma \ref{lem:p-norm}.
	We then have
	\[
	(I-\gamma Q^{\hat{\pi}_{K}})(J^{\widehat{\pi}_{K}}-J^{*})\leq\gamma(Q^{\pi_{K}}-Q^{\hat{\pi}_{K}})(\widehat{J}_{K}-J^{*}).
	\]
	Note that $(I-\gamma Q^{\hat{\pi}_{K}})$ is invertible
	and its inverse is a monotonic operator, and we have 
	\[
	J^{\widehat{\pi}_{K}}-J^{*}\leq\gamma(I-\gamma Q^{\hat{\pi}_{K}})^{-1}(Q^{\pi_{K}}-Q^{\hat{\pi}_{K}})(\widehat{J}_{K}-J^{*}).
	\]
	Using (\ref{lem8-1}) and (\ref{lem8-2}), and that fact that $\max\left\{ |a|,|b|\right\} \leq|a|+|b|$,
	we obtain
	\begin{equation*}		
	\begin{aligned}
	J^{\widehat{\pi}_{K}}-J^{*}\leq2(I-\gamma Q^{\hat{\pi}_{K}})^{-1}
	\left\{ \sum_{k=0}^{K-1}\gamma^{K-k}Q_{1}\varepsilon_{k}+\gamma^{K+1}Q_{2}\left(\widehat{J}_{0}-J^{*}\right)\right\} 
	\end{aligned}
	\end{equation*}
	where 
	$$Q_{1}=(Q^{\pi_{K}}Q^{\pi_{K-1}}\dots Q^{\pi_{k+1}}
	+Q^{\hat{\pi}_{K}}Q^{\pi_{K-1}^{*}}Q^{\pi_{K-2}^{*}}\dots Q^{\pi_{k+1}^{*}})/2$$ 
	and
	$$Q_{2}=(Q^{\pi_{K}}Q^{\pi_{K-1}}\dots Q^{\pi_{0}}
	+Q^{\hat{\pi}_{K}}Q^{\pi_{K-1}^{*}}Q^{\pi_{K-2}^{*}}\dots Q^{\pi_{0}^{*}})/2.$$
	
	Taking the absolute value of both sides, we obtain the desired bound. 
\end{IEEEproof}

\section*{Proof of Lemma \ref{lem:Lp bounds}}
\begin{IEEEproof}[Proof of Lemma \ref{lem:Lp bounds}]
	The proof follows the proof of \cite[Lemma 4]{munos2008finite}. From
	Lemma \ref{lem:Point error bounds}, we have 		
	\begin{equation*}
	\begin{aligned}
	J^{\widehat{\pi}_{K}}-J^{*}\leq&\frac{2\gamma\left(1-\gamma^{K+1}\right)}{\left(1-\gamma\right)^{2}}
	\left[\sum_{k=0}^{K-1}\alpha_{k}A_{k}\left|\varepsilon_{k}\right|+\alpha_{K}A_{K}|J^{*}-\widehat{J}_{0}|\right],
	\end{aligned}
	\end{equation*}
	with the positive coefficients 		
	\[
	\alpha_{k}=\frac{\left(1-\gamma\right)\gamma^{K-k-1}}{1-\gamma^{K+1}},\quad0\leq k<K,
	\]
	and		
	$$
	\alpha_{K}={[\left(1-\gamma\right)\gamma^{K}]}/{(1-\gamma^{K+1})},
	$$
	such that $\sum_{k=0}^{K}\alpha_{k}=1$ and the probability kernels
	\begin{equation*}		
	\begin{aligned}
	A_{k}  =  \frac{1-\gamma}{2}(I-\gamma Q^{\hat{\pi}_{K}})^{-1}
	[Q^{\pi_{K}}Q^{\pi_{K-1}}\dots Q^{\pi_{k+1}}
	+Q^{\hat{\pi}_{K}}Q^{\pi_{K-1}^{*}}Q^{\pi_{K-2}^{*}}\dots Q^{\pi_{k+1}^{*}}],
	\end{aligned}
	\end{equation*}
	for $0\leq k<K$ and
	\begin{equation*}		
	\begin{aligned}
	A_{K}=\frac{1-\gamma}{2}(I-\gamma Q^{\hat{\pi}_{K}})^{-1}[Q^{\pi_{K}}Q^{\pi_{K-1}}\dots Q^{\pi_{0}}
	+Q^{\hat{\pi}_{K}}Q^{\pi_{K-1}^{*}}Q^{\pi_{K-2}^{*}}\dots Q^{\pi_{0}^{*}}].
	\end{aligned}
	\end{equation*}		
	We have		
	\begin{equation*}		
	\begin{aligned}
	&\left\Vert J^{\widehat{\pi}_{K}}-J^{*}\right\Vert _{p,\varrho}^{p}\\
	= &\int\varrho\left(ds\right)|J^{\widehat{\pi}_{K}}\left(s\right)-J^{*}\left(s\right)|^{p}\\
	\leq & \left[\frac{2\gamma\left(1-\gamma^{K+1}\right)}{\left(1-\gamma\right)^{2}}\right]^{p}
	\int\varrho\left(ds\right)\left[\sum_{k=0}^{K-1}\alpha_{k}A_{k}\left|\varepsilon_{k}\right|+\alpha_{K}A_{K}\left|J^{*}-\widehat{J}_{0}\right|\right]^{p}\left(s\right)\\
	\leq&  \left[\frac{2\gamma\left(1-\gamma^{K+1}\right)}{\left(1-\gamma\right)^{2}}\right]^{p}
	\int\varrho\left(ds\right)\left[\sum_{k=0}^{K-1}\alpha_{k}A_{k}\left|\varepsilon_{k}\right|^{p}+\alpha_{K}A_{K}|J^{*}-\widehat{J}_{0}|^{p}\right]\left(s\right),
	\end{aligned}
	\end{equation*}
	by using two times Jensen's inequality (since $A_{k}$ are positive
	linear operators $A_{k}\mathbf{1}=\mathbf{1}$ and convexity of $x\rightarrow\left|x\right|^{p}$).
	The term $|J^{*}-\widehat{J}_{0}|$ is bounded by $J_{\max}.$
	Under Assumption \ref{assu:Absolute_continuity}, $\rho A_{k}\leq(1-\gamma)\sum_{M\geq0}\gamma^{M}c\left(M+K-k\right)\mu.$
	If the approximation error in all\textbf{ }iterations $k=0,\dots,K-1$
	falls below the tolerance $\left\Vert \varepsilon_{k}\right\Vert _{p,\mu}\leq\varepsilon$,
	we deduce		
	\begin{equation*}		
	\begin{aligned}
	\left\Vert J^{\widehat{\pi}_{K}}-J^{*}\right\Vert _{p,\varrho}^{p}\leq \left[\frac{2\gamma\left(1-\gamma^{K+1}\right)}{\left(1-\gamma\right)^{2}}\right]^{p}
	[(1-\gamma^{K+1})^{-1}C_{\varrho,\mu}\varepsilon^{p}
	+\gamma^{K}(1-\gamma)(1-\gamma^{K+1})^{-1}J_{\max}^{p}].
	\end{aligned}
	\end{equation*}		
	
	There exists $K$ that is linear in $\log\left(1/\eta\right)$ and
	$\log J_{\max}$ such that $\gamma^{K}\leq[{\eta(1-\gamma)^{2}}/{(2\gamma J_{\max})}]^{p}.$
	By this choice of $K,$ the second term is bounded by $\eta^{p}$
	and we have		
	\[
	\left\Vert J^{\widehat{\pi}_{K}}-J^{*}\right\Vert _{p,\varrho}^{p}\leq\left[\frac{2\gamma}{\left(1-\gamma\right)^{2}}\right]^{p}C_{\varrho,\mu}\varepsilon^{p}+\eta^{p}.
	\]
	
\end{IEEEproof}	

\section*{Proof of Lemma \ref{lem:Contraction}}
\begin{IEEEproof}[Proof of Lemma \ref{lem:Contraction}]
	We first write $T$ as
	\begin{equation*}
	\begin{aligned}
	\left[TJ\right]\left(s\right)
	=\min_{a\in\mathbb{A}}\left\{ c(s,a)+\gamma\max_{\mu\in\mathcal{Q}\left(s,a\right)}\int J\left(y\right)\mu\left(dy\right)\right\},\quad\forall s\in\mathbb{S},
	\end{aligned}
	\end{equation*}
	where each $\mathcal{Q}\left(s,a\right)\subset\mathcal{P}\left(\mathbb{S}\right)$
	via Fenchel duality. For any $s\in\mathbb{S}$, we have 
	\begin{equation*}
	\begin{aligned}
	&\left|\left[TJ_{1}\right]\left(s\right)-\left[TJ_{2}\right]\left(s\right)\right|\\
	\leq&\gamma\max_{a\in\mathbb{A}}\left|\max_{\mu\in\mathcal{Q}\left(s,a\right)}\int J_{1}\left(y\right)\mu\left(dy\right)-\max_{\mu\in\mathcal{Q}\left(s,a\right)}\int J_{2}\left(y\right)\mu\left(dy\right)\right|\\
	\leq&  \gamma\max_{a\in\mathbb{A}}\max_{\mu\in\mathcal{Q}\left(s,a\right)}\int|J_{1}\left(y\right)-J_{2}\left(y\right)|\mu\left(dy\right)\\
	\leq&\gamma\|J_{1}-J_{2}\|_{\infty},
	\end{aligned}
	\end{equation*}
	using Fact \ref{Fact}. 
\end{IEEEproof}

\section*{Proof of Lemma \ref{lem:Contraction operators error}}
\begin{IEEEproof}[Proof of Lemma \ref{lem:Contraction operators error}]
	Since$\|\widehat{T}J-TJ\|_{\infty}\leq\|\widehat{T}J-\widetilde{T}J\|_{\infty}+\|\widetilde{T}J-TJ\|_{\infty},$
	we need to bound terms $\|\widehat{T}J-\widetilde{T}J\|_{\infty}$
	and $\|\widetilde{T}J-TJ\|_{\infty},$ separately. First, we bound
	the term $\|\widehat{T}J-\widetilde{T}J\|_{\infty}$ in the following
	lemma.
	\begin{lemma}\label{lem5-1}
		Let $\varepsilon>0.$ Under Assumption \ref{assu:Risk_estimation}, we have
		\[
		\mathbb{P}\left(\|\widehat{T}J-\widetilde{T}J\|_{\infty}\leq\frac{\varepsilon}{2}\right)\geq1-n\left|\mathbb{A}\right|\theta\left(\frac{\varepsilon}{2\gamma},m\right).
		\]
	\end{lemma}
	\begin{IEEEproof}
		Fix $s\in\mathbb{S}$ and $J\in  B\left(\mathbb{S};J_{\max}\right),$
		we have
		\begin{equation*}
		\begin{aligned}
		&\left|[\widetilde{T}J]\left(s\right)-[\text{\ensuremath{\widehat{T}}}J]\left(s\right)\right|\\
		\leq &\left|\min_{a\in\mathbb{A}}\left\{ c(s',a)+\gamma\rho\left(J\left(Y^{s,a}\right)\right)\right\}-\min_{a\in\mathbb{A}}\left\{ c(s',a)+\gamma\hat{\rho}_{m}\left(\left\{J\left(Y_j^{s,a}\right)\right\}^m_{j=1}\right)\right\}\right|\\
		\leq & \gamma\max_{a\in\mathbb{A}}\left|\rho\left(J\left(Y^{s,a}\right)\right)-\hat{\rho}_{m}\left(\left\{J\left(Y_j^{s,a}\right)\right\}^m_{j=1}\right)\right|.
		\end{aligned}
		\end{equation*}
		where the first inequality follows from the definition of random operators
		$\widehat{T}$ and $\widetilde{T}$, and the second inequality is due
		to Fact \ref{Fact} 1). Under Assumption \ref{assu:Risk_estimation},
		we get
		\[
		\mathbb{P}\left(\left|\rho\left(J\left(Y^{s,a}\right)\right)-\hat{\rho}_{m}\left(\left\{J\left(Y_j^{s,a}\right)\right\}^m_{j=1}\right)\right|>\frac{\varepsilon}{2\gamma}\right)\leq\theta\left(\frac{\varepsilon}{2\gamma},m\right).
		\]
		Note that $|\mathcal{S}|=n$ and $$\max_{s\in\mathbb{S}}\left|[\widetilde{T}J]\left(s\right)-[\text{\ensuremath{\widehat{T}}}J]\left(s\right)\right|=\max_{s'\in\mathcal{S}}\left|[\widetilde{T}J]\left(s'\right)-[\text{\ensuremath{\widehat{T}}}J]\left(s'\right)\right|$$
		$\widetilde{T}J$
		is piecewise constant on $\left\{ B_{s}\right\} _{s\in\mathcal{S}}.$ We then obtain 
		\[
		\mathbb{P}\left(\|\widehat{T}J-\widetilde{T}J\|_{\infty}\leq\frac{\varepsilon}{2}\right)\geq1-n\left|\mathbb{A}\right|\theta\left(\frac{\varepsilon}{2\gamma},m\right)
		\]
		by a union bounding argument.
	\end{IEEEproof}
	
	Next we bound the term $\|\widetilde{T}J-TJ\|_{\infty}.$ 
	\begin{lemma}\label{lem5-2}
		Let $\varepsilon>0.$ Under Assumption \ref{assu:Epsilon_net_regularity} and \ref{assu:Epsilon_net_partition},
		if the $\epsilon-$net $\mathcal{S}$ is chosen such that
		\[
		\epsilon\leq\frac{\varepsilon}{2\left(\kappa_{c}+\gamma\kappa_{\mu}J_{\max}\right)},
		\]
		we have $$\|\widetilde{T}J-TJ\|_{\infty}\leq\frac{\varepsilon}{2}.$$
	\end{lemma}
	\begin{IEEEproof}
		We first show that $TJ$ is Lipschitz continuous with constant $\kappa_{c}+\gamma\kappa_{\mu}J_{\max}.$
		For $s,s'\in\mathbb{S}$ and $J\in  B\left(\mathbb{S};J_{\max}\right),$
		we have
		\begin{equation*}
		\begin{aligned}
			&   \left|\left[TJ\right]\left(s\right)-\left[TJ\right]\left(s'\right)\right|\\
			 \leq & \max_{a\in\mathbb{A}}\bigg|c(s,a)-c(s',a)+\gamma\max_{\mu\in\mathcal{Q}\left(s,a\right)}\int J\left(y\right)\mu\left(dy\right)
			 -\gamma\max_{\mu^{'}\in\mathcal{Q}\left(s',a\right)}\int J\left(y\right)\mu^{'}\left(dy\right)\bigg|\\
			 \leq & \max_{a\in\mathbb{A}}|c(s,a)-c(s',a)|
			 +\gamma\max_{a\in\mathbb{A}}\bigg|\max_{\mu\in\mathcal{Q}\left(s,a\right)}\int J\left(y\right)\mu\left(dy\right)
			 -\max_{\mu^{'}\in\mathcal{Q}\left(s',a\right)}\int J\left(y\right)\mu^{'}\left(dy\right)\bigg|\\
			 \leq & \kappa_{c}\|s-s'\|_\infty+\gamma\max_{a\in\mathbb{A}}\int|J\left(y\right)(\mu_{*}\left(dy\vert s,a\right)
			 -\mu_{*}^{'}\left(dy\vert s',a\right))|\\
			 \leq & \left(\kappa_{c}+\gamma\kappa_{\mu}J_{\max}\right)\|s-s'\|_{\infty}.
		\end{aligned}
		\end{equation*}
		The third inequality holds due to Assumption \ref{assu:Epsilon_net_regularity}
		1), $\mu_{*}\left(y\vert s,a\right)\in\arg\max_{\mu\in\mathcal{Q}\left(s,a\right)}\int J\left(y\right)\mu\left(dy\right)$
		and $\mu_{*}^{'}\left(y\vert s',a\right)\in\arg\max_{\mu^{'}\in\mathcal{Q}\left(s',a\right)}\int J\left(y\right)\mu^{'}\left(dy\right).$
		The last inequality is true because of Assumption \ref{assu:Epsilon_net_regularity}
		2) and Lemma \ref{lem:Bounded}. Recall that $\widetilde{T}J$
		is piecewise constant on $\left\{ B_{s}\right\} _{s\in\mathcal{S}}.$
		Under Assumption \ref{assu:Epsilon_net_partition}, we conclude
		\[
		\|\widetilde{T}J-TJ\|_{\infty}\leq\left(\kappa_{c}+\gamma\kappa_{\mu}J_{\max}\right)\epsilon.
		\]
		Upper bounding the RHS by $\varepsilon/2$
		yields the result. 
	\end{IEEEproof}
Combining Lemmas \ref{lem5-1} and \ref{lem5-2} gives the desired bound.
\end{IEEEproof}

\section*{Proof of Lemma \ref{lem:Stoc_approximation error}}
\begin{IEEEproof}[Proof of Lemma \ref{lem:Stoc_approximation error}]
	\begin{enumerate}
		\item First, we claim $\Vert \widehat{J}_{K}-J^{*}\Vert _{\infty}\leq\gamma^{K}J_{\max}+\sum_{k=0}^{K-1}\gamma^{K-k-1}\Vert\varepsilon_{k}\Vert _{\infty}$
		for $K\geq1.$ When $K=1$, we verify
		\begin{equation*}
		\begin{aligned}
		\|\widehat{J}_{1}-J^{*}\|_{\infty}&\leq\|T\widehat{J}_{0}-TJ^{*}+\varepsilon_{0}\|_{\infty}\\
		&\leq\gamma\|\widehat{J}_{0}-J^{*}\|_{\infty}+\|\varepsilon_{0}\|_{\infty}\\
		&\leq\gamma J_{\max}+\left\Vert \varepsilon_{0}\right\Vert _{\infty}.
		\end{aligned}
		\end{equation*}
		by Lemma \ref{lem:Contraction}. Assume that the claim holds for $K=t$
		\[
		\|\widehat{J}_{t}-J^{*}\|_{\infty}\leq\gamma^{t}J_{\max}+\sum_{k=0}^{t-1}\gamma^{t-k-1}\left\Vert \varepsilon_{k}\right\Vert _{\infty}.
		\]
		When $K=t+1,$ by induction, we have
		\begin{equation*}
		\begin{aligned}
		\|\widehat{J}_{t+1}-J^{*}\|_{\infty}&\leq\|T\widehat{J}_{t}-TJ^{*}+\varepsilon_{t}\|_{\infty}\\
		&\leq\gamma\|\widehat{J}_{t}-J^{*}\|_{\infty}+\|\varepsilon_{t}\|_{\infty}\\
		&\leq\gamma^{t+1}J_{\max}+\sum_{k=0}^{t}\gamma^{t-k}\left\Vert \varepsilon_{k}\right\Vert _{\infty}.
		\end{aligned}
		\end{equation*}
		
		Finally, if $\left\Vert \varepsilon_{k}\right\Vert _{\infty}\leq\varepsilon$
		for all $0\leq k<K,$ we obtain 
		\begin{equation*}
		\begin{aligned}
		\|\widehat{J}_{K}-J^{*}\|_{\infty}&\leq\gamma^{K}J_{\max}+\sum_{k=0}^{K-1}\gamma^{K-k-1}\varepsilon\\
		&\leq\gamma^{K}J_{\max}+\frac{\varepsilon}{1-\gamma}.
		\end{aligned}
		\end{equation*}
		\item From the proof of Lemma \ref{lem:Lp bounds}, under Assumption
		\ref{assu:Absolute_continuity} and if the approximation error in
		all\textbf{ }iterations $k=0,\dots,K-1$ falls below the tolerance
		$\left\Vert \varepsilon_{k}\right\Vert _{p,\mu}\leq\varepsilon$,
		we deduce
		\begin{equation*}
		\begin{aligned}
		&\Vert J^{\widehat{\pi}_{K}}-J^{*}\Vert _{p,\varrho}^{p}\\
		\leq & \left[\frac{2\gamma}{\left(1-\gamma\right)^{2}}\right]^{p}\big[(1-\gamma^{K+1})^{p-1}C_{\varrho,\mu}\varepsilon^{p}
		+\gamma^{K}(1-\gamma)(1-\gamma^{K+1})^{p-1}J_{\max}^{p}\big]\\
		\leq & \left[\frac{2\gamma}{(1-\gamma)^{2}}\right]^{p}\big[(C_{\varrho,\mu}^{1/p}\varepsilon)^{p}
		+(\gamma^{K/p}(1-\gamma)^{1/p}(1-\gamma^{K+1})^{1-1/p}J_{\max})^{p}\big].
		\end{aligned}
		\end{equation*}
		The second inequality follows from the fact that $(1-\gamma^{K+1})^{p-1}\leq1$
		for all $K\geq0$ (since $p\geq1$). Thus,
		\begin{equation*}
		\begin{aligned}
		\Vert J^{\widehat{\pi}_{K}}-J^{*}\Vert _{p,\varrho}
		\leq\frac{2\gamma}{(1-\gamma)^{2}}[C_{\varrho,\mu}^{1/p}\varepsilon
		+\gamma^{K/p}(1-\gamma)^{1/p}(1-\gamma^{K+1})^{1-1/p}J_{\max}].
		\end{aligned}
		\end{equation*}
	\end{enumerate}
\end{IEEEproof}

\section*{Proof Lemma \ref{lem:Stochastic dominance}}
\begin{IEEEproof}[Proof Lemma \ref{lem:Stochastic dominance}]
	Define a random variable 
	\[
	\mathfrak{Y}\left(\theta\right)=\begin{cases}
	\max\left\{ \theta-1,1\right\} , & \mbox{w.p. }p,\\
	K^{*}, & \mbox{w.p. }1-p,
	\end{cases}
	\]
	as a function of $\theta.$ It can be seen that $Y_{k+1}$ has the
	same distribution as $\left[\mathfrak{Y}\left(\Theta\right)|\Theta=Y_{k}\right].$
	Using \cite[Theorem 1.A.3(d)]{Shaked2007} and \cite[Theorem 1.A.6]{Shaked2007},
	the rest of the proof follows the proof of \cite[Theorem 4.1]{Haskell_EDP_2015}
	thus omitted.
\end{IEEEproof}

\section*{Proof of Lemma \ref{lem:Stationary distribution}}
\begin{IEEEproof}[Proof of Lemma \ref{lem:Stationary distribution} ]
	The stationary probabilities $\left\{ \mu\left(i\right)\right\} _{i=1}^{K^{*}}$
	satisfy the following set of equations
	\begin{eqnarray}
	&  & \mu\left(1\right)=p\mu\left(1\right)+p\mu\left(2\right),\label{eq:Stationary distribution-1}\\
	&  & \mu\left(i\right)=p\mu\left(i+1\right),\quad\forall i=2,\dots,K^{*}-1,\label{eq:Stationary distribution-2}\\
	&  & \sum_{i=1}^{K^{*}}\mu\left(i\right)=1.\label{eq:Stationary distribution-3}
	\end{eqnarray}
	From the recursive relation (\ref{eq:Stationary distribution-2}),
	we have 
	\[
	\mu\left(i\right)=p^{K^{*}-i}\mu\left(K^{*}\right),\quad\forall i=2,\dots,K^{*}-1,
	\]
	and from (\ref{eq:Stationary distribution-1}) we have
	\[
	\mu\left(1\right)=\frac{p}{1-p}\mu\left(2\right)=\frac{p^{K^{*}-1}}{1-p}\mu\left(K^{*}\right).
	\]	
	We can solve $\mu\left(K^{*}\right)$ using Equation (\ref{eq:Stationary distribution-3})
	\begin{eqnarray*}
		1 & = & \sum_{i=1}^{K^{*}}\mu\left(i\right)\\
		& = & \frac{p^{K^{*}-1}}{1-p}\mu\left(K^{*}\right)+\sum_{i=2}^{K^{*}}p^{K^{*}-i}\mu\left(K^{*}\right)\\
		& = & \left[\frac{p^{K^{*}-1}}{1-p}+\frac{1-p^{K^{*}-1}}{1-p}\right]\mu\left(K^{*}\right)\\
		& = & \frac{1}{1-p}\mu\left(K^{*}\right),
	\end{eqnarray*}
	which implies $\mu\left(K^{*}\right)=1-p.$ Therefore, 
	\[
	\mu\left(i\right)=p^{K^{*}-i}\mu\left(K^{*}\right)=\left(1-p\right)p^{K^{*}-i},\quad\forall i=2,\dots,K^{*}-1,
	\]
	and
	\[
	\mu\left(1\right)=p^{K^{*}-1}.
	\]
\end{IEEEproof}

\section*{Proof of Proposition \ref{prop:Asymptotic bound}}
\begin{IEEEproof}[Proof of Proposition \ref{prop:Asymptotic bound}]
	\begin{enumerate}
		\item From Lemma \ref{lem:Stochastic dominance} and \ref{lem:Stationary distribution},
		and the definition of $p$, we have
		\begin{equation*}
		\begin{aligned}
		\mathbb{P}\left(\|\widehat{J}_{k}-J^{*}\|_{\infty}>\epsilon_{g}\right)
		\leq&\mathcal{\mathcal{Q}}\left(Y>1\right)\\
		=&1-\mu\left(1\right)\\
		=&1-\left(1-\delta\right)^{K^{*}-1}.
		\end{aligned}
		\end{equation*}
		Let the RHS be less than or equal to $\delta_{1},$ and we have $1-\delta_{1}\leq\mu\left(1\right)=\left(1-\delta\right)^{K^{*}-1}\leq1-\delta.$
		Therefore, by Lemma \ref{lem:Contraction operators error}, we choose
		$\varepsilon<\epsilon_{g}.$ Furthermore, $\epsilon$ and $m$
		should be selected such that
		\[
		\epsilon\leq\frac{\varepsilon}{2\left(\kappa_{c}+\gamma\kappa_{\mu}J_{\max}\right)}
		\]
		and 
		\[
		\theta\left(\frac{\epsilon_{g}}{2\gamma},m\right)\leq\frac{\delta_{1}}{\left|\mathbb{A}\right|\left|\mathcal{S}\right|}.
		\]
		\item From Lemma \ref{lem:Stochastic dominance} and \ref{lem:Stationary distribution},
		and the definition of $p$, we have
		\[
		\mathbb{P}\left(\|J^{\widehat{\pi}_{k}}-J^{*}\|_{p,\varrho}>\epsilon_{g}\right)\leq\mathcal{\mathcal{Q}}\left(Y>1\right)=1-\mu\left(1\right).
		\]
		Let the RHS be less than or equal to $\delta_{1},$ and we have $1-\delta_{1}\leq\mu\left(1\right)=\left(1-\delta\right)^{K^{*}-1}\leq1-\delta.$
		Therefore, by Lemma \ref{lem:Error each iteration}, we choose $\varepsilon<\epsilon_{g}-d_{p,\mu}\left(T\mathcal{F},\mathcal{F}\right).$
		Furthermore, $n,m$ should be selected such that
		\[
		n>128\left(\frac{8J_{\max}}{\varepsilon}\right)^{2p}\left(\log\left(1/\delta_{1}\right)+\log\left(32\mathcal{N}_{0}\left(n\right)\right)\right)
		\]
		where $$\mathcal{N}_{0}\left(n\right)=\mathcal{N}\left(\frac{1}{8}\left(\frac{\varepsilon}{4}\right)^{p},\mathcal{F},n,\mu\right)$$
		and 
		\[
		\theta\left(\varepsilon/4,m\right)\leq\frac{\delta_{1}}{4n\left|\mathbb{A}\right|}.
		\]
	\end{enumerate}
\end{IEEEproof}

\section*{Proof of Lemma \ref{lem:Mixing time}}
\begin{IEEEproof}[Proof of Lemma \ref{lem:Mixing time}]
	The proof follows that of \cite[Lemma 5.1]{Haskell_EDP_2015}. The
	transition matrix $\mathfrak{Q}\in\mathbb{R}^{K^{*}\times K^{*}}$
	of the Markov chain $\{ Y_{k}\} _{k\geq0}$ has the form 
	\begin{equation*}
	\mathfrak{Q}=\left(\begin{IEEEeqnarraybox*}[][c]{,c/c/c/c/c/c,}
	p & 0 & 0 & \dots & 0 & 1-p\\
	p & 0 & 0 & \dots & 0 & 1-p\\
	0 & p & 0 & \dots & 0 & 1-p\\
	\vdots & \vdots & \vdots & \ddots & \vdots & \vdots\\
	0 & 0 & 0 & \dots & 0 & 1-p\\
	0 & 0 & 0 & 0 & p & 1-p %
	\end{IEEEeqnarraybox*}\right).
	\end{equation*}
	
	We claim the eigenvalues $\lambda$ of $\mathfrak{Q}$ are $0$ and
	$1$. To see this, suppose $\lambda\neq0$ and $\mathfrak{Q}x=\lambda x$
	for some nonzero $x=\left(x_{1},x_{2},\dots,x_{K^{*}}\right)\in\mathbb{R}^{K^{*}}.$
	The first and second equalities of linear system $\mathfrak{Q}x=\lambda x$
	are
	\begin{equation*}
	\begin{aligned}
		&   \lambda x_{1}=px_{1}+\left(1-p\right)x_{K^{*}},\\
		&   \lambda x_{2}=px_{1}+\left(1-p\right)x_{K^{*}}.
	\end{aligned}
	\end{equation*}
	This implies $x_{2}=x_{1}.$ The third equality of linear system $\mathfrak{Q}x=\lambda x$
	is
	\[
	\lambda x_{3}=px_{2}+\left(1-p\right)x_{K^{*}}=px_{1}+\left(1-p\right)x_{K^{*}}=\lambda x_{2},
	\]
	which implies $x_{3}=x_{2}.$ Continuing this reasoning inductively,
	we have $x_{1}=x_{2}=\dots=x_{K^{*}}$ for any eigenvector $x$ of
	$\mathfrak{Q}.$ Therefore, it is true that the eigenvalues $\lambda$
	of $\mathfrak{Q}$ are $0$ and $1$. By \cite[Theorem 12.3]{levin2009markov},
	we have
	\[
	t_{mix}\left(\delta_{2}\right)\leq\log\left(\frac{1}{\delta_{2}\mu_{\min}}\right)\frac{1}{1-\lambda_{*}},
	\]
	where $\lambda_{*}=\max\left\{ \left|\lambda\right|:\lambda\,\,\text{is an eigenvalue of }\mathfrak{Q},\,\lambda\neq1\right\}=0.$
	Plugging in $\lambda_{*}$ gives the desired result.
\end{IEEEproof}

\section*{Proof of Proposition \ref{prop:Non-asymptotic bound}}
\begin{IEEEproof}[Proof of Proposition \ref{prop:Non-asymptotic bound}]
	For $$k\geq\log\left(\frac{1}{\delta_{2}\mu_{\min}}\right)\geq t_{mix}\left(\delta_{2}\right),$$
	we have
	\[
	d\left(k\right)=\frac{1}{2}\sum_{i=1}^{K^{*}}\left|\mathcal{Q}_{k}\left(Y_{k}=i\right)-\mu\left(i\right)\right|\leq\delta_{2},
	\]
	which implies
	\[
	\mathcal{Q}_{k}\left(Y_{k}=1\right)\geq\mu\left(1\right)-2\delta_{2}.
	\]
	
	Therefore, from Lemma \ref{lem:Stochastic dominance}, we have
	\begin{equation*}
	\begin{aligned}
	\mathbb{P}\left(\|\widehat{J}_{k}-J^{*}\|_{\infty}>\epsilon_{g}\right)
	\leq&\mathcal{Q}_{k}\left(Y_{k}>1\right)\\
	=&1-\mathcal{Q}_{k}\left(Y_{k}=1\right)\\
	\leq&1+2\delta_{2}-\mu\left(1\right)
	\end{aligned}
	\end{equation*}
	and
	\begin{equation*}
	\begin{aligned}
	\mathbb{P}\left(\|J^{\widehat{\pi}_{k}}-J^{*}\|_{p,\varrho}>\epsilon_{g}\right)
	\leq&\mathcal{Q}_{k}\left(Y_{k}>1\right)\\
	=&1-\mathcal{Q}_{k}\left(Y_{k}=1\right)\\
	\leq&1+2\delta_{2}-\mu\left(1\right).
	\end{aligned}
	\end{equation*}
\end{IEEEproof}

%
%
%
%

\ifCLASSOPTIONcaptionsoff
  \newpage
\fi

\end{document}